\documentclass[preprint,11pt]{elsarticle}
\usepackage[hmargin=2.5cm,vmargin=2.5cm]{geometry}
\usepackage{array}

\usepackage{amssymb}
\usepackage{amsthm}
\usepackage{amsmath}
\usepackage{tcolorbox}
\usepackage{bm}
\usepackage{hyperref}
\usepackage{longtable}
\usepackage{caption}
\hypersetup{
    colorlinks=true,
    linkcolor=blue,
    filecolor=magenta,      
    urlcolor=cyan,
    pdftitle={Overleaf Example},
    pdfpagemode=FullScreen,
    }

\journal{Journal of Mechanics and Physics of Solids}

\begin{document}

\begin{frontmatter}



\title{A mechanically-derived contact model for adhesive elastic-perfectly plastic particles. Part I: Utilizing the method of dimensionality reduction}


\author[inst1]{William Zunker}

\affiliation[inst1]{organization={Massachusetts Institute of Technology},%Department and Organization
            addressline={77 Massachusetts Ave}, 
            city={Cambridge},
            postcode={02319}, 
            state={MA},
            country={USA}}

\author[inst1]{Ken Kamrin\corref{cor1}}
            \ead{kkamrin@mit.edu}
            \cortext[cor1]{Corresponding author.}

\begin{abstract}
In this two part series~\cite{zunker2023partII}, we present a contact model able to capture the response of interacting adhesive elastic-perfectly plastic particles under a variety of loadings. In Part I, we focus on elastic through fully-plastic contact with and without adhesion.  For these contact regimes the model is built upon the method of dimensionality reduction which allows the problem of a 3D axisymmetric contact to be mapped to a semi-equivalent problem of a 1D rigid indenter penetrating a bed of independent Hookean springs. Plasticity is accounted for by continuously varying the 1D indenter profile subject to a constraint on the contact pressure. Unloading falls out naturally, and simply requires lifting the 1D indenter out of the springs and tracking the force. By accounting for the incompressible nature of this plastic deformation, the contact model is able to capture multi-neighbor dependent effects such as increased force and formation of new contacts. JKR type adhesion is recovered seamlessly within the method of dimensionality reduction by simply allowing the springs to `stick’ to the 1D indenter's surface. Because of the mechanics-focused formulation of the contact model, only a few physical inputs describing the interacting particles are needed: particle radius, Young's modulus, Poisson ratio, yield stress, and effective surface energy. The contact model is validated against finite element simulations and analytic theory---including Hertz's contact law and the JKR theory of adhesion. These comparisons show that the proposed contact model is able to accurately capture plastic displacement, average contact pressure, contact area, and force as a function of displacement for contacts as well as particle volume within the elastic to fully-plastic regimes.
\end{abstract}

%\includegraphics[width=\textwidth, trim = 0cm  0cm 0cm 0cm]{grabs_I.pdf}

%    \item Contact model is developed for adhesive elastic-perfectly plastic spherical particles.
%    \item Model is mechanically-derived from the method of dimensionality reduction.
%    \item Model parameters are all physical (e.g. Young’s modulus, Poisson ratio etc.).
%    \item Force, area, and average-pressure at contacts as well as particle volume are all independently tracked.
%    \item Numerical implementation suitable for the DEM is outlined.

\begin{keyword}
 B. contact mechanics \sep A. powder compaction \sep B. elastic–plastic material \sep A. adhesion  \sep method of dimensionality reduction 
\PACS 0000 \sep 1111
\MSC 0000 \sep 1111
\end{keyword}

\end{frontmatter}


\section{Introduction} \label{Introduction}

Powder compaction is ubiquitous in a variety of applications: metallurgy~\cite{samal2015powder}, ceramics~\cite{sigmund2000novel}, additive manufacturing~\cite{meier2019modeling}, pharmaceutics~\cite{ccelik2016pharmaceutical}, and more. The objective is often to form powders into an (effectively) continuous medium with an acceptable green strength~\cite{laptev2005green}. This requires high pressures that lead to significant non-reversible plastic deformation~\cite{boudina2022insight} and large adhesive forces~\cite{mashadi1987characterization,wu2006predicting}. Due to its widespread relevance, understanding and predicting the behavior of compacted particles is paramount.     

Numerical simulation offers a low cost and flexible alternative to experimentation when it comes to understanding powder compaction. A pervasive numerical technique for simulating powders or other granular material is the discrete element method (DEM)~\cite{cundall1979discrete}; it evolves each particle in the system independently by integrating its Newton's equations of motion. The total force on each particle is found through the summation of body and surfaces forces. In many applications, the body force is gravity leading to a simple treatment. Surface forces, relating to contact between neighboring particles, on the other hand are complex, and have led to a myriad of models that attempt to account for various effects: friction~\cite{cundall1979discrete,walton1986viscosity,thornton1991impact}, rolling friction~\cite{iwashita1998rolling,zhou1999rolling}, torsion~\cite{dintwa2005torsion}, adhesion~\cite{johnson1971surface,maugis1992adhesion,soulie2006influence}, viscoelasticity~\cite{olsson2019contact} and plasticity~\cite{storaakers1997similarity}. 

In this work, we restrict our focus to modeling normal contact\footnote{Normal contact models typically describe the contact force as a function of relative overlap or displacement between the center points of the two particles.} between adhesive elastic-perfectly plastic spherical particles, setting aside other phenomena. The simplest normal contact models are the linear spring model and Hertz's contact law~\cite{hertz1882beruhrung}\footnote{For a modern revisit to the derivation of Hertz's contact law see~\cite{mowlavi2021contact}.}. Despite their utility, both are limited to elastic contacts under small deformations. Applications involving powder compaction routinely lead to significant plastic deformation and adhesive forces requiring a more robust contact model.  

\begin{figure*} [!htb]
 	\centering
 	%\raggedright
 	% Trim{LEFT LOWER RIGHT UPPER}
 	\includegraphics[width=\textwidth, trim = 0.5cm  4.5cm 0cm 0.5cm]{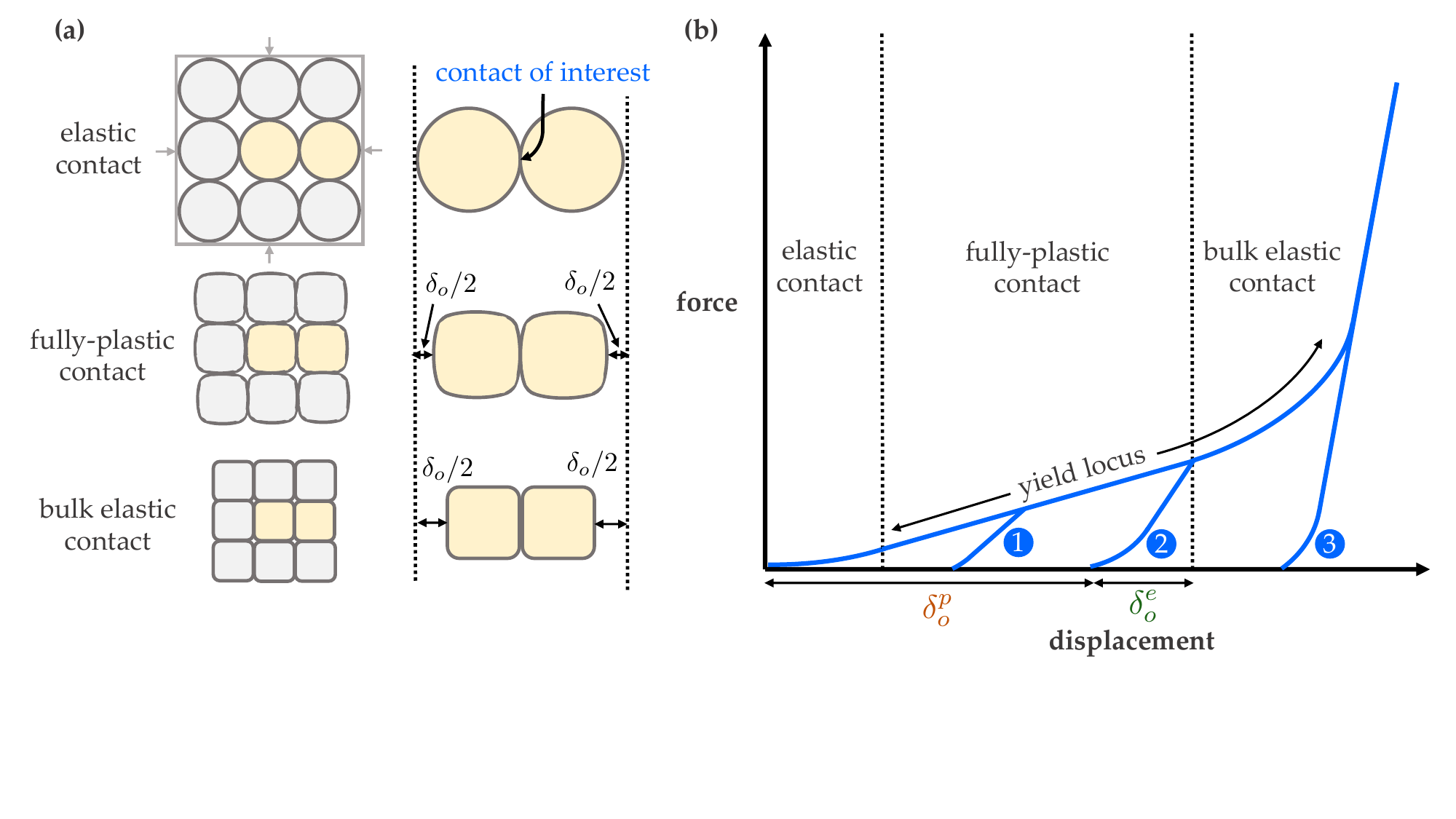}
 	\caption{Qualitative description of the compaction of a collection of elastic-plastic particles. (a) Uniform compression is applied and three loading regimes are shown: elastic, fully-plastic, and bulk elastic. (b) Generic force-displacement curve corresponding to the the contact of interest. Curves 1, 2, and 3 correspond to unloading. The partitioning of displacement between plastic and elastic components is shown for unloading curve 2.}
 	\label{contact_regimes}
 \end{figure*}

To appreciate the complexity of contact during powder compaction, consider a collection of spherical non-adhesive elastic-perfectly plastic particles subjected to uniform compaction as shown in  Fig.~\ref{contact_regimes}(a) with an associated force-displacement curve for one contacting pair shown in Fig.~\ref{contact_regimes}(b). 

\noindent Under uniform compression the particles will move through three loading regimes: elastic, fully-plastic, and bulk elastic\footnote{This loading regime classification is slightly simplified and neglects two smaller transitory regimes commonly identified in literature: the elastic-plastic regime~\cite{johnson1970correlation,kogut2002elastic} that comes after the elastic regime as well as the low compressibility regime~\cite{tsigginos2015force} that comes after the fully-plastic regime.}. Although not shown in Fig.~\ref{contact_regimes}(b) adhesion can also be included and is often present; this alters the unloading curves and causes a tensile force to develop before separation. 

In the elastic regime, no plastic deformation occurs and all deformations are recoverable.  Importantly, each contact can be approximated to have no affect on its neighbors. These conditions mean that the force at each contact is well described by Hertz's contact law.

The fully-plastic regime is characterized by blunting of the surface, significant plastic deformation, unique unloading curves from forward loading\footnote{Forward loading is intended to describe all loading paths in the elastic regime and only the loading paths along the yield locus in the fully plastic regime.} that depend on the accumulated plastic deformation, and partitioning of the displacement $\delta_o$ into an elastic $\delta^e_o$ and plastic $\delta^p_o$ component as shown in Fig.~\ref{contact_regimes}(b). Another important feature of the fully-plastic regime is that the assumption of independent contacts no longer holds, and the force at each contact becomes \textit{multi-neighbor dependent}---meaning it is dependent on the other active contacts. Secondary contacts can also become significant in this regime, caused by radial expansion of the particle even without relative displacement between particle centers.   

Further compaction leads to the bulk elastic regime caused by a drastic decrease in pore space as the material tends towards a continuous medium. This gives rise to a bulk elastic response causing a sharp uptick in the measured contact force. The behavior in this regime is the focus of Part II.

Many efforts have been made to create contact laws describing part, or all of, the force-displacement curve Fig.~\ref{contact_regimes}(b) as detailed in Table~\ref{previous_contact_laws}\footnote{Work focusing on impact between elastic-plastic particles~\cite{thornton1991impact,thornton1998theoretical,stronge2000contact,zhang2002modeling,du2009energy,jackson2010predicting,wu2022energy,thornton2017elastic} is omitted.}. In addition to the authors and publication date, there is information about important features of the contact model: the regimes it spans (E = elastic, FP = fully-plastic, and BE = bulk-elastic), the minimum ratio of Young's modulus to yield stress $E/Y$ actually validated (n.a. = not applicable and n.r. = not reported), ability to unload, ability to model adhesion, and if multi-neighbor dependent effects are accounted for. 

%~\cite{chang1987elastic,storaakers1997similarity,mesarovic2000adhesive,zhao2000asperity,jackson2003finite,jackson2005finite,etsion2005unloading,harthong2009modeling,zait2010unloading,brake2012analytical,gonzalez2012nonlocal,agarwal2018contact,olsson2013force,frenning2013towards,frenning2015towards,brodu2015multiple,rathbone2015accurate,garner2018study,edmans2020numerical,edmans2020unloading,giannis2021modeling}

 A wide array of capabilities in the models are displayed, with only two models, Garner et al.~\cite{garner2018study} and Giannis et al.~\cite{giannis2021modeling} covering all regimes in forward loading as well as including treatment of unloading, adhesion, and multi-neighbor dependent effects. However, both of these models are empirically fitted and require refitting of the parameters for each new material. The information the models produce is also limited to force-displacement---contact area, contact pressures, and particle volume are not tracked. Additionally, adhesion in both models is accounted for through simplistic spring constants that do not correspond to any well accepted theory of adhesion (e.g. JKR theory of adhesion~\cite{johnson1971surface}). 

One other broad observation is that the minimum $E/Y$ validated (excluding purely elastic contact models) for the the majority of models was $E/Y>100$. It will be demonstrated that beyond $E/Y>100$ there is rapidly diminishing changes in the overall contact behavior since it approaches the rigid-plastic limit. This leaves the space of $E/Y<100$ untapped by most models which shows markedly different behavior due to the elastic deformations being comparable to the plastic deformations.

It is clear from reviewing prior work that no existing contact model is able to fully capture the interaction of elastic-plastic contacts. Filling this gap is important to many industrial applications, making it a worthwhile pursuit to develop a contact model that includes the following features:

\begin{enumerate}
  \item Analytical formulation based primarily on principles from the continuum mechanics of solids.
  \item Model inputs are commonly measured material parameters: particle radius, Young's modulus, Poisson ratio, yield stress, and effective surface energy.
  \item Treats all three regimes: elastic, fully-plastic, and bulk elastic. 
  \item Captures unloading curves without additional or special treatment.
  \item Treats adhesive contact in a manner consistent with a well-established theory of adhesion.
  \item Accounts for multi-neighbor dependent effects, such as increased force and formation of new contacts, caused by interaction of existing contacts.
  \item Tracks the evolution of plastic deformation, contact area, volume of particles, and contact pressures.
  \item Allows for large variations in the ratio between the Young's modulus and yield stress (i.e. no assumption of nearly rigid-plastic).
  \item Allows for differing radii between contacting particles. 
  \item Accounts for material hardening effects.
\end{enumerate}

\begin{table}[ht]
\begin{center}
\begin{tabular}{|| m{11.5em} | m{4.5em} | c | c | c | >{\centering\arraybackslash}m{4.5em} ||}
\hline 
Authors & Regimes & min($E/Y$) & Unloading & Adhesion & Multi-neighbor dependent \\ [0.5ex] 
\hline\hline
Chang et al.~\cite{chang1987elastic}, 1987 & E, FP & 1000 & no & no & no \\ 
\hline
Stor{\aa}kers et al.~\cite{storaakers1997similarity}, 1997 & FP & rig. plas. & no & no & no  \\ 
\hline
Mesarovic and Johnson~\cite{mesarovic2000adhesive}, 2000 & FP & 10000 & yes & yes & no  \\
\hline
Zhao et al.~\cite{zhao2000asperity}, 2000 & E, FP & 1000 & no & no & no  \\
\hline
Jackson and Green~\cite{jackson2003finite, jackson2005finite}, 2003 and 2005 & E, FP & 1000 & no & no & no  \\
\hline
Etsion et al.~\cite{etsion2005unloading}, 2005 & E, FP & 300  & yes & no & no  \\
\hline
Harthong et al.~\cite{harthong2009modeling}, 2009 & E, FP, BE & 1000  & no & no & yes  \\
\hline
Zait et al.~\cite{zait2010unloading}, 2010 & E, FP & 500  & yes & no & no  \\
\hline
Brake~\cite{brake2012analytical}, 2012 & E, FP & 300  & yes & no & no  \\
\hline
Gonzalez et al.~\cite{gonzalez2012nonlocal, agarwal2018contact}, 2012 and 2018 & E & n.a.  & yes & no & yes  \\
\hline
Olsson and Larsson~\cite{olsson2013force}, 2013 & E, FP & 1000 & yes & yes & no  \\
\hline
Frenning~\cite{frenning2013towards,frenning2015towards}, 2013 and 2015 & E, FP, BE & 50  & no & no & yes  \\
\hline
Brodu et al.~\cite{brodu2015multiple}, 2015 & E & n.a.  & yes & no & yes  \\
\hline
Rathbone et al.~\cite{rathbone2015accurate}, 2015 & E, FP & 160 & yes & no & no  \\
\hline
Garner et al.~\cite{garner2018study}, 2018 & E, FP, BE & 100  & yes & yes & yes  \\
\hline
Gonzalez~\cite{gonzalez2019generalized}, 2019 & E, FP & 100 & yes & yes & no  \\
\hline
Edmans and Sinka~\cite{edmans2020numerical,edmans2020unloading}, 2020 & E, FP & 1 & yes & no & no  \\
\hline
Giannis et al.~\cite{giannis2021stress}, 2021 & E & n.a. & yes & no & yes  \\
\hline
Giannis et al.~\cite{giannis2021modeling}, 2021 & E, FP, BE & n.r. & yes & yes & yes  \\ 
\hline 
Zhang et al.~\cite{zhang2022research}, 2022 & E, FP & 79.4 & no & no & no  \\ 
\hline 
Xu and Zhu~\cite{xu2023new}, 2023 & E, FP & 400 & no & no & no  \\ 
\hline 
\end{tabular}
\end{center}

\captionsetup{skip=0.1pt} 
\caption{Previous contact models involving elastic or elastic plastic particles. A few important features of the contact models are highlighted beginning with the regimes it covers: elastic (E), fully-plastic (FP), and bulk elastic (BE). The minimum ratio of Young's modulus to yield stress $E/Y$ for which the contact model was actually validated either by direct comparison to FEM or experiment is also listed (n.a. = not applicable and n.r. = not reported). The final three columns detail the capabilities of the models to model unloading, adhesion, and multi-neighbor dependent effects. }
\label{previous_contact_laws}
\end{table}

Over this two part series~\cite{zunker2023partII}, we will develop a contact model that includes items 1-8. Item 9 \textbf{is not} directly addressed in this work since only single particle compressions are considered, but is the subject of active research. Item 10 \textbf{is not} considered in this work (only elastic-perfectly plastic materials are), but is envisioned to be treatable within the framework of the presented contact model. Part I will lay the primary foundation of the contact model from the Method of Dimensionality reduction (MDR) described in Section~\ref{The method of dimensionality reduction}. Finite element simulations involving contact between an elastic-perfectly plastic particle and rigid flats are presented in Section~\ref{FEM_section}. Using the simulation results for guidance it will be demonstrated in Sections~\ref{Non-adhesive elastic-plastic contact model}-\ref{Contact model summary} that the MDR contains an array of useful features that lead to the development of a comprehensive contact model only lacking a bulk elastic response---addressed in Part II. Direct comparison of the contact model to the finite element simulations is carried out in Section~\ref{Verification of MDR contact model against finite element simulations}. To assist in usage a sketch of the numerical implementation of the contact model is included in the~\ref{Sketch of numerical implementation}.    

\section{Essential physical ideas of the contact model}

The full unabbreviated development of the contact model requires significant discussion\footnote{The need for significant discussion is twofold: the method of dimensionality reduction is not commonly known necessitating a pedagogical introduction and some finite element results need to be presented to corroborate certain modeling decisions.}. It is therefore worthwhile to qualitatively summarize the important physical constituents of the contact model.   

The contact model is built upon the method of dimensionality reduction (MDR), a powerful contact modeling technique that allows 3D elastic axisymmetric contact problems to be mapped through a set of integral transforms to a semi-equivalent simplified problem in 1D space. Importantly, the MDR allows the combination of pressure profile and contact area in 3D space to be exactly correlated to the unloaded 3D shape of an elastic indenter. Thus, if the pressure profile and contact area are known at all stages during elastic-plastic contact, and we simply think of elastic-plastic contact as a sequence of elastic contacts, the MDR can be used to determine the unloaded shape allowing complete knowledge of the contact behavior at all times. Fundamental to this solution methodology is the fact that the pressure profile and contact area  evolution between contacting elastic-plastic spheres is relatively simple. In particular:  

\begin{itemize}
    \item Finite element simulations show that the contact pressure profile is reasonably approximated as uniform in the fully-plastic regime, meaning it is describable through a scalar value that is found to have an evolution that falls onto a hardening curve defined by the rigid-plastic limit. The initial value of the curve matches the expected pressure seen in a hardness test and the large indentation value is given simply as the yield stress since the contact begins to look more like uniaxial compression\footnote{Because of these two analytical limits \textit{the only} fitted information that is incorporated into our model is how the average pressure evolves between these two limits, which comes from the finite element simulations.}.
    \item The contact area evolution matches well with a completely analytic form that transitions from Hertz's law to a simple geometric deformation picture that respects the incompressibility of plastic deformation in the fully-plastic regime.
\end{itemize}

\noindent As stated, these simplified forms for the pressure profile and contact area evolution can then be fed to the MDR to determine the unloaded indenter shape at all instances even as plastic deformation is accumulated. With the unloaded indenter shape specified at all moments, it is straightforward using the MDR, yet again, to find the force response during unloading and in the presence of adhesion. Adhesion (and unadhesion) is straightforward to capture using MDR since the transformed 1D space simplifies the application of Griffiths-based fracture mechanics ideas that give rise to, for example, the JKR theory of adhesion. The actual formula for contact force is ultimately of closed form  and depends on only three parameters: the apparent overlap $\delta$, the maximum experienced apparent overlap $\delta_\textrm{max}$, and the apparent radius\footnote{The terminology \textit{apparent} radius is meant to indicate that the particle radius may grow during loading due to inelastic effects. This growth naturally leads to the definition of the apparent overlap and the maximum experienced apparent overlap. Formal definitions and further discussion will be given shortly.} $R$   

\begin{equation} \label{Fbox}
    \boxed{F = \hat{F}(\delta,\delta_\textrm{max}, R).} 
\end{equation}

\noindent This ability to capture the full force-displacement curve with one consistent analytical framework is the major benefit of utilizing the MDR. Moreover, because of the mechanics-focused development of the contact model, it is able to seamlessly handle a wide array of material properties all while providing greater information about the contact extending beyond just the force. For example, one can track the plastic displacement at the contact, average contact pressure, contact area, and particle volume as a function of displacement.      

\section{The method of dimensionality reduction for elastic axisymmetric contact} \label{The method of dimensionality reduction}

\subsection{Normal contact without adhesion} \label{Normal contact without adhesion}

The method of dimensionality reduction (MDR), developed by Popov and He{\ss} in 2013 \cite{popov2013methode,popov2015method}\footnote{Popov and He{\ss} attribute much of the mathematical machinery involved in the MDR to the earlier works of F{\"o}ppl and Schubert~\cite{foppl1941elastische,schubert1942frage}}, is a powerful technique that enables great simplification of a vast number of contact mechanics problems. Central to the MDR, is the framework it provides to transform suitable 3D contact problems to semi-equivalent 1D problems of a rigid indenter penetrating a bed of independent Hookean springs (Winkler foundation). A wide array of contact phenomena have been shown to admit transformation to this simplified space: normal contact, adhesion, tangential contact, viscoelastic contact, and contact of functionally graded materials. For those interested in a detailed discussion of all these applications see \cite{popov2019handbook}. Here, we present only an overview of the essential results pertinent to the development of the contact model. 

\begin{figure*} [!htb]
 	%\centering
 	\raggedright
 	% Trim{LEFT LOWER RIGHT UPPER}
        % scale = 0.28
 	\includegraphics[width=\textwidth, trim = 0.5cm  2cm 3cm 0.5cm]{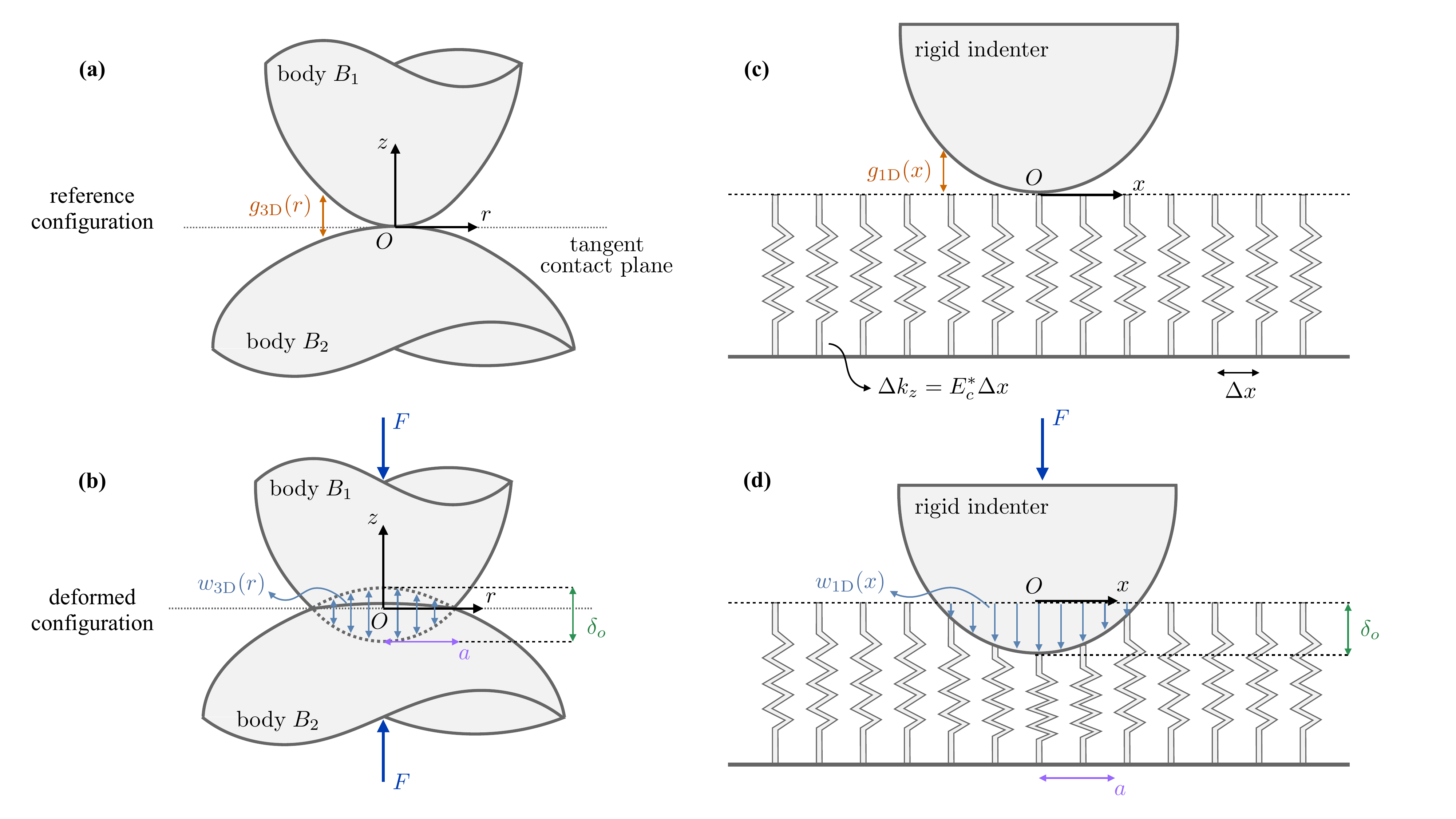}
 	\caption{General axisymmetric normal contact problem. The left column shows contact in the unmapped space between two 3D elastic bodies. The right column shows the corresponding semi-equivalent mapped problem of a plane rigid indenter penetrating a 1D bed of independent Hookean springs defined using the method of dimensionality reduction (MDR). (a) Reference configuration for the axisymmetric contact between two 3D elastic bodies $B_1$ and $B_2$. The initial gap function $g_\textrm{3D}(r)$ is taken as the distance between the individual profiles of each body. (b) In the deformed configuration, application of the force $F$ leads to a total relative displacement $\delta_o$ between the two bodies. Additionally, a finite contact radius $a$, and normal surface displacement profile $w_\textrm{3D}(r)$ are formed. (c) The reference configuration for the semi-equivalent 1D problem defined using the MDR. The transformed initial gap function $g_\textrm{1D}(x)$ is measured as the distance from the indenter's surface to the top of the spring foundation. (d) The same force $F$ is applied to the rigid indenter causing it to penetrate the springs a distance $\delta_o$ and leading to a finite contact radius $a$, where $\{\delta_o, a\}_\textrm{1D} = \{\delta_o, a\}_\textrm{3D}$. A normal displacement profile $w_\textrm{1D}$ also arises, however $w_\textrm{1D} \neq w_\textrm{3D}$.   }
 	\label{MDR_cheat_sheet}
 \end{figure*}

The MDR provides exact solutions for contact between any axially symmetric elastic bodies as shown in Fig.~\ref{MDR_cheat_sheet}(a). In outlining the solution process, we will restrict our attention to this class of contact problems. Initially, we will consider the non-adhesive case, but then soon after show how adhesion can be easily treated within the MDR framework. The standard entry point for the application of the MDR consists of two preparatory steps:

\begin{enumerate}
  \item The elastic bodies are replaced by a bed of independent Hookean springs as shown in Fig.~\ref{MDR_cheat_sheet}(c). The springs are separated by a distance $\Delta x$ and have a spring constant

  \begin{equation} \label{Deltakz}
	\Delta k_z = E^*_c \Delta x,
\end{equation}

where $E^*_c$ is the composite plane strain modulus of the two bodies $B_1$ and $B_2$, each with their respective Young's modulus $E_i$ and Poisson's ratio $\nu_i$

  \begin{equation} \label{Ecomposite}
	E^*_c = \left( \frac{1-\nu_{1}^2}{E_{1}} + \frac{1-\nu_{2}^2}{E_{2}} \right)^{-1}.
\end{equation}

\item The initial gap function $g_\textrm{3D}(r)$ is transformed to a plane profile $g_\textrm{1D}(x)$ via 

  \begin{equation} \label{g1Dtrans}
	g_\textrm{1D}(x) = |x|\int_{0}^{|x|} \frac{g^\prime_\textrm{3D}(r)}{\sqrt{x^2-r^2}} \,dr,
\end{equation}

where $\prime$ indicates the derivative. This is known as the `Abel transform'. For completeness we note that the inverse transform exists and is

  \begin{equation} \label{g3Dtrans}
	g_\textrm{3D}(r) = \frac{2}{\pi}\int_{0}^{r} \frac{g_\textrm{1D}(x)}{\sqrt{r^2-x^2}} \,dx.
\end{equation}

\end{enumerate}

\noindent These two steps complete the transformation of the 3D axisymmetric contact problem to that of a plane rigid indenter penetrating a linearly elastic foundation. Evolution of the contact problem can now be carried out entirely in the 1D space. 

We now imagine a force controlled experiment in which a force $F$ is applied to the plane rigid indenter. This force causes the indenter to penetrate into the foundation a distance $\delta_o$, creating a contact whose half-width (or ``contact radius'') is $a$ as shown in Fig.~\ref{MDR_cheat_sheet}(d). The normal surface displacement profile $w_\textrm{1D}(x)$ of the foundation is given by the difference of the displacement and the profile shape
  \begin{equation} \label{w1D}
	w_\textrm{1D}(x) = \delta_o - g_\textrm{1D}(x).
\end{equation}

\noindent The surface displacement profile has its maximum at $x=0$ where it is equal to the displacement, is required for non-adhesive problems to be zero at $x=\pm a$, and is defined to be zero outside of the contact radius $|x| > a$. These conditions gives a simple formula relating the contact radius and displacement: $0 = \delta_o - g_\textrm{1D}(a)$.

The normal displacement profile causes compression of the foundation's springs. At a given location $x$ the force in the spring is given as $\Delta F_s(x) = \Delta k_z w_\textrm{1D}(x) = E^*_c w_\textrm{1D}(x) \Delta x$. Balance of forces requires that the sum of forces in the springs be equal to the total applied force. In the limit where spring spacing becomes very small $\Delta x \rightarrow dx$ the sum becomes an integral. Thus, balance of forces is equivalent to
  \begin{equation} \label{Fw1D}
	F =  E^*_c\int_{-a}^{a} w_\textrm{1D}(x) \,dx.
\end{equation}

\noindent From (\ref{Fw1D}) it is natural and useful to define a 1D linear force density 
\begin{equation} \label{q1D}
    q_\textrm{1D}(x) = E^*_c w_\textrm{1D}(x), 
\end{equation} 

The remarkable property of the MDR is that the displacement and contact radius that arise from the applied force in the transformed space, are exactly equal to those that would arise in the 3D elastic problem for the same applied force. In the 3D elastic problem, $a$ is the true radius of the disk-shaped contact. Conversely, if we consider a kinematically controlled experiment by imposing a displacement, the force and contact radius that arise in the transformed space would again correspond exactly to those that would arise in the 3D space for the same displacement. In other words, the triplet of physical quantities: force, contact radius, and displacement are shared between the 3D and transformed spaces. This powerful result means that the contact problem after transformation can be treated entirely in the transformed space whilst still providing the essential contact properties that arise in 3D space.

Akin to the initial gap function transforms, two other sets of important transforms between the 3D and 1D spaces exist for both contact pressures and surface displacement profiles. The 3D contact pressure $p_\textrm{3D}(r)$ and linear force density $q_\textrm{1D}(x)$ are related through the following two transforms
  \begin{equation} \label{q1DtransANDp3Dtrans}
	q_\textrm{1D}(x) = 2\int_{x}^{\infty} \frac{r\, p_\textrm{3D}(r)}{\sqrt{r^2-x^2}} \,dr, \qquad p_\textrm{3D}(r) = -\frac{1}{\pi} \int_{r}^{\infty} \frac{q^\prime_\textrm{1D}(x)}{\sqrt{x^2-r^2}} \,dx.
\end{equation}

\noindent Similarly, the 3D normal displacement profile $w_\textrm{3D}(r)$ and the 1D normal displacement profile $w_\textrm{1D}(r)$ are related through

  \begin{equation} \label{w3DtransANDw1Dtrans}
	w_\textrm{1D}(x) = \frac{2}{E^*_c}\int_{x}^{\infty} \frac{r \, w_\textrm{3D}(r)}{\sqrt{r^2-x^2}} \,dr, \qquad w_\textrm{3D}(r) = \frac{2}{\pi} \int_{0}^{r} \frac{w_\textrm{1D}(x)}{\sqrt{r^2-x^2}} \,dx.
\end{equation}

The set of three transformations for the initial gap functions, contact pressures, and normal surface displacement profiles provide yet another powerful property of the MDR. Consider a contact problem that has been transformed to the 1D space and that is being evolved by application of a varying force on the indenter. At any given arbitrary state, the inverse transforms can be evaluated to give back the current initial gap function (which will be unchanged for an elastic body), contact pressure, and normal surface displacement profile in the 3D space. Thus, the MDR allows for an efficient solution of contact problem, whilst still providing a pathway back to the 3D problem if variation of the contact pressure or normal surface displacement is desired at a given state.    

A derivation of the MDR is given in~\ref{Derivation of the method of dimensionality reduction}.

\subsection{Normal contact with adhesion} \label{Normal contact with adhesion}

In 1971, Johnson, Kendall, and Roberts published their widely used (JKR) theory of adhesive contact \cite{johnson1971surface}. Fundamental to the theory, is the idea that adhesive contact can be seen as the superposition of two normal contact problems: a non-adhesive contact and an adhesive retraction. Given that the MDR provides exact solutions to axisymmetric normal contact problems, the superposition proposed in the JKR theory can be easily adapted to the MDR space---allowing adhesive contact to be modeled. The first two preparatory steps remain the same and consider the non-adhesive contact problem:

\begin{enumerate}
  \item The elastic bodies are replaced by a 1D linear elastic foundation with spring constants given as (\ref{Deltakz}).

  \item The initial gap function $g_\textrm{3D}(r)$ is transformed to a plane profile $g_\textrm{1D}(x)$ via (\ref{g1Dtrans}). 

\end{enumerate}

\noindent The plane indenter is then submerged into the 1D linear elastic foundation compressing the springs and giving rise to a finite contact radius $a$. Correspondingly, a force $F_\textrm{n.a.}(a)$ and displacement $\delta_{o,\textrm{n.a.}}(a) = g_\textrm{1D}(a)$ (where the subscript n.a. stands for non-adhesive) also arise. 

The superimposed adhesive retraction is considered in the third step and takes into account the adhesion:

\begin{enumerate}
\setcounter{enumi}{2}
  \item The springs in contact with the plane indenter are assumed to adhere to the the indenter's surface. The plane indenter is then lifted out of the foundation causing the springs at the outer edge of the contact radius to be pulled in tension. The retraction is stopped once the outer springs reach the maximum allowable elongation before separation (i.e. de-adhesion)

    \begin{equation} \label{Deltal}
	\Delta l(a) = \sqrt{\frac{2 \pi a \Delta \gamma}{E^*_c}},
    \end{equation}

which is related to the effective surface energy of the contacting bodies $\Delta\gamma$ and depends upon the contact radius $a$. Justification of this criterion by the principle of virtual work and an alternative, but harmonious, interpretation through the lens of linear elastic fracture mechanics can be found in~\ref{Critical extensional length}.

\end{enumerate}

 \begin{figure*} [!htb]
 	%\centering
 	\raggedright
 	% Trim{LEFT LOWER RIGHT UPPER}
 	\includegraphics[width=\textwidth, trim = 1cm  16cm 1cm 2.5cm]{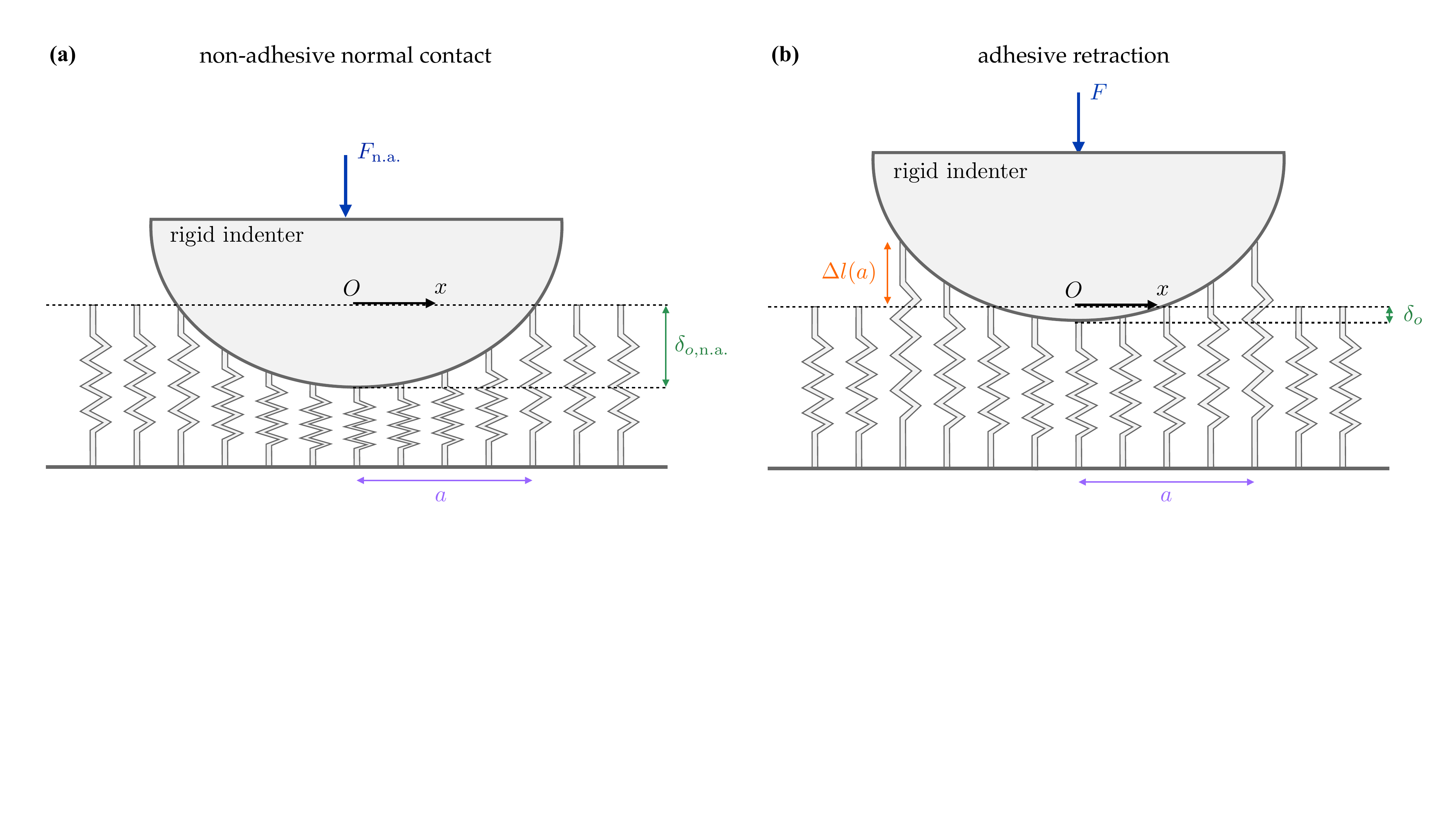}
 	\caption{Adapted from \cite{popov2019handbook}. Solution of elastic normal contact with adhesion using the method of dimensionality reduction (MDR). Two problems are superimposed to provide the complete solution: (a) non-adhesive (\textrm{n.a.}) normal contact without adhesion and (b) a subsequent adhesive retraction.}
 	\label{MDR_adhesion}
 \end{figure*}

\noindent This adhesive retraction corresponds to a uniform decompression of the springs by a total distance of $\Delta l$, implying that $w_\textrm{1D}(a) = -\Delta l(a)$. The force $F_\textrm{a.r.}(a)$ (where the subscript $\textrm{a.r.}$ stands for adhesive retraction) related to this decompression is calculated in a similar manner as before

    \begin{equation} \label{Frt_integral}
	F_\textrm{a.r.}(a) = -\int_{-a}^{a} E^*_c \Delta l(a) \,dx.
    \end{equation}

\noindent The argument of the integral has no dependence on $x$ therefore it simplifies to

    \begin{equation} \label{Frt}
	F_\textrm{a.r.}(a) = -2aE^*_c\Delta l(a).
    \end{equation}

\noindent Superposition of the two normal contact problems leads to the final equilibrium described by
        \begin{align} \label{superimpose_na_rt}
	    & a = a_\textrm{n.a.} = a_\textrm{a.r.}, \nonumber \\
            & F(a) = F_\textrm{n.a.}(a) + F_\textrm{a.r.}(a), \\     
            & \delta_o(a) = \delta_{o,\textrm{n.a.}}(a) - \Delta l(a) \nonumber.
        \end{align}

\noindent Here, the contact radius $a$ is shared by both contact problems, whereas the applied force $F(a)$ and displacement $\delta_o (a)$ are given by contributions from both the non-adhesive and adhesive retraction problems. Importantly, the triplet $\{ F,a,\delta_o \}$ defining equilibrium in the 1D adhesive problem \textit{coincides exactly} to that of the 3D adhesive problem. Thus, the JKR theory of adhesion has an exact mapping to the 1D space allowing for simplified analysis of adhesive contact. Importantly, the adhesive MDR framework is easily adaptable to non-spherical contact geometries.

Up to this point only equilibrium at a single instance has been considered, however no mention of the stability of this equilibrium has been made. From the discussion in~\ref{Stability Condition} the critical contact radius $a_c$ below which complete separation occurs is determined. In general, the critical contact radius is dependent on the type of loading: force or displacement controlled. The overall result is almost identical for both types of loadings only differing by a constant
    \begin{equation} \label{a_crtical}
        	\frac{dg_\textrm{1D}(a)}{da}\Bigr|_{a=a_c} = \xi \sqrt{\frac{\pi \Delta \gamma}{2 E^*_c a_c}}, \quad \quad \xi = \begin{cases}
      3, & \textrm{force-control},\\
      1, & \textrm{displacement-control}.
    \end{cases} 
    \end{equation}

% A derivation of this result is included in Appendix ???

A nice property of treating the adhesive contact problem with the MDR is that the transformation rules for the contact pressure (\ref{q1DtransANDp3Dtrans}) and displacements (\ref{w3DtransANDw1Dtrans}) still hold true.       

\section{Finite element contact simulations} \label{FEM_section}

\subsection{Geometric and material restrictions} \label{Geometric and material restrictions}

In our current study, all contacting particles are spherical and of equal initial radius $R_o$; they are made of the same homogeneous linear elastic-perfectly plastic isotropic material described by a Young's modulus $E$, Poisson's ratio $\nu$, and yield stress $Y$. Large variations in $\nu$ and the ratio of $E/Y$ are considered. Adhesion is described by an effective surface energy $\Delta \gamma$. 

\subsection{Finite element simulation description}

\begin{figure*} [!htb]
 	\centering
 	%\raggedright
 	% Trim{LEFT LOWER RIGHT UPPER}
 	\includegraphics[width=\textwidth, trim = 0cm  11cm 2cm 0.5cm]{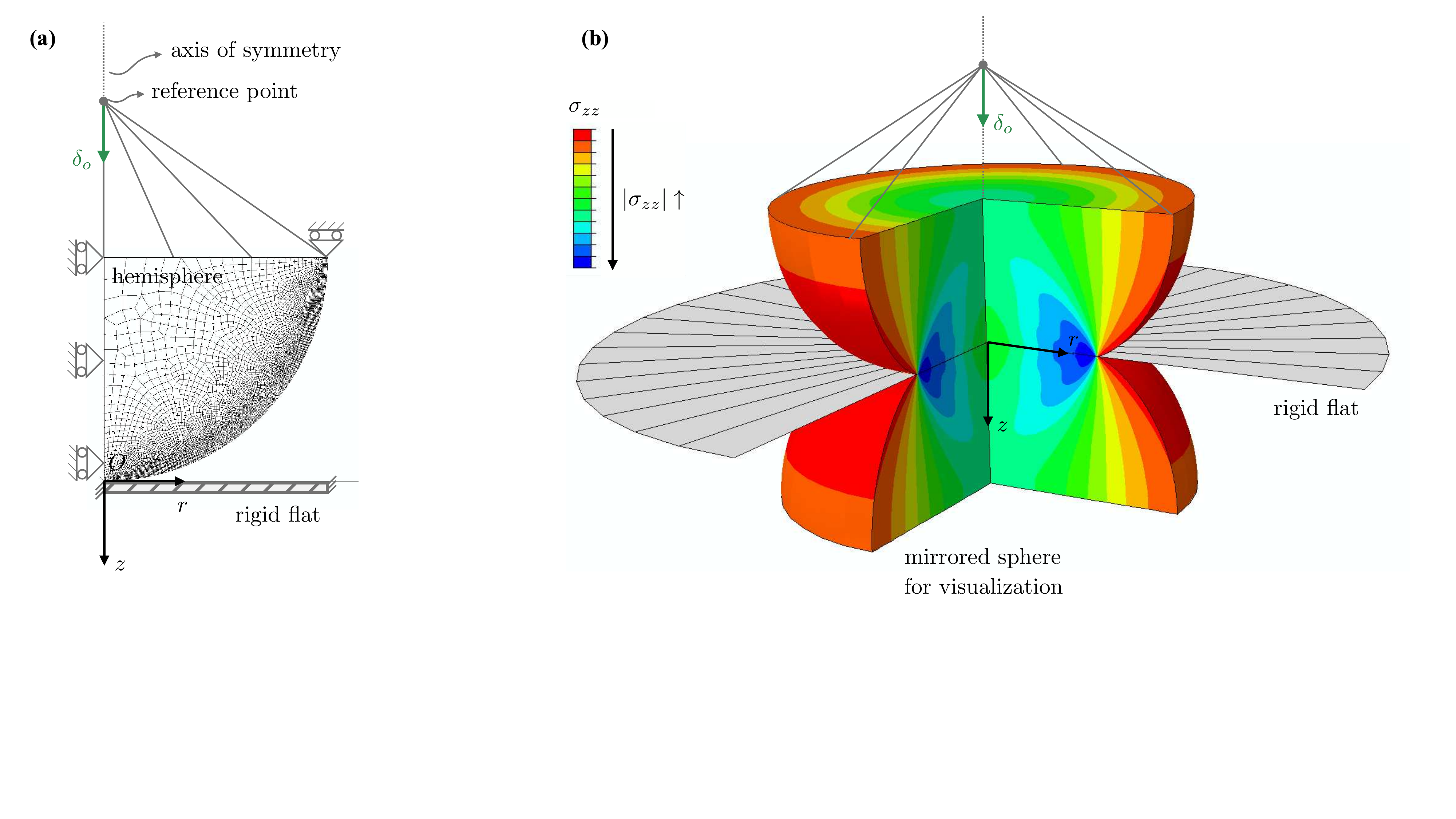}
 	\caption{Finite element set up used in the parameter study. (a) Schematic of the FEM simulation. (b) Plot of the $z$-component of the the stress field $\sigma_{zz}$ in the hemisphere in the fully-plastic regime. The simulation is mirrored to show the equivalence of the problem to that of two contacting deformable spheres.}
 	\label{fem_setup}
 \end{figure*}

To gain insight into the behavior of contact between two elastic-perfectly plastic spherical particles, a finite-element method (FEM) study is conducted using the software Abaqus. In this study and development of the contact model, we examine the equivalent problem of contact between a rigid flat and elastic-perfectly plastic sphere due to its simplified geometric properties. Transformation between the sphere-sphere and sphere-flat problems is simple and requires scaling only by constants. To understand the contact behavior of this benchmark problem, we perform a series of FEM simulations considering uniaxial compression of a spherical particle with varying material properties between two rigid flats. Following the assumptions from Section \ref{Geometric and material restrictions}, the particle material is homogeneous and isotropic, enabling a 2D axisymmetric simulation as shown in Fig.~\ref{fem_setup}(a).    

To drive the simulations, a $z$-displacement $\delta_o$ is applied uniformly to the top of the hemisphere. No constraint is imposed in the radial direction on the top edge to allow radial expansion as it would naturally occur in uniaxial compression. Contact between the flat and hemisphere is modeled as frictionless in the tangential direction and hard in the normal direction. In order to detect the onset of plastic deformation, a local plastically incompressible von Mises yield criterion is used. A mesh refinement study was conducted, leading to a mesh of 7,653 CAX4R (4-node bilinear axisymmetric quadrilateral, reduced integration, hourglass control) elements used to discretize the hemisphere, with significantly increased mesh refinement along the curved edge where contact occurs. The flat is an analytical rigid surface. Validation of the simulation in the limit of small deformation was carried out by comparison to Hertz's law. 

%When present, adhesion is modeled using a traction-separation law parameterized by an initial elastic stiffness, a damage initiation stress, and subsequent linear damage evolution specified by a critical fracture energy.

\subsection{Finite element parametric study}

With the simulation well-defined, a parametric study was conducted in which the ratio of the Young's modulus to yield stress $E/Y$ and Poisson ratios $\nu$ for the hemisphere are varied independently. For $E/Y$, the study range is $6.25 \leq E/Y \leq 200$, this approaches the non-negligible elastic deformation and rigid-plastic limits, respectively. For $\nu$ the study range is $0 \leq \nu \leq 0.45$, which captures no induced expansion and approaching incompressible, respectively. At this stage no adhesion is considered. The outputs of interest at each instant are the contact force $F$, contact area $A_C$, and average contact pressure $\bar{p}$ as a function of displacement $\delta_o$. For now, only monotonic increasing loading is considered, since we will see that this is sufficient to develop a contact model using the method of dimensionality reduction (MDR) as unloading falls out naturally. 

\begin{figure*} [!htb]
 	%\centering
 	\raggedright
 	% Trim{LEFT LOWER RIGHT UPPER}
 	\includegraphics[width=\textwidth, trim = 1cm  2cm 13cm 1cm]{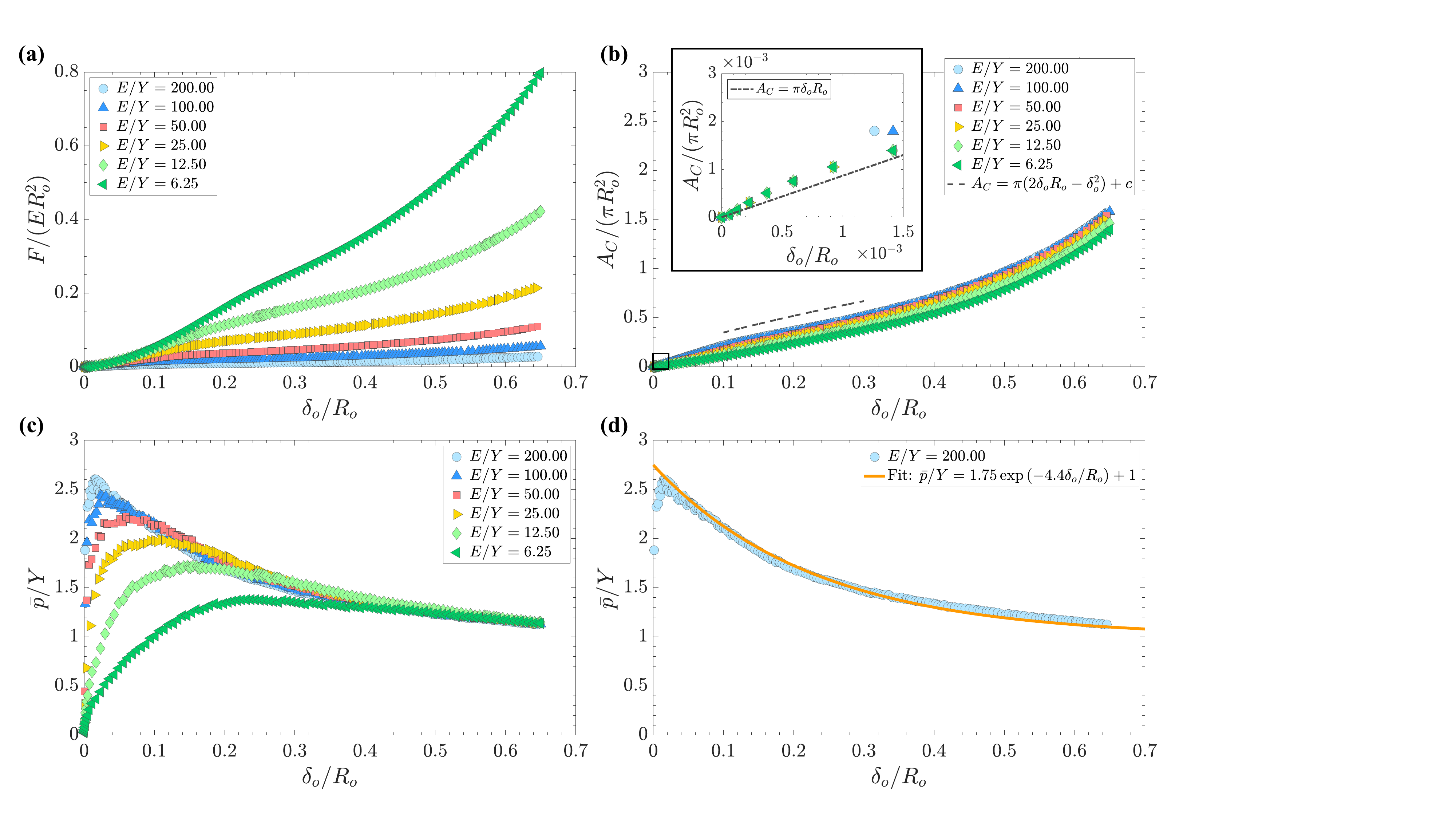}
 	\caption{Results from the FEM parametric study varying the ratio $E/Y$ within the range $6.25 \leq E/Y \leq 200.00$. (a) Force-displacement curves. (b) Contact area-displacement curves. The inset compares contact areas to Hertz law before plastic deformation occurs. (c) Average contact pressure-displacement curves. The limiting case of rigid plastic sphere is approximated by $E/Y = 200.00$. (d) Fitting of the $E/Y = 200.00$ curve to determine the hardening curve used to limit the allowable average pressure.}
 	\label{fem_EY_variation_results}
 \end{figure*}

The results of the FEM study for variations in $E/Y$ are shown in  Fig.~\ref{fem_EY_variation_results}(a)-(c), for each case $\nu = 0.3$. The results from the FEM study varying $\nu$ are given in~\ref{Variation of contact quantities with Poisson ratio} since varying the Poisson ratio is found to produce negligible changes in the resulting curves. A similar parametric studies were carried out in~\cite{edmans2020numerical,zhang2022research}, providing a basis for comparison.

\subsubsection{Variation of force with E/Y} \label{Variation of force with E/Y}

We begin by focusing our attention on the force displacement curves of Fig.~\ref{fem_EY_variation_results}(a), where the force is scaled by $ER^2_o$ and the displacement by $R_o$. %The scaling of $R_o$ is used over the commonly seen composite radius $R^*_c = (1/R_1 + 1/R_2)^{-1}$, (a quantity that arises naturally from the initial gap function) for two contacting bodies $B_1$ and $B_2$ with radii $R_1$ and $R_2$, since $R^*_c = R_o$ for sphere-flat contact. In all cases shown, $R_o$ is a constant since the geometry of the problem is unvarying. 
We see that as $E/Y$ is decreased\footnote{For this study $E$ is held constant, whereas $Y$ is varied.}, we get continually increasing force at the same displacement. This result arises because plasticity limits the allowable stress state, reducing it from what a purely elastic simulation would predict. As with all of these simulations, in the early stages of contact before the onset of plasticity we find nice agreement between the FEM results and Hertz's law.    

\subsubsection{Variation of area with E/Y} \label{Variation of area with E/Y}

Inspection of the contact area curves, scaled by $\pi R^2_o$, shows that for a given displacement higher $E/Y$ leads to a larger contact area. We note that most of the variation in contact area for different $E/Y$ is concentrated to the earlier stages of displacement; at large displacements all curves exhibit similar shapes merely shifted by an offset caused by the initial behavior. Focusing in on this initial behavior, we recall that Hertz's law gives the contact area between a rigid flat and spherical body as as $A_C = \pi \delta_o R_o$, whereas the contact area given by the purely geometric intersection of the sphere and flat is $A_C = \pi(2\delta_o R_o + \delta^2_o)$. For each case investigated, Hertz's expression is valid for a range of $\delta_o/R_o$ as shown in the inset of Fig.~\ref{fem_EY_variation_results}(b), however the exact range is affected by the ratio of $E/Y$. Interestingly, as $E/Y \rightarrow \infty$ the range of $\delta_o/R_o$ under which Hertz's area expression is valid shrinks to zero and we find that  the geometric approximation better describes the area evolution. On the other hand, as $E/Y \rightarrow 0$ the range of $\delta_o/R_o$ under which Hertz's area expression is valid increases, but deviations are seen either after the onset of plasticity or when the contact area radius is of similar magnitude to the particle radius---both violations of the assumptions of Hertzian contact. 

We observe that in large deformation neither the purely geometric picture nor Hertzian area formula suffice to describe the evolution of the contact area. However, there is a striking similarity between all curves during large deformation, hinting that they are all produced from the same underlying mechanism. Importantly, plastic deformation is taken as incompressible, meaning that the vertical confinement of the material between the rigid flat and equator of the hemisphere necessitates significant radial expansion. This radial expansion leads to areas that can exceed $\pi R_o^2$ and results in the curves seen in Fig.~\ref{fem_EY_variation_results}(b). By expanding the geometric overlap picture described here to account more precisely for plastic incompressibility, we will show that contact areas at these much larger deformations can also be successfully modeled (see equation (\ref{Areas}) ahead).   

\subsubsection{Variation of average pressure with E/Y} \label{Variation of average pressure with E/Y}

The final important quantity taken from the contact is the average pressure. Its variation with the scaled displacement for different $E/Y$ is plotted in Fig.~\ref{fem_EY_variation_results}(c). Here, the average pressure is scaled by $Y$, offering a direct connection to hardness. Most notably, we see that despite there being initially different evolutions for each $E/Y$ all the curves eventually coalesce. This master curve is defined by the rigid-plastic limit---approximated to be the case of $E/Y = 200$. This curve will play a key role in the development of the contact model, so we provide a fitted form as shown in Fig.~\ref{fem_EY_variation_results}(d)
    \begin{equation} \label{pbarfit}
        	\frac{\bar{p}}{Y} = 1.75\exp{(-4.4\delta_o/R_o)+1}.
    \end{equation}

\noindent In the limit of $\delta_o/R_o \rightarrow 0$, we recover a hardness $\bar{p}/Y$ between the accepted limits of $2.61$ to $2.84$ as seen experimentally and theoretically for perfectly plastic indentation of a nominally flat surface ~\cite{tabor2000hardness,ishlinsky1944problem,lee1972analysis}. In the other limit as $\delta_o/R_o \rightarrow 1$, we find that $\bar{p}/Y \rightarrow 1$.  Physically this implies that contact between two elastic-perfectly plastic hemispheres morphs into uniaxial compression of a cylinder for large relative displacements---a phenomenon caused by the significant radial expansion that was first noted in a pair of works by Jackson and Green~\cite{quicksall2004elasto,jackson2005finite}. 

Full pressure profiles for a number of different loading stages and material parameter combinations are given in~\ref{Pressure profile variation}.  It can be seen that in the fully plastic regime, these profiles look relatively uniform, excluding an annulus at the contact edge. This observation is reminiscent of the 2D analytical result that the pressure distribution under a flat indenter pressed into a half-space is exactly uniform~\cite{hill1998mathematical}.  The approximate uniformity of the pressure profiles will be exploited in the following model development.

\section{Development of non-adhesive elastic-plastic contact model} \label{Non-adhesive elastic-plastic contact model}

\subsection{Solution methodology}

The finite element parameter study gives insights into the evolution of the contact area and pressure profile with displacement. The force at the contact as a function of displacement---the ultimate objective of a contact model---is found by integrating the pressure profile over the contact area. A solution that could accurately track both of these quantities would be able to return the correct force at any stage of loading. Given that no known complete analytical solution exists for elastic-plastic contact between spherical bodies~\cite{ghaednia2017review}, we turn to the method of dimensionality reduction (MDR) to provide an efficient approximation. 

In a typical contact problem, we begin with knowledge of the initial gap function $g_\textrm{3D}(r)$ and then proceed from there to find a pressure distribution and area combination that satisfies the usual restrictions on the gap function (i.e. it is zero within the contact area and positive outside). This procedure works well for elastic contact, but in elastic-plastic contact, plastic deformation is continuously changing the unloaded shape of the indenter and, correspondingly, the unloaded gap function as shown in the sequence of contact snapshots in Fig.~\ref{regime_visualization}(a). It then becomes a formidable challenge to try and determine these unloaded gap functions, let alone their corresponding pressure profiles and contact area. The strength in the MDR is that it allows for a reverse entry point to solving axisymmetric contact problems. If the pressure profile $p_\textrm{3D}(r)$ and contact radius $a$ for an axisymmetric contact are known in the $\textrm{3D}$ space we can transform this to a linear force density using (\ref{q1DtransANDp3Dtrans}), by replacing the upper limit of integration with $a$

  \begin{equation} \label{q1Dtrans_a}
	q_\textrm{1D}(x) = 2\int_{x}^{a} \frac{r\, p_\textrm{3D}(r)}{\sqrt{r^2-x^2}} \,dr.
\end{equation}

\noindent Recalling (\ref{q1D}) and (\ref{w1D}), we see that simple algebraic manipulations give both the normal surface displacement and initial gap function

\begin{equation} \label{w1D_q1DoverEANDg1D}
    w_\textrm{1D}(x) = q_\textrm{1D}(x)/E^*_c, \qquad g_\textrm{1D}(x) = \delta_o - w_\textrm{1D}(x) 
\end{equation}

\noindent Transformation (\ref{g3Dtrans}) would complete the cycle and return $g_\textrm{3D}(r)$. 

In developing a contact model based on the MDR we take advantage of the reverse approach, exploiting the fact that the contact pressure profile at each instant in the fully-plastic regime can be reasonably approximated as uniform.  Combining the assumed uniform pressure profile with knowledge of how the contact area and average pressure change throughout the fully-plastic regime provides sufficient information to use the MDR to identify the plastically deformed shape of the indenter at any instant. With the shape of the indenter always known, it is a straightforward task to recover the unloading curve, which requires simply pulling the 1D indenter out of the bed of springs. Adhesion can also be treated easily using the methodology outlined in Section~\ref{Normal contact with adhesion}. This ability to capture forward loading, unloading, and adhesion with the same consistent framework is the major benefit of using the MDR to create a contact model.

\subsection{Functional dependencies of the force}

\begin{figure*} [!htb]
 	%\centering
 	\raggedright
 	% Trim{LEFT LOWER RIGHT UPPER}
 	\includegraphics[width=\textwidth, trim = 7.5cm  18cm 6cm 0cm]{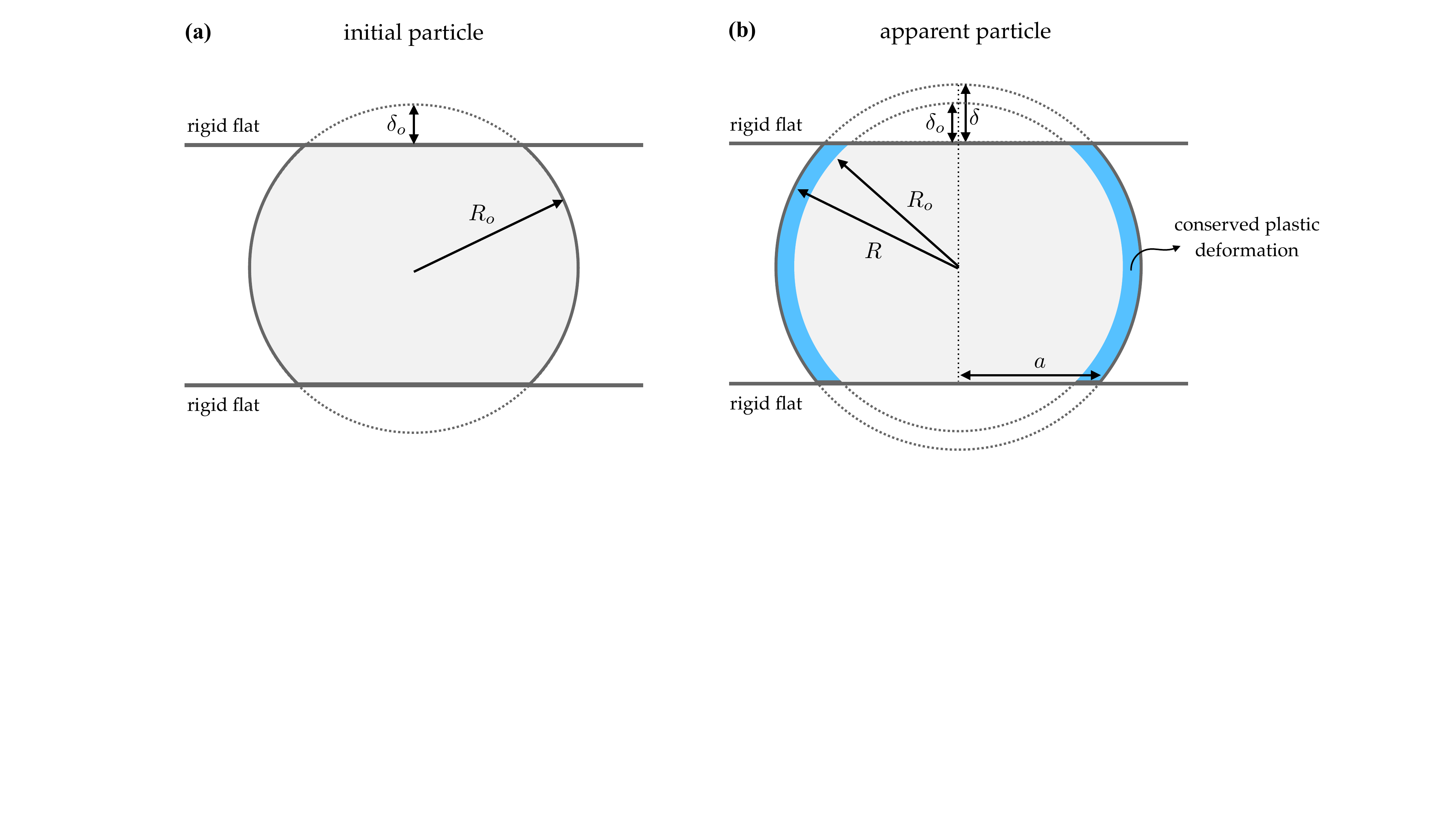}
 	\caption{Side view of uniaxial compression of a particle. (a) The initial particle of radius $R_o$ compressed between two rigid flats displaced by a distance $\delta_o$. (b) The apparent particle, defined by an apparent radius $R$, that is induced by respecting the incompressibility of plastic deformation. The apparent overlap $\delta$ is defined with respect to $R$ as well as the contact radius $a$.}
 	\label{kinematic_quantities}
 \end{figure*}

We again take our reference problem to be that of contact between an elastic-perfectly plastic sphere and rigid flats as shown in Fig.~\ref{kinematic_quantities}(a). The objective is to be able to write the force as a function of the apparent overlap $\delta$, maximum experienced apparent overlap $\delta_\textrm{max}$, and apparent radius $R$ leading to $F = \hat{F}(\delta,\delta_\textrm{max},R)$ as shown in (\ref{Fbox}). Here, $\delta_\textrm{max}$ is a contact state variable; it is a surrogate for the amount of plastic deformation that has occurred, enabling proper unloading and reloading behavior after plastic deformation. The apparent radius $R$ is a particle state variable; it grows to ensure that volume change due to plastic deformation is not lost. Geometrically, it is the radius of the sphere that most closely matches the deformed particle's free surface; that is, the part of the surface not in contact. Defining the apparent radius helps account for multi-neighbor dependent effects such as formation of new contacts and increased force at existing ones. The introduction of $R$ also has an important consequence in that it naturally leads to the definition of the apparent overlap and maximum experienced apparent overlap

\begin{equation} \label{apparent delta and apparent delta_max}
     \delta = \delta_o + R - R_o, \qquad \delta_\textrm{max} = \delta_{o,\textrm{max}} + R - R_o.
\end{equation}

\noindent As the definition suggests $\delta$ and $\delta_\textrm{max}$ are `displacements'\footnote{Apparent overlap is used in place of displacement since $\delta$ and $\delta_\textrm{max}$ can grow without relative motion between particles since they can change simply from $R$ growing.} measured with respect to $R$, whereas $\delta_o$ and $\delta_{o,\textrm{max}}$ are displacements measured with respect to $R_o$ as shown in Fig.~\ref{kinematic_quantities}(b). In the development of the contact model the apparent overlap measures are predominately used, however in plots $\delta_o$ or $\delta_{o,\textrm{max}}$ will still often appear as the independent variable. \textit{Care should therefore be taken to ensure the right displacement measure is being considered when interpreting both equations and results}.     

\subsection{Pressure profiles in the elastic and fully-plastic regimes}

The first step in using the reverse entry point for solving contact problems provided by the MDR is to determine the pressure profiles. For the deformations considered here, in Part I, there are two clear regimes to consider: elastic (i.e. Hertzian) and fully-plastic. The elastic regime is characterized by an elliptical pressure profile of the following form

\begin{equation} \label{p3DHertz}
    p_\textrm{3D}(r;\delta,R) = \frac{2E^*_c}{\pi\sqrt{R}} \left( \delta - \frac{r^2}{R} \right)^{1/2}. 
\end{equation}

\noindent The apparent radius $R$ and apparent overlap $\delta$ are used here to reflect the fact that the contact may form after $R$ has increased from $R_o$.

The pressure profile in the fully-plastic regime\footnote{The criterion for transition between the elastic and fully-plastic regime will be clearly defined in Section~\ref{transition_from_elastic_to_fully_plastic}.}, on the other hand, is taken to be perfectly uniform across the whole contact for sake of modeling simplicity---an assumption justified in Section~\ref{Variation of average pressure with E/Y}. Thus a single scalar value, the average contact pressure, characterizes the pressure profile at each instant. As noted in Section~\ref{Variation of average pressure with E/Y} the average contact pressure for the rigid-plastic case forms a hardening curve that all test cases coalesce with once they have reached the fully-plastic state. We assume that this hardening curve $p_Y$ can be written as a function of the current maximum experienced apparent overlap and apparent radius, leading to the following expression for the pressure profile in the fully-plastic regime
\begin{equation} \label{p3DFP}
    p_\textrm{3D}(r) = \hat{p}_Y(\delta_\textrm{max},R). 
\end{equation}

Following the procedure laid out in (\ref{q1Dtrans_a}) and (\ref{w1D_q1DoverEANDg1D}) we can connect each of these pressure profiles to their corresponding transformed normal surface displacements in the elastic foundation, and therefore, the force. 

 \begin{figure*} [b!!]
 	%\centering
 	\raggedright
 	% Trim{LEFT LOWER RIGHT UPPER}
 	\includegraphics[width=\textwidth, trim = 1cm  11.5cm 0cm 1cm]{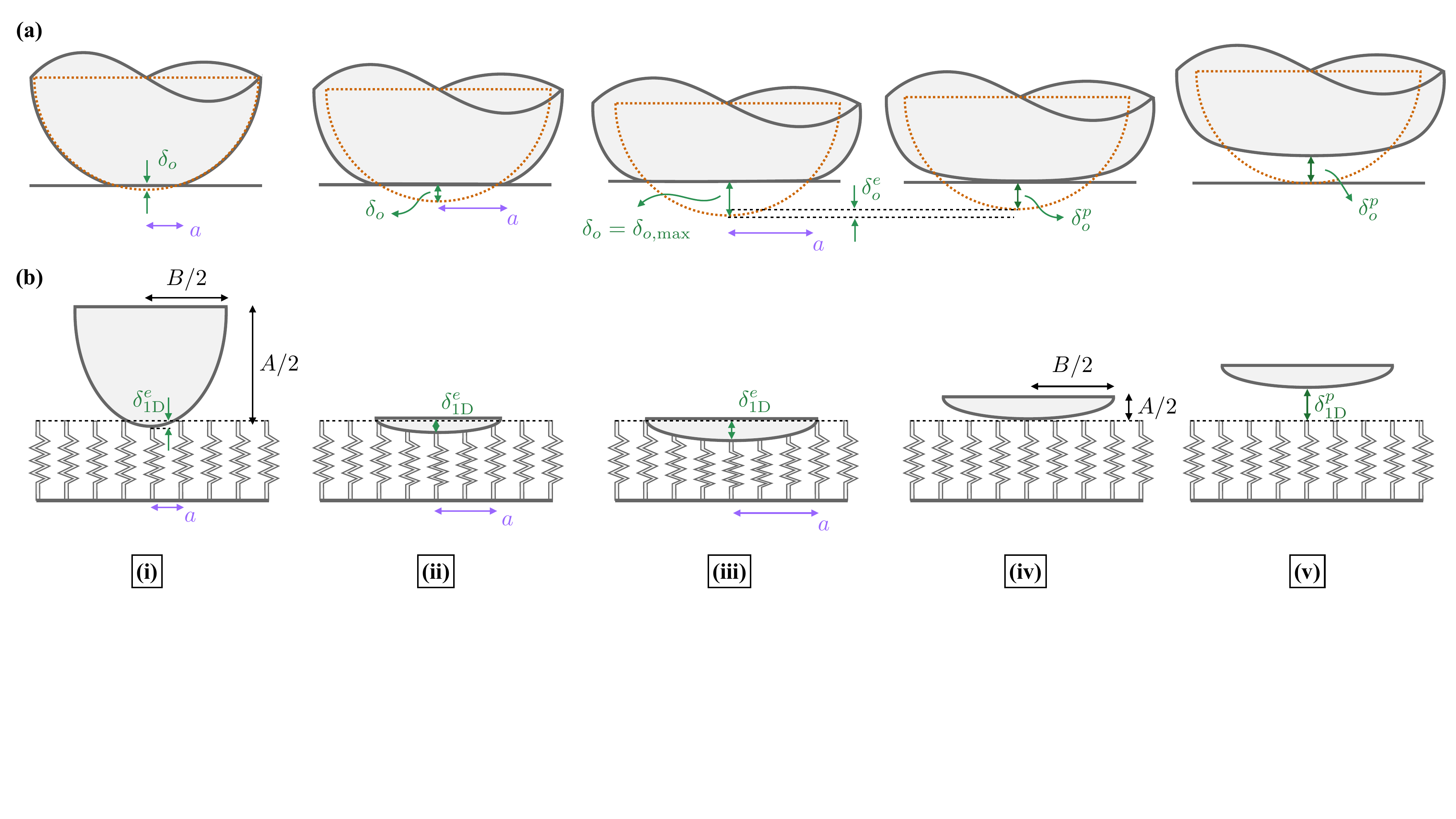}
 	\caption{Sketch of contact evolution in the physical 3D space (a) and transformed 1D space (b). In 3D space, the reference configuration begins as contact between an elastic-perfectly plastic sphere and rigid flat. Transformed to the 1D space, this corresponds to an elliptical indenter with an aspect ratio of $A/B = 2$. In snapshot (i), the displacement $\delta_o$ is such that the contact remains purely elastic, meaning the deformations are all recoverable. Applying the displacement to the 1D space leads to a penetration $\delta^e_\textrm{1D} = \delta_o$, which gives rise to the same contact radius $a$. In snapshot (ii), plastic deformation has occurred which blunts the 3D sphere surface. Correspondingly, the height $A/2$ and width $B/2$ of the still elliptical indenter in the transformed space changes to reflect this plastic deformation. In this case, $\delta^e_\textrm{1D}$ no longer equals $\delta_o$. In snapshot (iii), the sphere accumulates more plastic deformation and once again the transformed indenter's height and width change. Prior to unloading, the displacement is registered as $\delta_{o,\textrm{max}}$ to store the state of plastic deformation. The recovered elastic displacement $\delta^e_o$ is less than $\delta_o$. At snapshot (iv) the sphere has been unloaded, plastic deformation leads to a new rest configuration and a definable plastic displacement that is the difference between the original and updated profile. Unloading is an elastic process therefore the height and width of the transformed indenter from snapshot (iii) to (iv) remains fixed. At snapshot (v), the sphere is raised further to highlight once again the new rest configuration. The transformed indenter is also lifted allowing $\delta^p_\textrm{1D}$ to be defined, which does not equal $\delta^p_o$. }
 	\label{regime_visualization}
 \end{figure*}

\subsection{Displacement decomposition in the 3D and transformed space}

Before formulating the contact law it is necessary to discuss the displacement decomposition. As shown in Fig.~\ref{regime_visualization}(a) and (b) the displacement can be additively decomposed for the physical 3D space: $\delta_o = \delta^e_o + \delta^p_o$ and the transformed 1D MDR space: $\delta_o = \delta^e_\textrm{1D} + \delta^p_\textrm{1D}$. At first glance, equivalence between the elastic and plastic components of the displacement in the two spaces might be assumed; however, this is not always correct. The crux of the issue is that the MDR assumes the half-space approximation\footnote{The half-space approximation requires that the contact area is small with respect to the principal radii of curvature at the contacting point and the overall dimensions of the body; thus, displacements can be found by applying pressure distributions to elastic half-spaces and then projecting the half space displacements back onto the curved body} is valid for the contact. For the large deformations being considered, where the contact area is on the order of the principal radii of curvature, the half-space approximation needs to be amended by adding a displacement correction $\delta_R$---in similar spirit to the corrections made in~\cite{brodu2015multiple,hui2000accuracy}. This adjusts the displacement so that it is measured from a finite point, coinciding with the distance to the particle's center, rather than at infinity within the half-space. In other words, $\delta_R$ tracks the often times non-negligible displacement of the particle center in the half-space, allowing the elastic and plastic components between the two spaces to be related
    \begin{equation} \label{deltae3D_deltae1D}
        	 \delta^e_o = \delta^e_\textrm{1D} + \delta_R, \qquad \delta^p_o = \delta^p_\textrm{1D} - \delta_R.
    \end{equation}

\noindent A formula for $\delta_R$ will be given shortly.

\subsection{General formulation of the MDR contact law} 

\subsubsection{Elastic regime}

We begin with the elastic regime and note that the contact radius is given as 
    \begin{equation} \label{aHertz}
        a = \sqrt{\delta R}. 
    \end{equation}

\noindent The apparent radius $R$ and apparent overlap $\delta$ are used once again to reflect the fact that the contact may form after $R$ has increased from $R_o$. Using (\ref{aHertz}) for the upper limit of integration in (\ref{q1Dtrans_a}) and the elastic pressure profile (\ref{p3DHertz}) as the integrand as well as dividing the result by $E^*_c$ we arrive at
    \begin{equation} \label{w1Dhertz}
        	 w_\textrm{1D}(x;\delta,R) = \delta^e_\textrm{1D} - \frac{x^2}{R}.
    \end{equation}

\noindent Here, $\delta^e_\textrm{1D} = \delta$ due to the small deformations and the half-space approximation holding, making them interchangeable variables. In the transformed space we can visualize this contact as a rigid plane parabolic indenter whose initial gap function is given as $g_\textrm{1D}(x) = x^2/R$ contacting an elastic foundation. With the normal surface displacement profile defined it is a straightforward task to determine the force by using (\ref{Fw1D}) leading us directly to the celebrated Hertz's contact law

    \begin{equation} \label{Fhertz}
        	 F(\delta) = \frac{4}{3}E^*_c\sqrt{R}\delta^{3/2}. 
    \end{equation}

\subsubsection{Fully-plastic regime}

The contact law in the fully-plastic regime needs to allow for an evolving plane indenter shape to reflect changes in the unloaded 3D profile caused by plastic deformation.  This amounts to correctly changing the width and height of the 1D ellipse continuously as plastic deformation accumulates. 
Plastic flow will be dictated by never allowing the contact pressure to exceed the hardening curve. Upon unloading at any point along the hardening curve the problem is reduced back to an elastic one with the indenter shape parameterized by the contact state variable $\delta_\textrm{max}$ and particle state variable $R$, which are both directly related to the amount of plastic deformation accumulated. In this sense, one can view the fully-plastic regime simply as a sequence of elastic contact problems.     

In the fully-plastic regime, we distinguish between the contact radius $a$ that tracks the current value, and the maximum experienced contact radius $a_\textrm{max}$ that is a function of $\delta_\textrm{max}$ and $R$
\begin{equation}
a_\textrm{max} = \hat{a}_\textrm{max}(\delta_\textrm{max},R). 
\end{equation}

\noindent This distinction is necessary since $a_\textrm{max}$ will be used to parameterize the shape of the indenter and needs to be unvarying during unloading. Using $a_\textrm{max}$, we carry out the same procedure of finding the transformed normal surface displacement for the assumed \textit{uniform} pressure profile (\ref{p3DFP}) in 3D space. The upper limit of integration in (\ref{q1Dtrans_a}) is set to $a_\textrm{max}$, resulting in the following maximum normal surface displacement profile
    \begin{equation} \label{w1Dplastic}
        	 w_\textrm{1D,max}(x;\delta_\textrm{max},R) = \frac{A}{B} \sqrt{\frac{B^2}{4} - x^2},
    \end{equation}

\noindent where 
  \begin{equation} \label{A and B}
	A(\delta_\textrm{max},R) = \frac{4 p_Y}{E^*_c} a_\textrm{max}, \qquad B(\delta_\textrm{max},R) = 2a_\textrm{max}.
\end{equation}

\noindent The maximum normal surface displacement profile given in (\ref{w1Dplastic}) is that of an ellipse submerged up to its equator (i.e. $A/2$) whose full width is given by $A(\delta_\textrm{max},R)$ and full height by $B(\delta_\textrm{max},R)$. %This means that an elliptical plane rigid indenter, parameterized by (\ref{A and B}), submerged to its equator in the 1D space is the shape that exactly corresponds back to the flat pressure profile in the 3D space. This provides justification for selecting the plane indenter in the elastic regime to also be elliptical in shape, hence both regimes can be unified under one indenter shape family in the 1D space. The key differences in the fully-plastic regime is that $A$ and $B$ are defined differently and vary as $\delta_\textrm{max}$ changes. 

With the shape of the transformed plane indenter identified for the fully-plastic regime we can now write the expression for the normal displacement profile

    \begin{equation} \label{w1Dfullyplastic}
        	 w_\textrm{1D}(x;\delta,\delta_\textrm{max},R) = \delta^e_{\textrm{1D}}(\delta,\delta_\textrm{max}) - \left( \frac{A}{2} - \frac{A}{B} \sqrt{\frac{B^2}{4} - x^2} \right).
    \end{equation}

\noindent To determine the force we once again use (\ref{Fw1D}) resulting in

    \begin{equation} \label{Fcontactlaw}
        	 F(\delta,\delta_\textrm{max},R) = \frac{E^*_c AB}{4}\left[ \arccos\left( {1 - \frac{2\delta^e_{\textrm{1D}}}{A}}\right) - \left( 1 - \frac{2\delta^e_{\textrm{1D}}}{A} \right) \sqrt{\frac{4\delta^e_{\textrm{1D}}}{A} - \frac{4(\delta^e_{\textrm{1D}})^2}{A^2}}\right]
    \end{equation}

\subsubsection{Unification of the elastic and fully-plastic regimes under one plane indenter profile}

We opt to make a slight adjustment that allows unification of the elastic and fully-plastic regime under one family of plane indenter shapes. In particular, we choose to approximate the parabolic indenter (\ref{w1Dhertz}) of the elastic case with an ellipse whose local second-order expansion about $x = 0$ coincides with that of the parabolic indenter. This leads to a normal displacement profile identical in form to (\ref{w1Dfullyplastic}), but parameterized by different $A$ and $B$\footnote{The expressions for $A$ and $B$ in the elastic regime are not unique, it is the aspect ratio $A/B = 2$ that is critical to ensure that the displacement locally around $x=0$ reduces to (\ref{w1Dhertz}). Scaling $A$ and $B$ by $R$ is done for ease of implementation and so that the local expansion remains a good approximation during any deformation within the elastic regime.}

\begin{equation} \label{Ahertz}
	A = 4R, \qquad B = 2R.
\end{equation}

\noindent The force in the elastic regime can then also be represented by (\ref{Fcontactlaw}) with (\ref{Ahertz}) substituted in place of (\ref{A and B}). By design the ellipse in the elastic regime looks locally parabolic, thus for small deformations, which implies $\delta^e_\textrm{1D} = \delta$, we recover (\ref{Fhertz}).

\subsection{Closure of the contact law} \label{Closure of the contact law}

 To close the formulation of the contact law, our last task is to determine expressions for $p_Y$, $\delta^e_{\textrm{1D}}$\footnote{Unlike the elastic regime, in the fully-plastic regime $\delta \neq \delta^e_\textrm{1D}$.}, $a_\textrm{max}$, and $R$, which we shall do now.
\subsubsection{Pressure along hardening curve}

The contact pressure along the hardening curve $p_Y = \hat{p}_Y(\delta_\textrm{max},R)$ is simply taken to be (\ref{pbarfit}) with a minor adjustment of replacing $\delta_o/R_o$ with $\delta_\textrm{max}/R$.

\subsubsection{Transformed elastic displacement} 

Determining $\delta^e_\textrm{1D}$ requires determining the displacement correction $\delta_R$. The first step in doing this is to calculate the force caused by $\delta^e_{1D}=\delta^e_\textrm{1D,max} = A/2$

    \begin{equation} \label{Fplasticmax}
        	 F_\textrm{max}(\delta_\textrm{max},R) = \frac{E^*_c AB}{4}\left[ \arccos\left( {1 - \frac{2\delta^e_{\textrm{1D,max}}}{A}}\right) - \left( 1 - \frac{2\delta^e_{\textrm{1D,max}}}{A} \right) \sqrt{\frac{4\delta^e_{\textrm{1D,max}}}{A} - \frac{4(\delta^e_{\textrm{1D,max}})^2}{A^2}}\right].
    \end{equation}

\noindent We then calculate the position (or `depth') of the center point of the particle in relation to the undeformed surface of the 3D elastic half-space
    \begin{equation} \label{z_c}
        	 z_R = R - (\delta_\textrm{max} - \delta^e_\textrm{1D,max}). 
    \end{equation}

\noindent The displacement of the center point, also known as the correction displacement, is then found by approximating $F_\textrm{max}$ to be distributed evenly over the contact area on the 3D half-space and using superposition of Boussinesq solutions to compute the corresponding displacement at the point $z_R$ as 
    \begin{equation} \label{delta_R}
        	 \delta_R = \frac{F_\textrm{max}}{\pi a_\textrm{max}^2} \left[  \frac{2a^2_\textrm{max} (\nu - 1) - z_R(2\nu - 1)( \sqrt{a^2_\textrm{max} + z^2_R} - z_R)} {2G\sqrt{a_\textrm{max}^2 + z_R^2}} \right]. 
    \end{equation}

\noindent Upon unloading or reloading in the fully-plastic regime we need to account for the fact that the reference point  used to measure displacement also is changing. To do this we assume that the reference point unloads or reloads from its max value of $\delta_R$ to $0$ linearly as $\delta$ unloads or reloads. This leads to the following adjusted equation for $\delta^e_\textrm{1D}$ 
    \begin{equation} \label{deltae1D}
        	 \delta^e_\textrm{1D}(\delta,\delta_\textrm{max},R) = \frac{\delta - \delta_\textrm{max} + \delta^e_\textrm{1D,max} + \delta_R}{1 + \delta_R/\delta^e_\textrm{1D,max}}.
    \end{equation}

\subsubsection{Contact radius in fully-plastic limit}

To define the maximum experienced contact radius $a_\textrm{max}$, we recall from Section~\ref{Variation of area with E/Y} that the contact area $A_C$ in the beginning of the fully-plastic regime is found to be well approximated by the geometric intersection between the sphere and rigid flat (up to a constant offset): $A_C = \pi(2\delta_o R_o - \delta^2_o)$. For large deformations $\delta_o/R_o > 0.3$, significant deviations from this geometric picture occur, after which it underpredicts the total contact area. The primary factor leading to this underprediction is the lack of accounting for the incompressibility of the plastic deformation. Since the contact is confined vertically, the volume conservation required by plastic deformation necessitates radial expansion, leading to a larger contact area than would be given by the geometric intersection of a sphere with radius $R_o$. To account for this effect, we allow the sphere radius to change as plastic deformation accumulates. This naturally leads to the introduction of the apparent radius $R$ and, consequently, $\delta$ and $\delta_\textrm{max}$. Thus, we write the maximum experienced contact radius as
        \begin{equation} \label{amax}
        	 a_\textrm{max} = \sqrt{(2\delta_\textrm{max} R - \delta^2_\textrm{max}) + c_A/\pi}.
    \end{equation}

\noindent A constant $c_A$, that will be defined shortly, has been introduced as an offset to make the contact area continuous when switching from the Hertizan to fully-plastic regimes. 

\subsubsection{Apparent particle radius update}

For conciseness we simply report the differential update form of $R$ 

    \begin{equation} \label{eqn_for_R}
    \Delta R = \textrm{max} \left[ \frac{\Delta V^e - \sum_{i=1}^{N} \pi\Delta \delta_{o,i}(2\delta_{o,i} R_o - \delta^2_{o,i} + R^2 - R^2_o) }{2 \pi R \sum_{i=1}^{N}(\delta_{o,i} + R - R_o) - 4 \pi R^2}, \; 0 \right],
    \end{equation}

\noindent that allows the apparent radius to be explicitly updated provided the change in elastic volume $\Delta V^e$ (positive for a compressive increment), number of contacts $N$, change in displacement $\Delta \delta_o$ at each contact, displacement $\delta_o$ at each contact, initial radius $R_o$, and apparent radius $R$ are all known. A detailed derivation of (\ref{eqn_for_R}) is included in~\ref{Deriving the differential update form of the apparent radius}, where it is shown that the expression arises geometrically from modeling the deformed particle as a truncated sphere of radius $R$ and enforcing that all volume change comes strictly from elastic deformation due to plastic incompressibility. 

To determine the elastic volume change for the whole particle, the relation between the mean normal stress and volumetric strain is used: $\textrm{tr}(\bar{\bm{\epsilon}}) = \textrm{tr}(\bar{\bm{\sigma}})/(3\kappa)$, where $\bar{\bm{\sigma}}$ is the volume-averaged Cauchy stress tensor, $\bar{\bm{\epsilon}}$ the volume-averaged infinitesimal strain tensor, and $\kappa$ the bulk modulus. To determine $\bar{\bm{\sigma}}$, the formulation proposed by Christoffersen et al.~\cite{christoffersen1981micromechanical} is used. It constructs $\bar{\bm{\sigma}}$ through the summation of the tensor product between the force vectors at each contact $\bm{f}$ and their corresponding branch vectors $\bm{b}$ over all contacts $c$ on the particle with volume $V$. This allows the volumetric strain to be expressed as 
        \begin{equation} \label{volumetric strain}
    	\textrm{tr}(\bar{\bm{\epsilon}}) =  \frac{1}{3\kappa V} \textrm{tr} \left( \sum_{c \; \in \; V} \bm{f}^c \otimes \bm{b}^c \right).
        \end{equation}

\noindent Here, $\bm{b}$ is the vector that points from the particle center to the center of the contact, with a magnitude equal to $\delta_o$. The current volume $V$ is then given by
        \begin{equation} \label{current volume}
    	V = V_o( 1 + \textrm{tr}(\bar{\bm{\epsilon}})).
        \end{equation}

\noindent where the initial volume is given as $V_o = 4/3\pi R^3_o$. The elastic volume change for a given update $\Delta V^e $ can then be found explicitly from the difference of (\ref{current volume}) between steps. 

Some caveats to (\ref{eqn_for_R}) at very high confinement will be discussed in Part II. 

\subsubsection{Transition from the elastic to fully-plastic regime} \label{transition_from_elastic_to_fully_plastic}

With contact laws for the elastic and fully-plastic regimes developed what remains is to discuss the transition between the two regimes. We elect for a simple sharp transition based on the average pressure at the contact. Namely, once the average pressure described by Hertz's contact law

    \begin{equation} \label{pbarhertz}
        	 \bar{p}_H = \frac{4E^*_c}{3 \pi \sqrt{R}}\sqrt{\delta},
    \end{equation}

\noindent meets that of the hardening curve $p_Y$ the formulation switches from elastic to fully-plastic. This guarantees that the contact pressure is continuous, however to maintain continuity of the force, the contact radius must also be continuous. To accomplish this we solve for the displacement at which the $\bar{p}_H = p_Y$ denoted as the yield displacement $\delta_Y$

    \begin{equation} \label{delta_Y_solve}
        	 \frac{4E^*_c}{3 \pi \sqrt{R}}\sqrt{\delta_Y} = Y\left( 1.75\exp{(-4.4\delta_Y/R)+1} \right).
    \end{equation}

\noindent We recall that the contact areas $A_C$ in the two separate regimes during forward loading are given as     
    \begin{equation} \label{Areas}
        	 A_C = \begin{cases}
      \pi \delta R, & \textrm{elastic},\\ \\
      \pi(2\delta R - \delta^2) + c_A, & \textrm{fully-plastic}.
    \end{cases} 
    \end{equation}

\noindent To make them continuous we solve for $c_A$ such that they intersect at $\delta_Y$, leading to
    \begin{equation} \label{c_A}
        	 c_A = \pi(\delta_Y^2 - \delta_Y R).
    \end{equation}

\noindent Continuity of both the average contact pressure and area ensures continuity of the force.

\section{Adding in adhesion} \label{Adding in adhesion}

Appending the contact law to include adhesion is a straightforward process that follows the framework given in Section~\ref{Normal contact with adhesion}. The key idea is to allow the springs to adhere to the indenter surface. Upon decompression these springs stretch in tension until they reach a critical extensional length $\Delta l$ (\ref{Deltal}). Guided by experimental observation, the JKR theory of adhesion predicts that two bodies, when brought close enough, will leap into contact creating a finite contact area despite no external load being applied. The MDR with adhesion, designed to recover the JKR theory of adhesion, also reflects this behavior. To make the adhesion formulation slightly simpler we neglect this ability of material to leap into contact and only allow material to adhere once it has contacted due to compression. However, once decompression takes place and the outer springs reach the maximum extensional length the simplified approach coincides with the JKR theory of adhesion. The main purpose of this simplification is to allow usage of the previous non-adhesive formulation during forward loading without modification.

During unloading, springs detach and the current contact radius needs to be continually compared to the the critical contact radius $a_c$ to determine whether full separation should occur. The critical criterion is based on whether the contact is force or displacement-controlled and is given by solving for $a_c$ in (\ref{a_crtical}). For the contact model at hand $g_\textrm{1D}(a)$ takes on the form
    \begin{equation} \label{g1Dellipse}
        	 g_\textrm{1D}(a) = \frac{A}{2} - \frac{A}{B}\sqrt{\frac{B^2}{4} - a^2}.
    \end{equation}

\noindent During the loading sequence of a contact it is necessary to monitor the normal surface displacement at the edge of the contact 

    \begin{equation}
	w_\textrm{1D}(a) = \delta^e_\textrm{1D} - g_\textrm{1D}(a),
    \end{equation}

\noindent and compare its value to $\Delta l (a)$. Adhesion at the contact can be broken into three states based on this comparison:

    \begin{enumerate}
        \item \textbf{No tensile springs:} $w_\textrm{1D}(a) = 0$

        In this case, all springs are in a compressive state and the tensile adhesive force is zero. The non-adhesive formulation from Section~\ref{Non-adhesive elastic-plastic contact model} to calculate the force holds without modification.

        \item \textbf{Tensile springs, but not exceeding critical length:} $ 0 < w_\textrm{1D}(a) \leq \Delta l(a)$

        In this case, a portion of the contact is in a tensile state with no springs exceeding the critical length---hence no detachment. As described in Section~\ref{Normal contact with adhesion} the total force can be found by considering the superposition of two problems: a non-adhesive contact and a adhesive retraction. The non-adhesive contribution to the force $F_\textrm{n.a.}$ is treated directly by simply setting $\delta = g_\textrm{1D}(a)$ in (\ref{Fcontactlaw}). The adhesive retraction corresponds to a uniform decompression of the springs by a distance $\delta^e_{1D} - g_\textrm{1D}(a)$, using (\ref{Fw1D}) to determine the force leads to 
        \begin{equation} \label{Frt2}
    	F_\textrm{a.r.} = 2 E^*_c(\delta^e_{1D} - g_\textrm{1D}(a))a.
        \end{equation}

        Summation of the two forces gives the total force
        \begin{equation} \label{Fna plus Frt}
    	F = F_\textrm{n.a.} + F_\textrm{a.r.} 
        \end{equation}

        \item \textbf{Tensile springs exceed critical length:} $ w_\textrm{1D}(a) > \Delta l(a)$

        In this case, the extensional length of the outer springs has exceeded the critical length and the springs must detach. To determine if a new stable equilibrium exists or if full separation occurs the contact radius that satisfies the following equation must be found  
        \begin{equation} \label{find_new_a}
    	\delta^e_\textrm{1D} + \Delta l(a) - g_\textrm{1D}(a) = 0. 
        \end{equation}

        If the contact radius that satisfies this equation is greater than the critical contact radius $a > a_c$ the contact remains intact and the process described in Case 2 is used to find the total force with the updated contact radius. If the new contact radius is less than the critical contact radius $a < a_c$ the contact separates entirely.  
        
    \end{enumerate}

  Adhesion between contacting surfaces is allowed to redevelop after separation provided the surfaces contact in compression again. 

\section{MDR contact model summary} \label{Contact model summary}

The MDR contact model is summarized below. The force is given as a function of the apparent overlap $\delta$, maximum experienced apparent overlap $\delta_\textrm{max}$, and apparent radius $R$

    \begin{equation*} 
        	 F(\delta,\delta_\textrm{max},R) = \frac{E^*_c AB}{4}\left[ \arccos\left( {1 - \frac{2\delta^e_{\textrm{1D}}}{A}}\right) - \left( 1 - \frac{2\delta^e_{\textrm{1D}}}{A} \right) \sqrt{\frac{4\delta^e_{\textrm{1D}}}{A} - \frac{4(\delta^e_{\textrm{1D}})^2}{A^2}}\right].
    \end{equation*}

\noindent In the elastic regime, before the onset of plasticity, $\delta^e_\textrm{1D} = \delta$. The composite plane strain modulus of the two contacting bodies $B_1$ and $B_2$, each with their respective Young's modulus $E_i$ and Poisson's ratio $\nu_i$ is given as
  \begin{equation*}
	E^*_c = \left( \frac{1-\nu_{1}^2}{E_{1}} + \frac{1-\nu_{2}^2}{E_{2}} \right)^{-1}.
\end{equation*}

\noindent The apparent overlap and maximum experienced apparent overlap are `displacements' measured from a spherical surface of radius $R$, and are related to the displacement $\delta_o$ and maximum experienced displacement $\delta_{o,\textrm{max}}$, which are displacements measured with respect to the initial radius $R_o$, through the following relations
\begin{equation*} 
    \delta = \delta_o + R - R_o, \qquad \delta_\textrm{max} = \delta_{o,\textrm{max}} + R - R_o. 
\end{equation*}

\noindent The parameters $A$ and $B$ define the elliptical indenter in the transformed space and are given as
    \begin{equation*}
        \begin{split}
            & A = 4R, \qquad \qquad \; \; B = 2R, \qquad \quad \; \textrm{elastic} \\
            & A = \frac{4p_Y}{E^*_c}a_\textrm{max}, \qquad B = 2a_\textrm{max}, \qquad \textrm{fully-plastic}.
        \end{split}
    \end{equation*}

\noindent The average pressure along the hardening curve $p_Y$ is defined by the average pressure for the limiting case of a rigid-plastic sphere
    \begin{equation*} 
        	p_Y = Y\left( 1.75\exp{(-4.4\delta_\textrm{max}/R)+1} \right).
    \end{equation*}

\noindent The transition between the elastic and fully-plastic regimes is taken to be sharp and is based on the average pressure at the contact. Namely, once the average pressure described by Hertz's contact law
    \begin{equation*} 
        	 \bar{p}_H = \frac{4E^*_c}{3 \pi \sqrt{R}}\sqrt{\delta},
    \end{equation*}

\noindent meets that of the hardening curve $p_Y$ the formulation switches from elastic to fully-plastic. The transformed 1D elastic displacement is given by 
    \begin{equation*} 
        	 \delta^e_\textrm{1D} = \frac{\delta - \delta_\textrm{max} + \delta^e_\textrm{1D,max} + \delta_R}{1 + \delta_R/\delta^e_\textrm{1D,max}},
    \end{equation*}

\noindent where
    \begin{equation*} 
        	 \delta^e_\textrm{1D,max} = A/2,
    \end{equation*}

\noindent the displacement correction is given as
    \begin{equation*} 
        	 \delta_R = \frac{F(\delta_\textrm{max},\delta_\textrm{max},R)}{\pi a_\textrm{max}^2} \left[  \frac{2a^2_\textrm{max} (\nu - 1) - z_R(2\nu - 1)( \sqrt{a^2_\textrm{max} + z^2_R} - z_R)} {2G\sqrt{a_\textrm{max}^2 + z_R^2}} \right], 
    \end{equation*}

\noindent and the distance from the free surface center of the sphere is
    \begin{equation*}
        	 z_R = R - (\delta_\textrm{max} - \delta^e_\textrm{1D,max}). 
    \end{equation*}

\noindent The maximum experienced contact radius is  
        \begin{equation*}
        	 a_\textrm{max} = \sqrt{(2\delta_\textrm{max} R - \delta^2_\textrm{max}) + c_A/\pi},
    \end{equation*}

\noindent where $c_A$ is found to enforce continuity of the contact radius between the elastic and plastic regimes

    \begin{equation*}
        	 c_A = \pi(\delta_Y^2 - \delta_Y R).
    \end{equation*}

\noindent The yield displacement $\delta_Y$ is found by equating $\bar{p}_H$ and $p_Y$ 

    \begin{equation*} 
        	 \frac{4E^*_c}{3 \pi \sqrt{R}}\sqrt{\delta_Y} = Y\left( 1.75\exp{(-4.4\delta_Y/R)+1} \right).
    \end{equation*} 

\noindent Finally, the radius is allowed to grow to respect the incompressible nature of the plastic deformation 

\begin{equation*}
    \Delta R = \textrm{max} \left[ \frac{\Delta V^e - \sum_{i=1}^{N} \pi\Delta \delta_{o,i}(2\delta_{o,i} R_o - \delta^2_{o,i} + R^2 - R^2_o) }{2 \pi R \sum_{i=1}^{N}(\delta_{o,i} + R - R_o) - 4 \pi R^2}, \; 0 \right].
 \end{equation*}

\noindent The elastic volume change $\Delta V^e$ (taken as positive for a compressive increment) is determined from the volume change between steps, where the volume $V$ is calculated as

        \begin{equation*} 
    	V = V_o \left( 1 + \frac{1}{3\kappa V}\textrm{tr} \left( \sum_{c \; \in \; V} \bm{f}^c \otimes \bm{b}^c \right) \right).
        \end{equation*}

Adhesion can be appended to this contact law using the methodology outlined in Section~\ref{Adding in adhesion}.

\section{Verification of MDR contact model against finite element simulations} \label{Verification of MDR contact model against finite element simulations}

\subsection{Verification of MDR contact model in non-adhesive case} \label{Verification of MDR contact model in non-adhesive case}

A numerical implementation of the contact law was created following the blueprint given in~\ref{Sketch of numerical implementation}. For reference the key MDR contact model parameters are tabulated in~\ref{Parameters of the MDR contact model}.   

\subsubsection{Force-displacement comparison}

\begin{figure*} [!htb]
 	%\centering
 	\raggedright
 	% Trim{LEFT LOWER RIGHT UPPER}
 	\includegraphics[width=\textwidth, trim = 1cm  2.5cm 13cm 1cm]{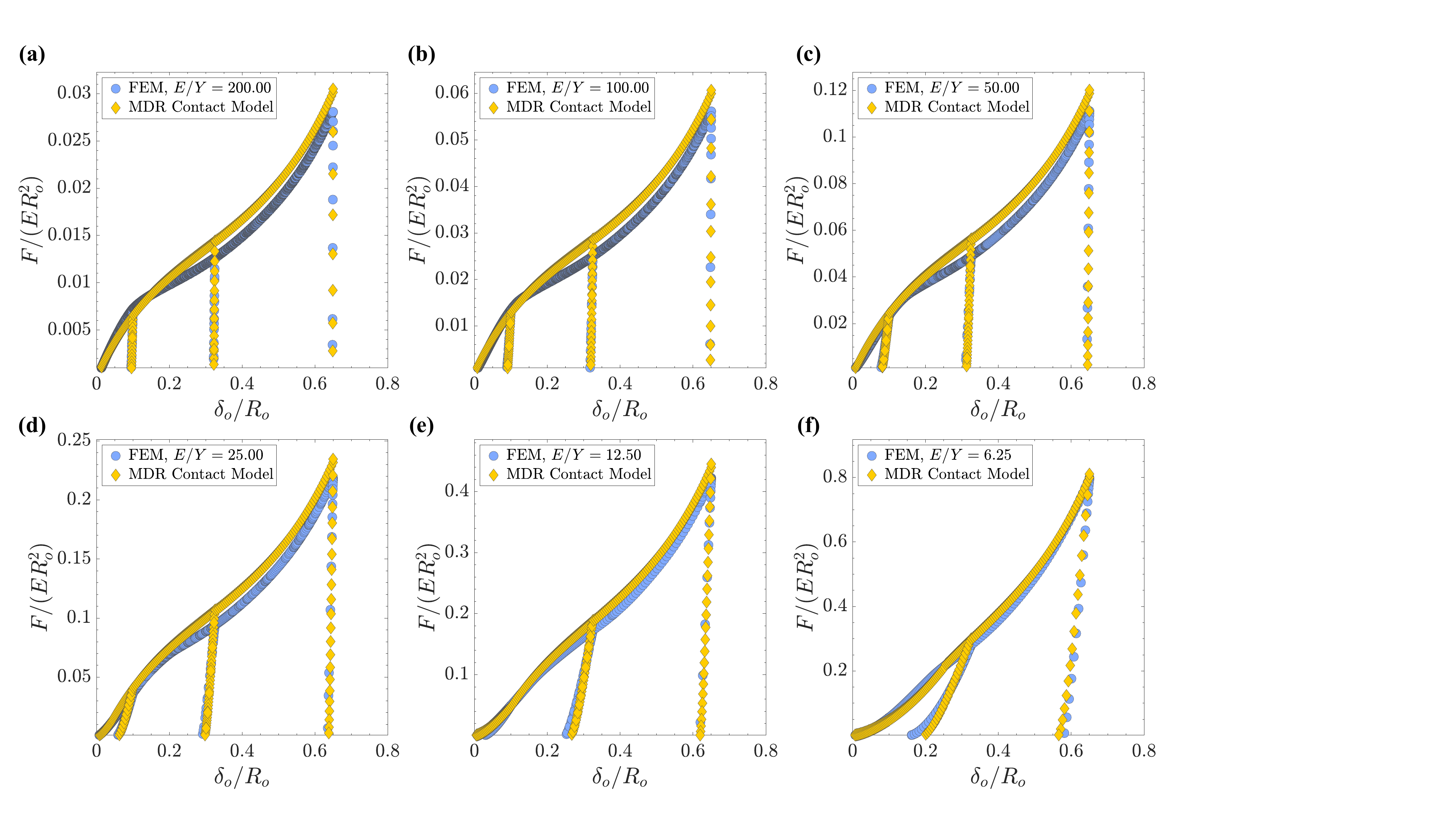}
 	\caption{Force-displacement curves for the FEM simulations and the MDR contact model. (a) to (f) show decreasing values of $E/Y$ from 200.00 to 6.25.}
 	\label{FEM_MDRCM_comparison_force}
 \end{figure*}

Initially, contact without adhesion was considered so comparison of the contact model to the force-displacement curves of Fig.~\ref{fem_EY_variation_results}(a) could be made. The results of this comparison for a range of different $E/Y$ values is contained in Fig.~\ref{FEM_MDRCM_comparison_force}(a)-(f). Inspection of cases (a)-(e) reveals a semi-quantitative agreement between the FEM results and the MDR contact model. Agreement in the early stages is most striking for all cases. For cases (a)-(c), where $E/Y = 200.00, 100.00,$ and $50.00$ respectively, the elastic regime is almost negligible and the early stages of deformation are all captured by the formulation for the fully-plastic regime. For cases (d) and (e), $E/Y =25.00$ and $E/Y=12.50,$ respectively, the elastic regime is more significant. For all cases (a)-(e) by $\delta_o/R_o > 0.15$ the fully-plastic formulation has taken over. In this large deformation regime the contact model recovers the nonlinear trends in the force-displacement behavior. This reaffirms the selection of the fit for the average contact pressure and the radius growth scheme. Importantly, we find almost an exact agreement in the unloading curves between the FEM simulations and the MDR contact model. This means that the contact model is accurately predicting the total plastic displacement at every moment in time. We emphasize that these unloading curves fall out of the formulation with no additional effort; each one is the byproduct of simply pulling the transformed 1D rigid indenter out the springs and tracking the force.  Case (f), $E/Y = 6.25$, shows nice agreement as well, but it should be noted that in this case the elastic strains are beginning to exceed the typically acceptable values of (3-5\%) of linear elasticity. Therefore caution should be used when applying the contact model to $E/Y$ ratios this low since it assumes all elastic deformations are within the usual bounds of linear elasticity.   

\subsubsection{Contact area-displacement comparison}

\begin{figure*} [!htb]
 	%\centering
 	\raggedright
 	% Trim{LEFT LOWER RIGHT UPPER}
 	\includegraphics[width=\textwidth, trim = 1cm  2.5cm 13cm 1cm]{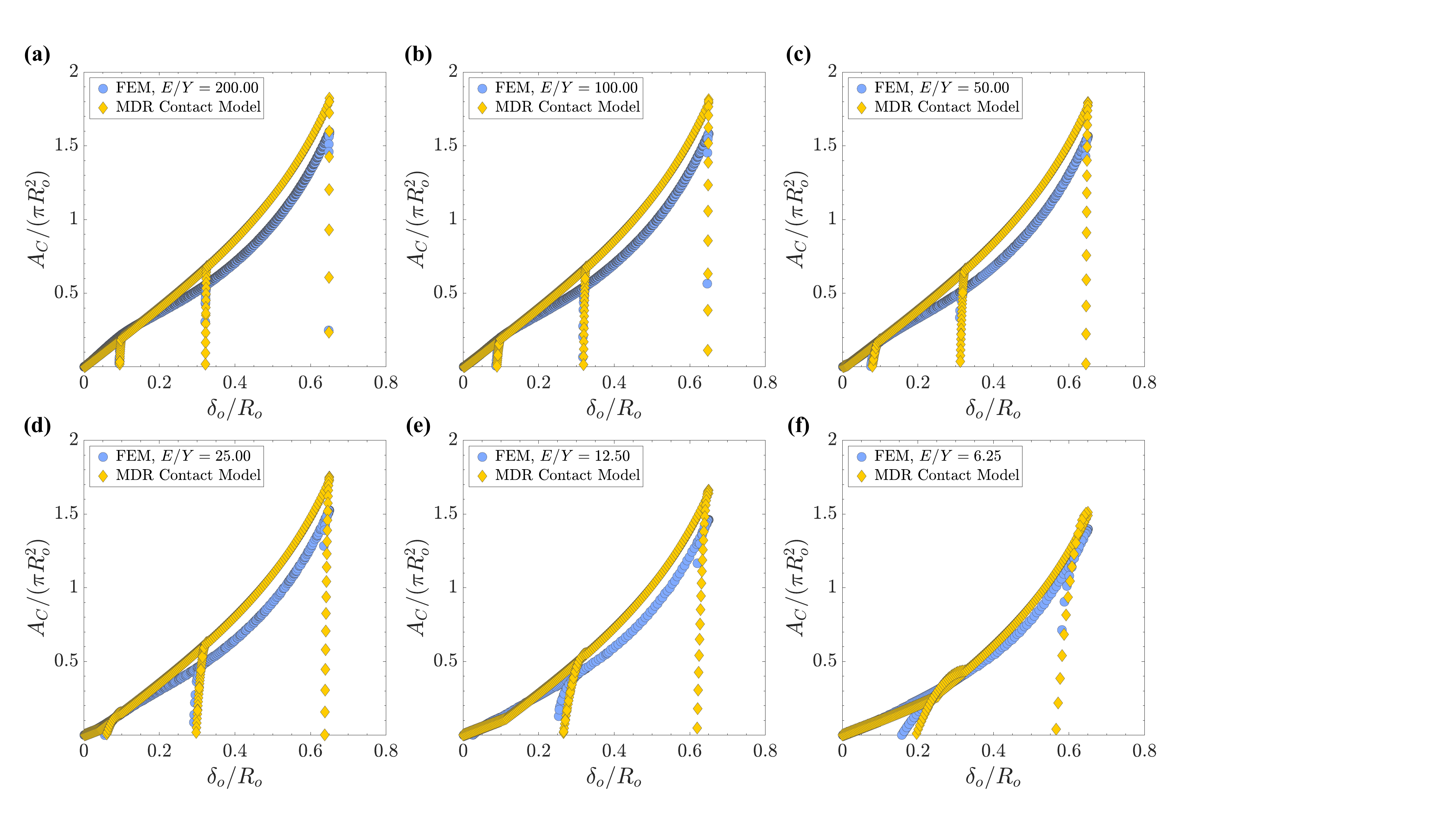}
 	\caption{Area-displacement curves for the FEM simulations and the MDR contact model. (a) to (f) show decreasing values of $E/Y$ from 200.00 to 6.25.}
 	\label{FEM_MDRCM_comparison_area}
 \end{figure*}

The contact area comparison is shown in Fig.~\ref{FEM_MDRCM_comparison_area}(a)-(f)  for the same values of $E/Y$. As is expected, and necessary for matching the force-displacement curves, we find nice agreement between the area-displacement curves with similar deviation at large displacements. Once again the match of the unloading behavior is nicely captured by the MDR contact model.

\subsubsection{Average contact pressure-displacement comparison}

\begin{figure*} [!htb]
 	%\centering
 	\raggedright
 	% Trim{LEFT LOWER RIGHT UPPER}
 	\includegraphics[width=\textwidth, trim = 1cm  2.5cm 13cm 2cm]{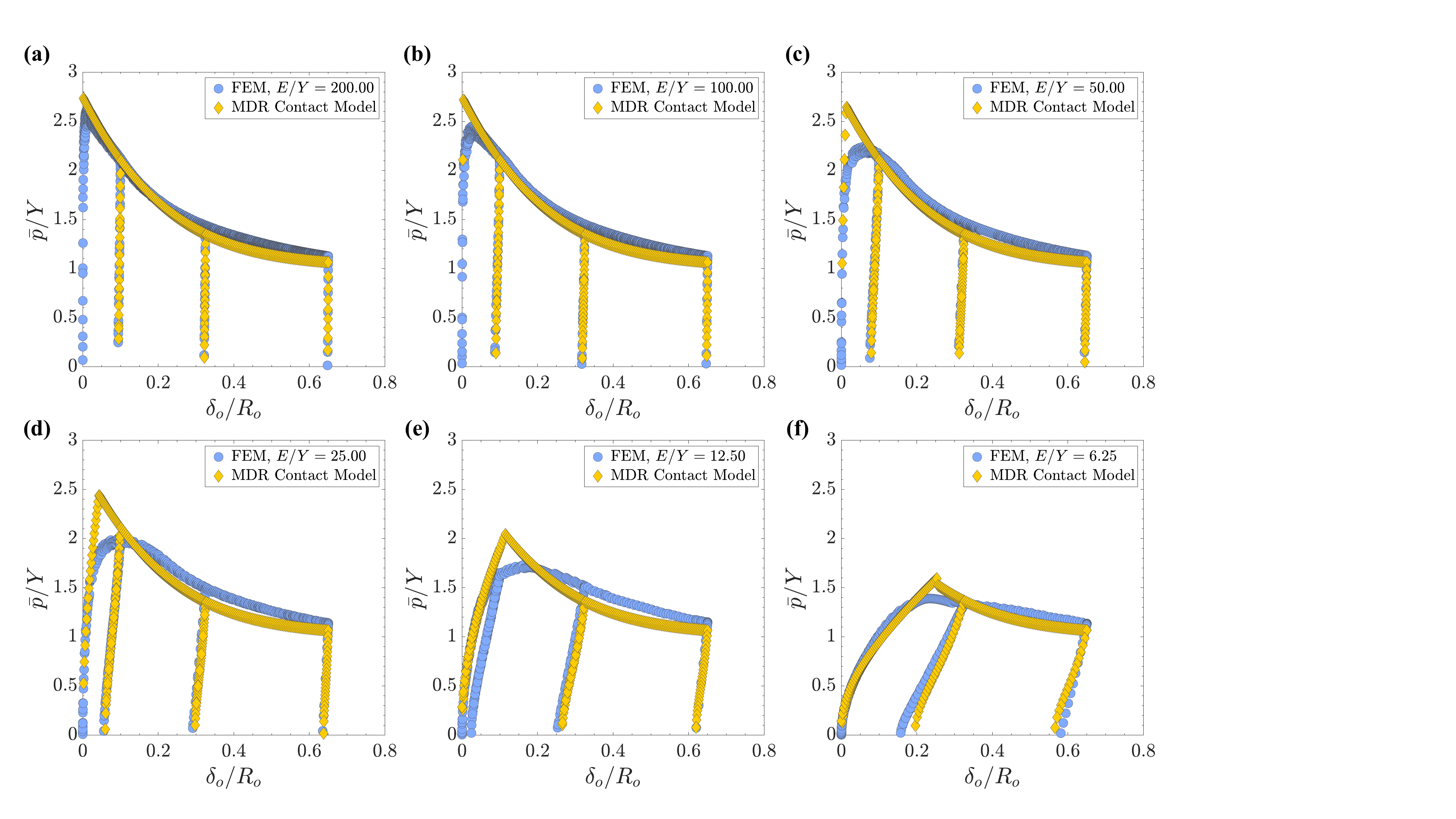}
 	\caption{Average contact pressure-displacement curves for the FEM simulations and the MDR contact model. (a) to (f) show decreasing values of $E/Y$ from 200.00 to 6.25.}
 	\label{FEM_MDRCM_comparison_pbar}
 \end{figure*}

The average contact pressure comparison Fig.~\ref{FEM_MDRCM_comparison_pbar}(a)-(f) best highlights the sharp transition from the elastic to fully-plastic regimes. We see clearly in, for example case (e) $E/Y = 12.50$ that the average pressure initially rises as Hertz would predict, but then switches to following the hardening curve upon intersection with it. For cases (a)-(c), $E/Y = 200.00, 100.00,$ and $50.00$ respectively, the elastic regime is small in comparison to the fully-plastic regime, therefore the hardening curve is effectively followed for the entire displacement history.

\subsubsection{Volume-displacement comparison}

\begin{figure*} [!htb]
 	%\centering
 	\raggedright
 	% Trim{LEFT LOWER RIGHT UPPER}
 	\includegraphics[width=\textwidth, trim = 1cm  2.5cm 13cm 2cm]{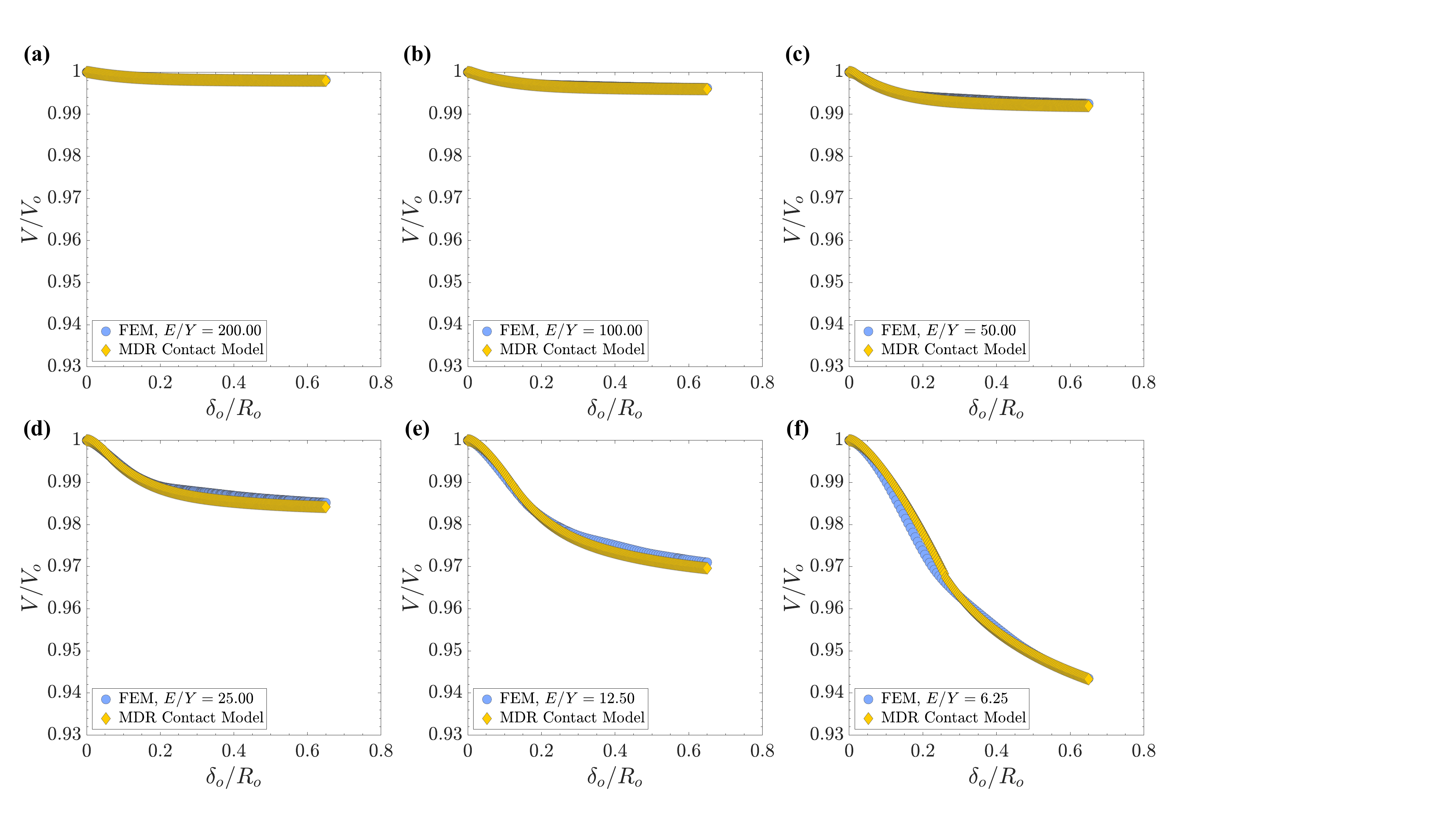}
 	\caption{Particle volume-displacement curves for the FEM simulations and the MDR contact model. (a) to (f) show decreasing values of $E/Y$ from 200.00 to 6.25.}
 	\label{FEM_MDRCM_comparison_vol}
 \end{figure*}

As part of the radius growth scheme the particle volume is continually tracked. Comparison between the FEM and MDR contact model volume predictions is given in Fig.~\ref{FEM_MDRCM_comparison_vol}(a)-(f), with nice agreement seen in all cases. Importantly, the MDR contact model is able to correctly predict the decreasing volume change in the higher $E/Y$ cases. This is reflective of the fact that for higher $E/Y$  most deformations are plastic (i.e. incompressible). 

\subsection{Preview of die compaction comparison}

\begin{figure*} [!htb]
 	%\centering
 	\raggedright
 	% Trim{LEFT LOWER RIGHT UPPER}
 	\includegraphics[width=\textwidth, trim = 1cm  2.5cm 13cm 2cm]{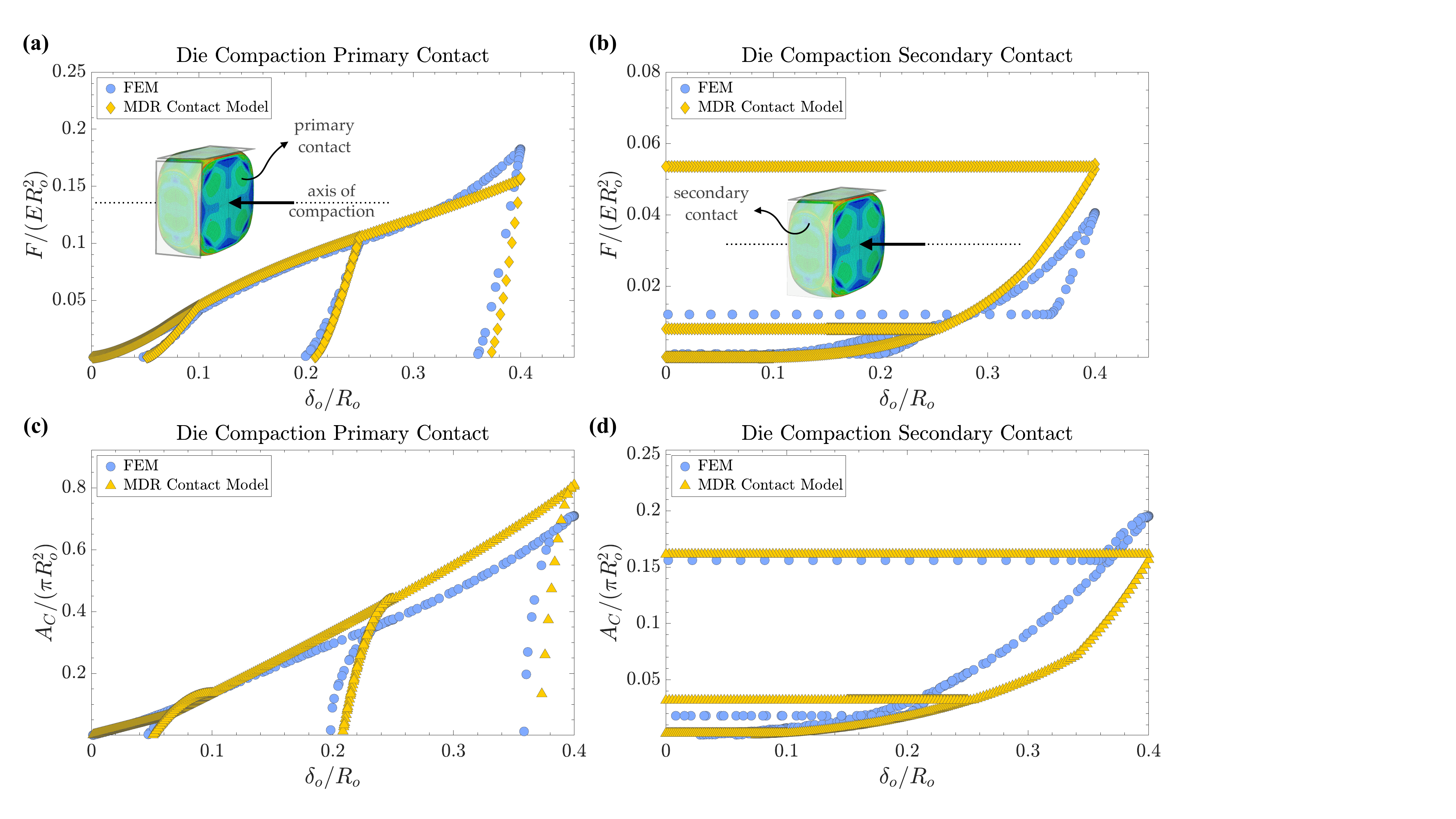}
 	\caption{Force-displacement curves for die compaction including (a) primary contact and (b) secondary contact behavior. Contact area-displacment curves for die compaction including (c) primary contact and (d) secondary contact behavior. }
 	\label{die_compaction_only_force}
 \end{figure*}

Throughout the development of the MDR contact model only one loading type, uniaxial compression of an elastic-plastic sphere between rigid flats, has been considered. To demonstrate the robustness of the contact model formulation, we consider a loading mode known as die compaction, which will be discussed in much further depth in Part II. A small schematic of the loading for a single, initially spherical, particle is shown in both Fig.~\ref{die_compaction_only_force}(a) and (b); it is identical to uniaxial compression between rigid flats along one axis with the addition of four other static rigid flats along the two other orthogonal axes that provide confinement. The static rigid flats do not displace during the loading and are initially placed so that they contact the particle only at a single point. During the loading two types of contacts arise: i) two primary contacts along the axis of compaction and ii) four secondary contacts at the static rigid flats. The primary contacts are similar in nature to those for uniaxial compression, but have an evolution that deviates slightly due to multi-neighbor dependent effects from the presence of the secondary contacts. The secondary contacts are quite distinct from any contact thus far investigated and are not caused by relative displacement between the sphere and a rigid flat, but rather arise due to radial expansion induced by the primary contact, making them often outright ignored by traditional contact models.

\textit{Without any modification to the formulation} the MDR contact model can be compared to a die compaction FEM simulation whose setup is detailed in Part II. The results of the force comparison for a particle with $E/Y = 20$ and $\nu = 0.3$ are shown in Fig.~\ref{die_compaction_only_force}(a) and (b) for the primary and secondary contact\footnote{Because the secondary contacts do not displace, the displacement plotted in Fig.~\ref{die_compaction_only_force}(b) and (d) is that of the primary contact.}, respectively. Focusing on the primary contact first, we note that the FEM force displacement evolution is of similar form to uniaxial compression and is well captured by the MDR contact model. The secondary contact force evolution is notably different. Nevertheless, the MDR contact model detects and evolves the force with reasonable accuracy. Upon unloading the secondary contact, the MDR contact model predicts a persistent non-zero force; this aligns physically with the fact that these contacts are formed by plastic deformation inducing radial expansion, a non-reversible process. It is important to emphasize that no information about the location or the response of these secondary contacts is pre-programmed into the contact model, rather they are detected and evolved entirely by the radius growth scheme. 

A comparison of the contact area-displacement curves is shown in Fig.~\ref{die_compaction_only_force}(c) and (d) again for the primary and secondary contact, respectively. As with the force, the FEM evolution of the primary contact area is similar to that of uniaxial compression. The MDR contact model correspondingly predicts the contact area evolution to a good degree of accuracy. The secondary contact area evolution is also reasonable well detected  by the MDR contact model, with an excellent match of the residual area upon unloading once in the fully-plastic regime. 

The force and contact area results for die compaction help highlight the flexibility of the MDR contact model formulation in capturing drastically different contact evolution. Further results confirming the MDR contact model for a variety of loadings are contained in Part II. 

\subsection{Verification of MDR contact model in adhesive case}

Verification of the adhesive MDR contact model is split into two cases: the elastic regime and the fully-plastic regime. In both cases, comparison is made between the MDR contact model result and the JKR theory of adhesion.     

\subsubsection{Adhesion in the elastic regime} \label{Adhesion in the elastic regime}

Verification of the adhesive MDR contact model in the elastic regime is carried out by comparison to the JKR theory of adhesion for contacting spheres. Using $E^*_c R_o/\Delta\gamma = 2.2\textrm{e}4$, a loading sequence is carried out within the elastic deformation range. The resulting force-displacement curve plotted alongside the JKR theory of adhesion and Hertz's contact law is shown in Fig.~\ref{seperation_force_two_examples}(a).

By design, we see that during forward loading the MDR contact model returns a non-adhesive normal contact behavior following Hertz's contact law precisely. This results from ignoring the ability of material to leap into contact when brought close enough together---a phenomena that the JRK theory of adhesion accounts for. During unloading there is an initial discrepancy between the MDR contact model and the JKR theory of adhesion, which arises because the outer springs have not reached the critical extensional length $\Delta l$. However, once the outer springs reach the critical extensional length, the approximate adhesion theory agrees exactly with the JKR theory of adhesion validating the adhesive MDR contact model for the elastic regime.  

\subsubsection{Adhesion in the fully-plastic regime} \label{Adhesion in the fully-plastic regime}

\begin{figure*} [!htb]
 	\centering
 	%\raggedright
 	% Trim{LEFT LOWER RIGHT UPPER}
 	\includegraphics[width=\textwidth, trim = 1cm  1.5cm 13cm 1cm]{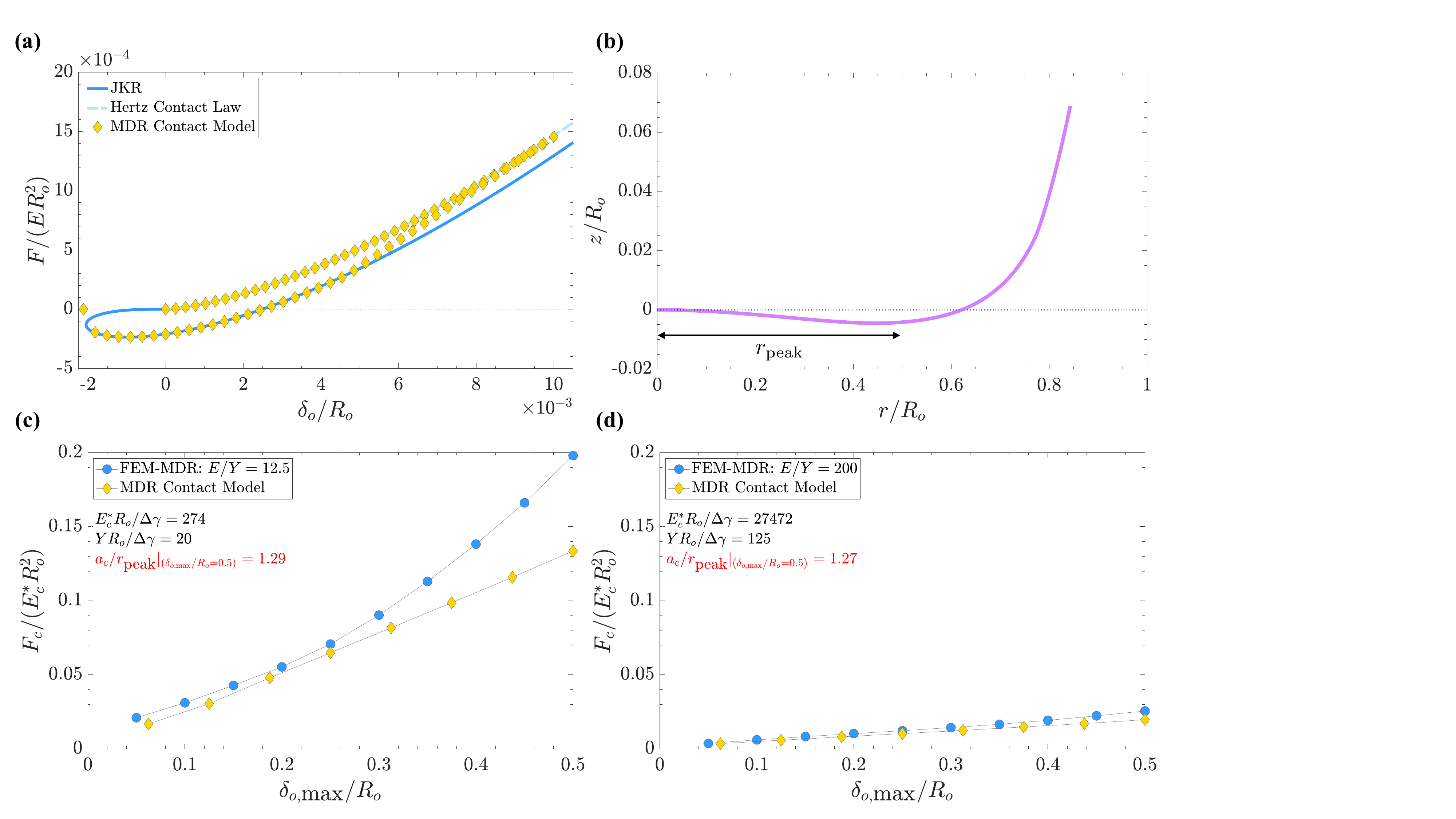}
 	\caption{Adhesive MDR contact model verification. (a) Force-displacement curves for the JKR theory of adhesion between contacting spheres, Hertz's contact law, and the adhesive MDR contact model. (b) Normalized relaxed profile for $E/Y = 12.5$ truncated at $r=a_\textrm{max}$ after loading to $\delta_{o,\textrm{max}}/R_o = 0.4$. (c)-(d) Plots of the normalized critical force versus the normalized maximum displacement. Two curves are shown for comparison with the FEM-MDR and the adhesive MDR contact model. Plot (c) corresponds to $E/Y = 12.5$ and (d) to $E/Y = 200$. The ratio $a_c/r_\textrm{peak}$ (evaluated at $\delta_{o,\textrm{max}}/R_o$ = 0.5) is indicated on each plot as well as the dimensionless groups $Y R_o/\Delta \gamma$ and $E^*_c R_o/\Delta\gamma$. }
 	\label{seperation_force_two_examples}
 \end{figure*}

Adhesive contact behavior is directly affected by the relaxed profiles of the contacting bodies as outlined in Section \ref{Normal contact with adhesion}. In the fully-plastic regime this introduces complexity because the relaxed profile is non-spherical and continuously changing as more plastic deformation accumulates. To appreciate this point we consider a variety of simulations, described in detail in~\ref{Concave-convex behavior of the relaxed profiles}, involving the \textit{non}-adhesive contact problem of an elastic perfectly plastic sphere contacting a rigid flat. From this study, it is revealed that the unloaded profile shapes change dramatically in nature from plastic deformation. Most notably, after sufficient loading $\delta_{o,\textrm{max}}/R_o > 0.15$ it is observed that the unloaded profiles are no longer purely convex, but rather, become concave near the point of first contact before returning to convex as shown in Fig.~\ref{seperation_force_two_examples}(b). Of the reviewed prior contact models for elastic-plastic particles with a mechanically-derived adhesive scheme~\cite{mesarovic2000adhesive,olsson2013force,gonzalez2019generalized} all of them validate their adhesive models for $\delta_{o,\textrm{max}}/R_o < 0.1$, meaning the concave-convex behavior is not considered. For applications involving powder compaction $\delta_{o,\textrm{max}}/R_o > 0.1$ is common, making a mechanically-derived treatment of adhesion at these larger displacements an open and industrially relevant problem.

This property of changing curvature significantly alters the adhesive behavior. Despite this, the adhesive MDR contact model is still able to reliably predict the strength of the adhered contact for a wide range of conditions. To demonstrate the predictive capabilities of the adhesive MDR contact model a series of tests are carried out that measure the separation force $F_c$ as a function of $\delta_{o,\textrm{max}}$ for both the FEM profiles (the ground truth) and MDR contact model as shown in Fig.~\ref{seperation_force_two_examples}(a) and (b). In these tests, the particle is loaded into the fully-plastic regime before being unloaded and detached. Specifics of the test are described in~\ref{Conditions for accuracy of the adhesive MDR contact model}. Significantly different material parameters are used for Fig.~\ref{seperation_force_two_examples}(a) and (b) and the important dimensionless groups are shown directly on the plots. We see that the two curves in both plots show relatively good agreement confirming the validity of the adhesive MDR contact model into deformations as large as $\delta_{o,\textrm{max}}/R_o = 0.5$; this is a notable enhancement to the previous upper bound of $\delta_{o,\textrm{max}}/R_o < 0.1$, expanding the space of problems capable of being modeled.  

 The validity of the adhesive MDR model relies on specific criteria being satisfied; we close this section by elaborating on them. In the presented MDR contact model it is assumed that the mapped 1D indenter is always elliptical in shape, mapping this back to the 3D space gives an indenter that shares a convex profile throughout. Based on this, it is natural to pose the following question:

\begin{itemize}
    \item \textit{Can the purely convex MDR contact model profile ever display similar adhesive behavior to the true concave-convex profile?} 
\end{itemize}

\noindent  From the results just discussed the answer must be yes, but two conditions are required to be satisfied

\begin{equation} \label{adhesive conditions}
a_c/r_\textrm{peak} > 1, \qquad 2a_c/r_\textrm{Ip} \gg 1,
\end{equation}

\noindent where $a_c$ is the critical contact radius of the 1D purely convex indenter, $r_\textrm{peak}$ is the distance from the contact center to the change in curvature as shown in Fig.~\ref{seperation_force_two_examples}(b), and $r_\textrm{Ip} = E^*_c \Delta \gamma/(\pi Y^2)$ is the plastic zone size. In words, condition (\ref{adhesive conditions}) states that the critical contact radius of the 1D purely convex indenter must exceed $r_\textrm{peak}$\footnote{Expressions describing the behavior of $r_\textrm{peak}$ are given in~\ref{Concave-convex behavior of the relaxed profiles}.} and that small scale yielding must prevail. 

To gain intuition for the left condition of (\ref{adhesive conditions}) we note that the critical contact radius of the 3D concave-convex indenter is always greater than or equal to $r_\textrm{peak}$, therefore for $a_c$ of  1D purely convex indenter to ever agree it must also exceed $r_\textrm{peak}$. Next, the more the critical radius exceeds the peaks the less important the specific details of the profile within the contact area become. The right condition of (\ref{adhesive conditions}) is a result of enforcing small-scale yielding and is taken directly from linear elastic fracture mechanics~\cite{anand2023introduction}. It is present to ensure that the plastic zone size is small compared to the critical contact area, which allows solutions from linear elasticity to still be applied; this criterion was similarly identified in~\cite{mesarovic2000adhesive}.

Conditions (\ref{adhesive conditions}) are sufficient to understand whether the adhesive MDR contact model will predict reasonable results. However, because $a_c$ and $r_\textrm{peak}$ both depend on the amount of plastic deformation it becomes a difficult task to predict a priori, before simulating, whether the adhesive contact model will be valid. A derivation of rules for a priori validity of the adhesive contact model and a more rigorous development of the conditions in (\ref{adhesive conditions}) can be found in ~\ref{Conditions for accuracy of the adhesive MDR contact model}.   

\section{Conclusions and perspectives}

In this paper, we have presented a multi-neighbor dependent contact model for adhesive elastic-perfectly plastic spherical particles easily implementable into a DEM code. Importantly, the model relies minimally on fitted parameters, and is driven primarily off of a few physical inputs: the initial radius $R_o$, Young's modulus $E$, Poisson ratio $\nu$, yield stress $Y$, and effective surface energy $\Delta \gamma$. The model is built upon the method of dimensionality reduction (MDR), that allows mapping of 3D axisymmetric contact problems to a simplified 1D problem involving a plane indenter penetrating a bed of independent Hookean springs.  Using MDR transforms, the elastic-plastic contact problem can be entirely evolved in the simplified 1D space while still having a pathway back to the 3D space at all instants. Emphasis was placed on the fact that the force, contact radius, and displacement are shared between the 3D and 1D space provided that the 3D contact is appropriately transformed to 1D space. Treatment of adhesion within the MDR was also added, and it was shown that the JKR theory of adhesion was easily recovered by allowing the MDR springs to stick to the corresponding plane rigid indenter with a separation criterion related to the effective surface energy.  

The elastic regime and fully-plastic regime in the contact model were both treated with an elliptical indenter in the 1D transformed space providing a unified treatment. The transition between the two regimes was constructed to ensure continuity of the force and contact area. Plasticity was accounted for in the fully-plastic regime by changing the aspect ratio of the elliptical indenter subject to a constraint on the pressure determined from the finite element study involving an elastic-perfectly plastic sphere and rigid flat. Importantly, unloading falls naturally out of the the contact model and simply requires lifting the plane indenter out of the 1D bed of springs. A JKR type adhesion was shown to be easily added to the contact model given its relation to the MDR and that it can be accurately applied even after significant plastic deformation has occurred. Multi-neighbor dependent effects were accounted for through requiring the plastic volume change of the particle to be conserved. This resulted in the ability of the particle radius to grow, which in turn helped predict increased force and area at existing contacts. A bulk elastic response is lacking in the current formulation of the contact model and is the subject of Part II.    

Detailed numerical implementations of the contact model were provided, allowing for easy adaptation. Verification of the contact model was carried out by comparison to a variety of finite element simulations representing a wide range of material properties and loading configurations. Satisfactory agreement was found between the two, showing that the contact model is able to accurately capture the evolution of the plastic displacement at the contact, force, contact area, average contact pressure, and particle volume as a function of displacement. 

The ratio $E/Y \approx 10$ was discussed as a conservative lower bound for usage of applicability of the contact model, after which elastic strains exceed typically accepted values for linear elasticity to hold. No upper bound on $E/Y$ exists and the proposed model can be used through the rigid plastic limit. Conditions for the applicability of adhesion were also discussed. 

The presented contact model makes progress towards a comprehensive contact model for interacting elastic-plastic particles. Importantly, it sets up a rigorous framework that is adaptable to new effects such as interacting particles with varying radii and hardening. The latter would require a new suite of finite element simulations to understand the effects on contact radius and average pressure. The adhesive model could also be improved by adding higher order terms in the pressure profile which would result in a slightly different indenter shape in the 1D space. Each of these improvements is considered a worthwhile pursuit and would lead to an even more complete contact model for elastic-perfectly plastic spherical particles.  

\section*{Declaration of competing interest}

The authors declare that they have no known competing financial interests or personal relationships that could have appeared to influence the work reported in this paper.

\section*{Acknowledgements}

The authors acknowledge the support of the International Fine Particle Research Institute (IFPRI) (Grant ARR-109-01). The authors would like to thank the members of the IFPRI for useful discussions related to the development of the contact model.

 \bibliographystyle{elsarticle-num} 
 \bibliography{bib}

\begin{thebibliography}{10}
\expandafter\ifx\csname url\endcsname\relax
  \def\url#1{\texttt{#1}}\fi
\expandafter\ifx\csname urlprefix\endcsname\relax\def\urlprefix{URL }\fi
\expandafter\ifx\csname href\endcsname\relax
  \def\href#1#2{#2} \def\path#1{#1}\fi

\bibitem{zunker2023partII}
W.~Zunker, K.~Kamrin, A mechanically-derived contact model for adhesive elastic-perfectly plastic particles. {P}art {II}: {C}ontact under high compaction---adding a bulk elastic response, arXiv preprint (2023).

\bibitem{samal2015powder}
P.~Samal, J.~Newkirk, Powder metallurgy methods and applications, ASM handbook of powder metallurgy 7 (2015).

\bibitem{sigmund2000novel}
W.~M. Sigmund, N.~S. Bell, L.~Bergstr{\"o}m, Novel powder-processing methods for advanced ceramics, Journal of the American Ceramic Society 83~(7) (2000) 1557--1574.

\bibitem{meier2019modeling}
C.~Meier, R.~Weissbach, J.~Weinberg, W.~A. Wall, A.~J. Hart, Modeling and characterization of cohesion in fine metal powders with a focus on additive manufacturing process simulations, Powder technology 343 (2019) 855--866.

\bibitem{ccelik2016pharmaceutical}
M.~{\c{C}}elik, Pharmaceutical powder compaction technology, CRC Press, 2016.

\bibitem{laptev2005green}
A.~Laptev, O.~Vyal, M.~Bram, H.~Buchkremer, D.~St{\"o}ver, Green strength of powder compacts provided for production of highly porous titanium parts, Powder Metallurgy 48~(4) (2005) 358--364.

\bibitem{boudina2022insight}
I.~Boudina, E.~Rondet, S.~Nezamabadi, T.~Sharkawi, Insight into tableted pellets by combining x-ray micro-computed tomography and experimental compaction, Powder Technology 397 (2022) 117083.

\bibitem{mashadi1987characterization}
A.~B. Mashadi, J.~Newton, The characterization of the mechanical properties of microcrystalline cellulose: a fracture mechanics approach, Journal of pharmacy and pharmacology 39~(12) (1987) 961--965.

\bibitem{wu2006predicting}
C.-Y. Wu, S.~M. Best, A.~C. Bentham, B.~C. Hancock, W.~Bonfield, Predicting the tensile strength of compacted multi-component mixtures of pharmaceutical powders, Pharmaceutical research 23 (2006) 1898--1905.

\bibitem{cundall1979discrete}
P.~A. Cundall, O.~D. Strack, A discrete numerical model for granular assemblies, geotechnique 29~(1) (1979) 47--65.

\bibitem{walton1986viscosity}
O.~R. Walton, R.~L. Braun, Viscosity, granular-temperature, and stress calculations for shearing assemblies of inelastic, frictional disks, Journal of rheology 30~(5) (1986) 949--980.

\bibitem{thornton1991impact}
C.~Thornton, K.~Yin, Impact of elastic spheres with and without adhesion, Powder technology 65~(1-3) (1991) 153--166.

\bibitem{iwashita1998rolling}
K.~Iwashita, M.~Oda, Rolling resistance at contacts in simulation of shear band development by dem, Journal of engineering mechanics 124~(3) (1998) 285--292.

\bibitem{zhou1999rolling}
Y.~Zhou, B.~Wright, R.~Yang, B.~H. Xu, A.-B. Yu, Rolling friction in the dynamic simulation of sandpile formation, Physica A: Statistical Mechanics and its Applications 269~(2-4) (1999) 536--553.

\bibitem{dintwa2005torsion}
E.~Dintwa, M.~V. Zeebroeck, E.~Tijskens, H.~Ramon, Torsion of viscoelastic spheres in contact, Granular Matter 7~(2) (2005) 169--179.

\bibitem{johnson1971surface}
K.~L. Johnson, K.~Kendall, A.~D. Roberts, Surface energy and the contact of elastic solids, Proceedings of the royal society of London. A. mathematical and physical sciences 324~(1558) (1971) 301--313.

\bibitem{maugis1992adhesion}
D.~Maugis, Adhesion of spheres: the jkr-dmt transition using a dugdale model, Journal of colloid and interface science 150~(1) (1992) 243--269.

\bibitem{soulie2006influence}
F.~Soulie, F.~Cherblanc, M.~S. El~Youssoufi, C.~Saix, Influence of liquid bridges on the mechanical behaviour of polydisperse granular materials, International journal for numerical and analytical methods in geomechanics 30~(3) (2006) 213--228.

\bibitem{olsson2019contact}
E.~Olsson, D.~Jelagin, A contact model for the normal force between viscoelastic particles in discrete element simulations, Powder Technology 342 (2019) 985--991.

\bibitem{storaakers1997similarity}
B.~Stor{\aa}kers, S.~Biwa, P.-L. Larsson, Similarity analysis of inelastic contact, International Journal of Solids and Structures 34~(24) (1997) 3061--3083.

\bibitem{hertz1882beruhrung}
H.~Hertz, {\"U}ber die ber{\"u}hrung fester elastischer k{\"o}rper, Journal f{\"u}r die reine und angewandte Mathematik 92~(156-171) (1882) 22.

\bibitem{mowlavi2021contact}
S.~Mowlavi, K.~Kamrin, Contact model for elastically anisotropic bodies and efficient implementation into the discrete element method, Granular Matter 23~(2) (2021) 1--29.

\bibitem{johnson1970correlation}
K.~L. Johnson, The correlation of indentation experiments, Journal of the Mechanics and Physics of Solids 18~(2) (1970) 115--126.

\bibitem{kogut2002elastic}
L.~Kogut, I.~Etsion, Elastic-plastic contact analysis of a sphere and a rigid flat, J. Appl. Mech. 69~(5) (2002) 657--662.

\bibitem{tsigginos2015force}
C.~Tsigginos, J.~Strong, A.~Zavaliangos, On the force--displacement law of contacts between spheres pressed to high relative densities, International Journal of Solids and Structures 60 (2015) 17--27.

\bibitem{thornton1998theoretical}
C.~Thornton, Z.~Ning, A theoretical model for the stick/bounce behaviour of adhesive, elastic-plastic spheres, Powder technology 99~(2) (1998) 154--162.

\bibitem{stronge2000contact}
W.~Stronge, Contact problems for elasto-plastic impact in multi-body systems, in: Impacts in Mechanical Systems: Analysis and Modelling, Springer, 2000, pp. 189--234.

\bibitem{zhang2002modeling}
X.~Zhang, L.~Vu-Quoc, Modeling the dependence of the coefficient of restitution on the impact velocity in elasto-plastic collisions, International journal of impact engineering 27~(3) (2002) 317--341.

\bibitem{du2009energy}
Y.~Du, S.~Wang, Energy dissipation in normal elastoplastic impact between two spheres, Journal of Applied Mechanics 76~(6) (07 2009).

\bibitem{jackson2010predicting}
R.~L. Jackson, I.~Green, D.~B. Marghitu, Predicting the coefficient of restitution of impacting elastic-perfectly plastic spheres, Nonlinear Dynamics 60 (2010) 217--229.

\bibitem{wu2022energy}
L.~Wu, D.~Huang, Energy dissipation study in impact: From elastic and elastoplastic analysis in peridynamics, International Journal of Solids and Structures 234 (2022) 111279.

\bibitem{thornton2017elastic}
C.~Thornton, S.~J. Cummins, P.~W. Cleary, On elastic-plastic normal contact force models, with and without adhesion, Powder Technology 315 (2017) 339--346.

\bibitem{garner2018study}
S.~Garner, J.~Strong, A.~Zavaliangos, Study of the die compaction of powders to high relative densities using the discrete element method, Powder Technology 330 (2018) 357--370.

\bibitem{giannis2021modeling}
K.~Giannis, C.~Schilde, J.~H. Finke, A.~Kwade, Modeling of high-density compaction of pharmaceutical tablets using multi-contact discrete element method, Pharmaceutics 13~(12) (2021) 2194.

\bibitem{chang1987elastic}
W.~R. Chang, I.~Etsion, D.~B. Bogy, An elastic-plastic model for the contact of rough surfaces, Journal of Tribology 109~(2) (1987) 257--263.

\bibitem{mesarovic2000adhesive}
S.~D. Mesarovic, K.~L. Johnson, Adhesive contact of elastic--plastic spheres, Journal of the Mechanics and Physics of Solids 48~(10) (2000) 2009--2033.

\bibitem{zhao2000asperity}
Y.~Zhao, D.~M. Maietta, L.~Chang, An asperity microcontact model incorporating the transition from elastic deformation to fully plastic flow, J. Trib. 122~(1) (2000) 86--93.

\bibitem{jackson2003finite}
R.~L. Jackson, I.~Green, A finite element study of elasto-plastic hemispherical contact, in: International Joint Tribology Conference, Vol. 37068, 2003, pp. 65--72.

\bibitem{jackson2005finite}
R.~L. Jackson, I.~Green, A finite element study of elasto-plastic hemispherical contact against a rigid flat, J. Trib. 127~(2) (2005) 343--354.

\bibitem{etsion2005unloading}
I.~Etsion, Y.~Kligerman, Y.~Kadin, Unloading of an elastic--plastic loaded spherical contact, International Journal of Solids and Structures 42~(13) (2005) 3716--3729.

\bibitem{harthong2009modeling}
B.~Harthong, J.-F. J{\'e}rier, P.~Dor{\'e}mus, D.~Imbault, F.-V. Donz{\'e}, Modeling of high-density compaction of granular materials by the discrete element method, International Journal of Solids and Structures 46~(18-19) (2009) 3357--3364.

\bibitem{zait2010unloading}
Y.~Zait, Y.~Kligerman, I.~Etsion, Unloading of an elastic--plastic spherical contact under stick contact condition, International Journal of Solids and Structures 47~(7-8) (2010) 990--997.

\bibitem{brake2012analytical}
M.~Brake, An analytical elastic-perfectly plastic contact model, International Journal of Solids and Structures 49~(22) (2012) 3129--3141.

\bibitem{gonzalez2012nonlocal}
M.~Gonzalez, A.~M. Cuiti{\~n}o, A nonlocal contact formulation for confined granular systems, Journal of the Mechanics and Physics of Solids 60~(2) (2012) 333--350.

\bibitem{agarwal2018contact}
A.~Agarwal, M.~Gonzalez, Contact radius and curvature corrections to the nonlocal contact formulation accounting for multi-particle interactions in elastic confined granular systems, International Journal of Engineering Science 133 (2018) 26--46.

\bibitem{olsson2013force}
E.~Olsson, P.-L. Larsson, On force--displacement relations at contact between elastic--plastic adhesive bodies, Journal of the Mechanics and Physics of Solids 61~(5) (2013) 1185--1201.

\bibitem{frenning2013towards}
G.~Frenning, Towards a mechanistic model for the interaction between plastically deforming particles under confined conditions: A numerical and analytical analysis, Materials Letters 92 (2013) 365--368.

\bibitem{frenning2015towards}
G.~Frenning, Towards a mechanistic contact model for elastoplastic particles at high relative densities, Finite Elements in Analysis and Design 104 (2015) 56--60.

\bibitem{brodu2015multiple}
N.~Brodu, J.~A. Dijksman, R.~P. Behringer, Multiple-contact discrete-element model for simulating dense granular media, Physical Review E 91~(3) (2015) 032201.

\bibitem{rathbone2015accurate}
D.~Rathbone, M.~Marigo, D.~Dini, B.~van Wachem, An accurate force--displacement law for the modelling of elastic--plastic contacts in discrete element simulations, Powder technology 282 (2015) 2--9.

\bibitem{gonzalez2019generalized}
M.~Gonzalez, Generalized loading-unloading contact laws for elasto-plastic spheres with bonding strength, Journal of the Mechanics and Physics of Solids 122 (2019) 633--656.

\bibitem{edmans2020numerical}
B.~D. Edmans, I.~C. Sinka, Numerical derivation of a normal contact law for compressible plastic particles, Mechanics of Materials 146 (2020) 103297.

\bibitem{edmans2020unloading}
B.~D. Edmans, I.~C. Sinka, Unloading of elastoplastic spheres from large deformations, Powder Technology 374 (2020) 618--631.

\bibitem{giannis2021stress}
K.~Giannis, C.~Schilde, J.~H. Finke, A.~Kwade, M.~Celigueta, K.~Taghizadeh, S.~Luding, Stress based multi-contact model for discrete-element simulations, Granular matter 23~(2) (2021) 1--14.

\bibitem{zhang2022research}
W.~Zhang, J.~Chen, C.~Wang, D.~Liu, L.~Zhu, Research on elastic--plastic contact behavior of hemisphere flattened by a rigid flat, Materials 15~(13) (2022) 4527.

\bibitem{xu2023new}
J.~Xu, J.~Zhu, A new contact model of sphere asperity in the fully plastic regime considering strain hardening, Journal of Applied Mechanics (2023) 1--25.

\bibitem{popov2013methode}
V.~L. Popov, M.~He{\ss}, Methode der Dimensionsreduktion in Kontaktmechanik und Reibung, Springer, 2013.

\bibitem{popov2015method}
V.~L. Popov, M.~He{\ss}, Method of dimensionality reduction in contact mechanics and friction, Springer, 2015.

\bibitem{foppl1941elastische}
L.~F{\"o}ppl, Elastische beanspruchung des erdbodens unter fundamenten, Forschung auf dem Gebiet des Ingenieurwesens A 12~(1) (1941) 31--39.

\bibitem{schubert1942frage}
G.~Schubert, Zur frage der druckverteilung unter elastisch gelagerten tragwerken, Ingenieur-Archiv 13~(3) (1942) 132--147.

\bibitem{popov2019handbook}
V.~L. Popov, M.~He{\ss}, E.~Willert, Handbook of contact mechanics: exact solutions of axisymmetric contact problems, Springer Nature, 2019.

\bibitem{tabor2000hardness}
D.~Tabor, The hardness of metals, Oxford university press, 2000.

\bibitem{ishlinsky1944problem}
A.~Ishlinsky, The problem of plasticity with axial symmetry and brinell’s test, J. Appl. Math. Mech 8 (1944) 201--224.

\bibitem{lee1972analysis}
C.~Lee, S.~Masaki, S.~Kobayashi, Analysis of ball indentation, International Journal of Mechanical Sciences 14~(7) (1972) 417--426.

\bibitem{quicksall2004elasto}
J.~J. Quicksall, R.~L. Jackson, I.~Green, Elasto-plastic hemispherical contact models for various mechanical properties, Proceedings of the Institution of Mechanical Engineers, Part J: Journal of Engineering Tribology 218~(4) (2004) 313--322.

\bibitem{hill1998mathematical}
R.~Hill, The mathematical theory of plasticity, Vol.~11, Oxford university press, 1998.

\bibitem{ghaednia2017review}
H.~Ghaednia, X.~Wang, S.~Saha, Y.~Xu, A.~Sharma, R.~L. Jackson, A review of elastic--plastic contact mechanics, Applied Mechanics Reviews 69~(6) (2017).

\bibitem{hui2000accuracy}
C.~Hui, Y.~Lin, J.~Baney, A.~Jagota, The accuracy of the geometric assumptions in the jkr (johnsonkendallroberts) theory of adhesion, Journal of adhesion science and technology 14~(10) (2000) 1297--1319.

\bibitem{christoffersen1981micromechanical}
J.~Christoffersen, M.~M. Mehrabadi, S.~Nemat-Nasser, {A Micromechanical Description of Granular Material Behavior}, Journal of Applied Mechanics 48~(2) (1981) 339--344.

\bibitem{anand2023introduction}
L.~Anand, K.~Kamrin, S.~Govindjee, Introduction to Mechanics of Solid Materials, OXFORD University Press, 2023.

\bibitem{johnson1987contact}
K.~L. Johnson, Contact mechanics, Cambridge university press, 1987.

\bibitem{barber2018contact}
J.~R. Barber, Contact mechanics, Springer, 2018.

\bibitem{bracewell1986fourier}
R.~N. Bracewell, The Fourier transform and its applications, Vol. 31999, McGraw-Hill New York, 1986.

\bibitem{hess2011exakte}
M.~He{\ss}, {\"U}ber die exakte Abbildung ausgew{\"a}hlter dreidimensionaler Kontakte auf Systeme mit niedrigerer r{\"a}umlicher Dimension, Cuvillier Verlag, 2011.

\bibitem{tabor1948simple}
D.~Tabor, A simple theory of static and dynamic hardness, Proceedings of the Royal Society of London. Series A. Mathematical and Physical Sciences 192~(1029) (1948) 247--274.

\bibitem{ashby1995mechanical}
M.~F. Ashby, L.~Gibson, U.~Wegst, R.~Olive, The mechanical properties of natural materials. i. material property charts, Proceedings of the Royal Society of London. Series A: Mathematical and Physical Sciences 450~(1938) (1995) 123--140.

\end{thebibliography}





% \end{thebibliography}

%%%%%%%%% APPENDIX %%%%%%%%%%% 
\appendix
\clearpage
\section{Details of the method of dimensionality reduction}

\subsection{Derivation of the method of dimensionality reduction} \label{Derivation of the method of dimensionality reduction}

\begin{figure*} [!htb]
 	\centering
 	%\raggedright
 	% Trim{LEFT LOWER RIGHT UPPER}
 	\includegraphics[scale=0.32, trim = 0cm  16cm 40cm 0cm]{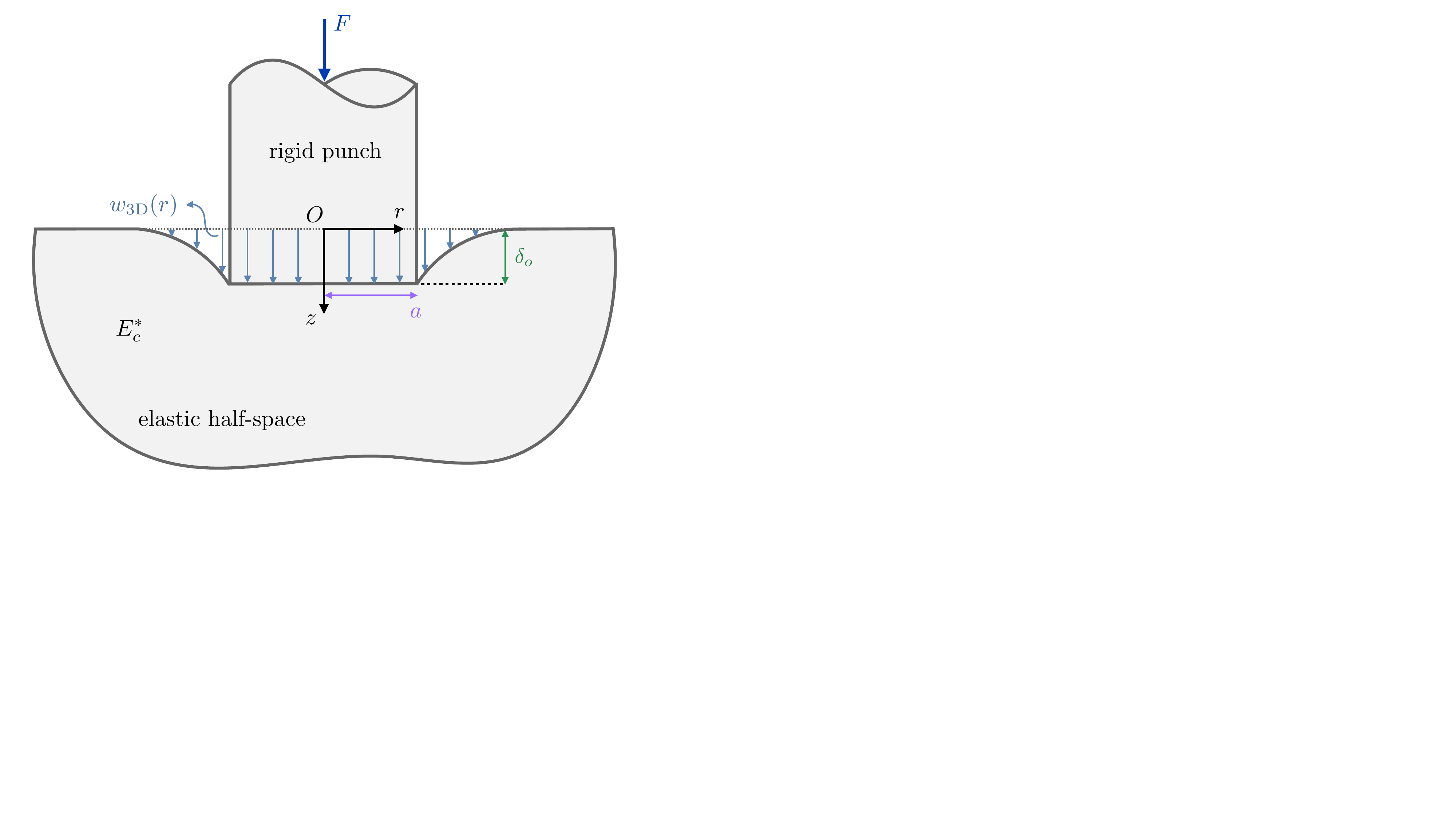}
 	\caption[Contact between a rigid cylindrical punch and elastic half-space.]{Deformed configuration for contact between a rigid cylindrical punch and elastic half-space characterized by a plane strain modulus $E^*_c$. In the reference configuration the two bodies initially touch with the surface of the rigid punch flush with the elastic half-space. A cylindrical coordinate system with origin $O$ is placed at the initial point of contact and on the rotational axis of symmetry of the rigid punch.  The radial coordinate $r$ is oriented parallel to the half space and the $z$ coordinate directed into the elastic half-space along the rotational axis of symmetry of the rigid punch. The contact is driven by a force $F$ that gives rise to a normal displacement profile $w_\textrm{3D}(r)$ and displacement $\delta_o$. The contact radius $a$ is given trivially as the radius of the rigid punch.}
 	\label{rigid_punch}
 \end{figure*}

Central to deriving the method of dimensionality reduction (MDR) is the solution to a flat rigid cylindrical punch contacting an infinite half-space as shown in Fig.~\ref{rigid_punch}. We recall the important results of the problem without a complete derivation, readers interested in the full solution to this classic problem can find it in most texts on the subject of contact mechanics (e.g ~\cite{johnson1987contact,barber2018contact}). 

In this problem, a force $F$ is applied to the punch of radius $a$ causing the elastic half-space characterized by a composite plain strain modulus $E^*_c$ to deform. The axisymmetry of the problem leads to the natural use of a cylindrical coordinate system with the radial coordinate $r$ oriented parallel to the half-space and the $z$ coordinate oriented normal and directed into the half-space. The total depth of penetration is given by $\delta_o$ and the contact radius trivially by $a$. The normal displacement profile of the elastic half-space $w_\textrm{3D}(r)$ is given by the piecewise function

    \begin{equation} \label{w3d punch}
   w_\textrm{3D}(r) = \begin{cases}
      \delta_o, & r < a,\\
      \frac{2}{\pi}\delta_o \arcsin{(a/r)}, & r \geq a.
    \end{cases} 
    \end{equation}

\noindent The pressure distribution at the contact interface $p_\textrm{3D}(r)$ is of the form

    \begin{equation} \label{p3d punch}
        p_\textrm{3D}(r) = \frac{1}{\pi}\frac{E^*_c \delta_o}{\sqrt{a^2-r^2}}.
    \end{equation}

\noindent Integrating the pressure distribution over the contact area leads to an expression for the force $F$

    \begin{equation} \label{F punch}
        F = 2E^*_c a \delta_o.
    \end{equation}

\noindent The contact stiffness $k_z$ is defined as the proportionality constant between the force and displacement

    \begin{equation} \label{kz punch}
        k_z = 2E^*_c a.
    \end{equation}

Equations (\ref{w3d punch}-\ref{kz punch}) summarize the key results from the solution of a flat rigid punch contacting an infinite half-space needed to derive the MDR. We now turn to the similar, but more general problem of a  axisymmetric rigid indenter contacting an elastic half space\footnote{The true generality of this problem can be seen by noting the fact that in the limit of small deformation contact between any two arbitrary axisymmetric bodies can be transformed to an equivalent problem of a rigid indenter contacting an elastic half-space. The transformation is simple and merely requires that the gap between the rigid indenter and half-space be the same as that of the two arbitrary axisymmetric bodies}. We restrict our attention to indenters with simply connected profiles containing no concave regions that give rise to a compact (i.e. continuous) contact area. The profile of the rigid indenter is given as $g_\textrm{3D}(r)$ and the elastic half-space is once again characterized by $E^*_c$. A normal force $F$ is applied to the rigid indenter giving rise to a unique displacement and contact radius pair: $\delta_o$ and $a$, respectively. Unlike in the case of a rigid cylindrical punch the displacement is now a function of the contact radius

    \begin{equation} \label{delta_o = g(a)}
        \delta_o = g(a)
    \end{equation}

\noindent Instead of only considering the final deformed configuration we consider now the continuous sequence of equilibrium states that are passed through from first contact to the final displacement $\delta_o$. We denote any one of these given equilibrium states by the triplet $\{ \tilde{F},\tilde{a},\tilde{\delta}_o \}$ where

    \begin{equation}
        \{0,0,0\} \leq \{ \tilde{F},\tilde{a},\tilde{\delta}_o \} \leq \{F,a,\delta_o \}. 
    \end{equation}

\noindent Integrating the incremental change in $\tilde{F}$ over the loading sequence results in the total force

    \begin{equation}
        	F = \int_{0}^{F} \,d\tilde{F} = \int_{0}^{a} \frac{d\tilde{F}}{d\tilde{\delta}_o} \frac{d\tilde{\delta}_o}{d\tilde{a}}\,d\tilde{a}.
    \end{equation}

\noindent Recalling (\ref{kz punch}) we have that $d\tilde{F}/d\tilde{\delta}_o = 2E^*_c\tilde{a}$, additionally (\ref{delta_o = g(a)}) gives $\tilde{\delta}_o = g(\tilde{a})$. The combination of these results allows us to rewrite the above integral as

    \begin{equation}
        	F = 2E^*_c\int_{0}^{a} \tilde{a} \frac{dg(\tilde{a})}{d\tilde{a}}\,d\tilde{a}.
    \end{equation}

\noindent Applying integration by parts leads to

\begin{equation} \label{F MDR derivation}
\begin{split}
 F & = 2E^*_c\left[ag(a) - \int_{0}^{a} g(\tilde{a})\,d\tilde{a}\right], \\
 & = 2E^*_c\left[ \int_{0}^{a} g(a) - g(\tilde{a})\,d\tilde{a} \right], \\
  & = 2E^*_c\left[ \int_{0}^{a} \delta_o - g(\tilde{a})\,d\tilde{a} \right]. \\
\end{split}
\end{equation}

With an expression for the force derived, we now turn to determining the pressure distribution at the contact interface. We begin by taking an incremental form of the pressure distribution (\ref{p3d punch}) caused by the flat rigid punch of radius $\tilde{a}$  

    \begin{equation}
        dp_\textrm{3D}(r) = \frac{1}{\pi}\frac{E^*_c}{\sqrt{\tilde{a}^2-r^2}}d\tilde{\delta}_o, \quad \textrm{for } r<\tilde{a}.
    \end{equation}

\noindent The sum of the incremental changes in pressure over all increments in displacement gives the total pressure distribution

    \begin{equation}
    \begin{split}
        p_\textrm{3D}(r) & = \int_{\delta_o(r)}^{\delta_o} \frac{1}{\pi}\frac{E^*_c}{\sqrt{\tilde{a}^2-r^2}}\,d\tilde{\delta}_o, \\
        & = \int_{r}^{a} \frac{1}{\pi}\frac{E^*_c}{\sqrt{\tilde{a}^2-r^2}}\frac{d\tilde{\delta}_o}{d\tilde{a}} \,d\tilde{a}.
    \end{split}    
    \end{equation}

\noindent Again noting that $\tilde{\delta}_o = g(\tilde{a})$, we can write

    \begin{equation} \label{p3d MDR derivation}
        p_\textrm{3D}(r) = \int_{r}^{a} \frac{1}{\pi}\frac{E^*_c}{\sqrt{\tilde{a}^2-r^2}}\frac{dg(\tilde{a})}{d\tilde{a}} \,d\tilde{a}.
    \end{equation}

Inspection of (\ref{F MDR derivation}) and (\ref{p3d MDR derivation}) shows that the function $g(a)$ uniquely defines both the force and pressure distribution. The task therefore becomes determining the function $g(a)$. To accomplish this we again return to the the solution of a flat rigid punch. Taking the incremental form of (\ref{w3d punch}) we can relate the infinitesimal normal surface displacement at the point $r=a$ to an infinitesimal indentation of a rigid punch of radius $\tilde{a} < a$

    \begin{equation}
   dw_\textrm{3D}(a) = \frac{2}{\pi} \arcsin{(\tilde{a}/a)} d\tilde{\delta}_o
    \end{equation}

Summing incremental changes of the vertical displacement over all increments in displacement gives the total vertical displacement

    \begin{equation}
    \begin{split}
        w_\textrm{3D}(a) & = \frac{2}{\pi}\int_{0}^{\delta_o} \arcsin{(\tilde{a}/a)}\,d\tilde{\delta}_o, \\
        & = \frac{2}{\pi}\int_{0}^{a}\arcsin{(\tilde{a}/a)}\frac{d\tilde{\delta}_o}{d\tilde{a}} \,d\tilde{a}.
    \end{split}    
    \end{equation}

\noindent Again noting that $\tilde{\delta}_o = g(\tilde{a})$, we can write

    \begin{equation}
        w_\textrm{3D}(a)  = \frac{2}{\pi}\int_{0}^{a}\arcsin{(\tilde{a}/a)}\frac{dg(\tilde{a})}{d\tilde{a}} \,d\tilde{a}.
    \end{equation}

\noindent Clearly, the vertical displacement must be equal to $w_\textrm{3D}(a) = \delta_o - g_\textrm{3D}(a)$

    \begin{equation}
        \delta_o - g_\textrm{3D}(a)  = \frac{2}{\pi}\int_{0}^{a}\arcsin{(\tilde{a}/a)}\frac{dg(\tilde{a})}{d\tilde{a}} \,d\tilde{a}.
    \end{equation}

\noindent Integration by parts and usage of $\tilde{\delta}_o = g(\tilde{a})$ gives the following form

    \begin{equation}
        g_\textrm{3D}(a)  = \frac{2}{\pi}\int_{0}^{a}\frac{g(\tilde{a})}{\sqrt{a^2 - \tilde{a}^2}} \,d\tilde{a}.
    \end{equation}

\noindent The above integral can be recognized as Abel's integral equation which has a known method for solving for $g(a)$~\cite{bracewell1986fourier}

    \begin{equation} \label{g(a) MDR derivation}
        g(a)  = a\int_{0}^{a}\frac{g^\prime_{\textrm{3D}}(\tilde{a})}{\sqrt{a^2 - \tilde{a}^2}} \,d\tilde{a}.
    \end{equation}

Comparison of (\ref{F MDR derivation}), (\ref{p3d MDR derivation}), and (\ref{g(a) MDR derivation}) with the results given in Section 3 shows that they exactly coincide, justifying the framework of the MDR.

\subsection{Critical extensional length} \label{Critical extensional length}

Adhesive normal contact can be seen as the superposition of two normal contact problems: a non-adhesive normal contact problem and a adhesive retraction. For the adhesive contact between axisymmetric bodies each of these problems can be independently mapped using the method of dimensionality reduction (MDR). Therefore the superposition of the two using the MDR is also valid. In the transformed MDR space the superposition can be visualized as follows. First, we consider the non-adhesive ($\textrm{n.a.}$) problem in which a force $F_\text{n.a.}$ is applied to the plane rigid indenter causing it to penetrate the bed of springs by an displacement $\delta_{o,\textrm{n.a.}}$ and form a finite contact radius $a$. We now assume that the springs adhere to the indenter surface and that a vertical translation occurs that lifts the indenter out of the bed of springs. The adhesive retraction is stopped once the outer springs reach the maximum allowable elongation before separation $\Delta l(a)$. During this translation the decompression of the springs causes the force to reduce to $F$ and the displacement to $\delta_o$, however the contact radius $a$ remains unchanged since no springs have detached. The equilibrium state given by $\{F,a,\delta_o \}$ corresponds exactly to that of the unmapped adhesive problem.

What remains is to determine the the maximum allowable elongation or critical length $\Delta l(a)$. To do this, we use the principle of virtual work, which states that a system is at equilibrium only when the internal and external work are equal for any conceivable virtual displacement. Applied to the problem at hand of adhesive contact we consider a perturbation that results in a small reduction of the contact radius $a \rightarrow a-\Delta x$. This results in a change in both the elastic energy and surface energy at the contact. To be at equilibrium the changes in these two energies must equal one another. Stated another way, we search for the configuration at which there is an indifference between adhesion and detachment of the spring. The change in surface energy in the unmapped space is given in terms of the area that has separated multiplied by the effective surface energy $\Delta \gamma$

    \begin{equation}
        \Delta E_\textrm{adh} = 2\pi a \Delta x \Delta \gamma.
    \end{equation}

\noindent The elastic energy can be expressed in terms of the mapped MDR space since the force-displacement relation is captured exactly by the MDR. The detachment of each of spring leads to a total loss of elastic energy given by  

    \begin{equation}
        \Delta E_\textrm{elas} = E^*_c \Delta x \Delta l^2.
    \end{equation}

\noindent Equating the two changes in energy and solving for $\Delta l$ recovers the expression for the critical length given previously without justification,

    \begin{equation} \label{critical l appendix}
	\Delta l(a) = \sqrt{\frac{2 \pi a \Delta \gamma}{E^*_c}}.
    \end{equation}

\noindent This result for the critical length was first given in~\cite{hess2011exakte}.

Although expression (\ref{critical l appendix}) suffices to characterize adhesion in the MDR space, it is informative to recast the criterion through the lens of linear elastic fracture mechanics (LEFM)~\cite{anand2023introduction}. To accomplish this we again note that adhesive contact can be seen as the superposition of two problems: a non-adhesive contact and a adhesive retraction through some distance $l$. Correspondingly, the normal stress $\sigma_\textrm{3D}$ (where we take tension as positive) along an adhesive contact, is the summation of a non-adhesive (n.a.) and an adhesive retraction (a.r.) contribution 

\begin{equation} \label{adhesive stress decomp}
    \sigma_\textrm{3D}(r;a) = \sigma_\textrm{3D,n.a.}(r;a) + \sigma_\textrm{3D,a.r.}(r;a).
\end{equation}

\noindent The adhesive retraction portion of the stress is exactly that of the normal stress profile beneath an adhesive cylindrical punch of radius $a$ displaced in tension by a distance $l$

\begin{equation} \label{punch stress field}
    \sigma_\textrm{3D,a.r.}(r;a) = \frac{E^*_c l}{\pi\sqrt{a^2-r^2}}.
\end{equation}

\noindent Here, what is of interest is the asymptotic stress field near the edge of the adhesive contact as $r\rightarrow a$ (i.e. nearby the \textit{crack} tip). By expanding (\ref{punch stress field}) about $r=a$ the asymptotic stress field can be written as

\begin{equation} 
    \sigma^\textrm{(asy)}_\textrm{3D}(r;a) = \sigma_\textrm{3D,n.a.}(r;a) + \frac{E^*_c l}{\sqrt{\pi a} \sqrt{2\pi (a - r)}} + \{ \textrm{bounded terms} \}.
\end{equation}

\noindent Noting that the non-adhesive portion of the stress goes to zero as $r \rightarrow a$ we arrive at our familar K-field result

\begin{equation} 
    \sigma^\textrm{(asy)}_\textrm{3D}(r;a) = \frac{K_I}{\sqrt{2\pi (a - r)}} \qquad \textrm{as } r \rightarrow a,
\end{equation}

\noindent where the stress intensity factor is

\begin{equation} \label{K_I}
    K_I = \frac{E^*_c l}{\sqrt{\pi a}}.
\end{equation}

\noindent Transforming the K-field to its corresponding 1D force density we find

\begin{equation} 
q_\textrm{1D}(x) = 2\int_{x}^{\infty} \frac{r\, \sigma^\textrm{(asy)}_\textrm{3D}}{\sqrt{r^2-x^2}} \,dr \qquad \Rightarrow \qquad q^{\textrm{(asy)}}_\textrm{1D}(a) = K_I\sqrt{\pi a}.
\end{equation}

\noindent From (\ref{critical l appendix}) we know the critical extensional length before separation, this allows us to rewrite (\ref{K_I}) as

\begin{equation} \label{K_Ic}
    K_{Ic} = \sqrt{2 \Delta \gamma E^*_c},
\end{equation}

\noindent which is known as the critical stress intensity factor. Finally, we arrive at the alternative LEFM adhesive criterion which states that fracture or de-adhesion occurs when the 1D linear force density reaches the following value at the edge of the contact

\begin{equation}
q^{\textrm{(asy)}}_\textrm{1D}(a) = K_{Ic}\sqrt{\pi a}.
\end{equation}

\subsection{Stability condition} \label{Stability Condition}

If we imagine an experiment where we continuously vary through different equilibrium states, the attachment and detachment of springs must be considered. Naturally, the stability of each of these states must be evaluated, in other words, is it more energetically favorable for the contacting material to stay adhered or separate entirely. The stability of an adhesive contact can be determined from the sign of the derivative of $\delta_o(a) = g_\textrm{1D}(a) - \Delta l(a)$ with respect to $a$

        \begin{align} \label{adhesive stability appendix}
	    & \frac{d}{da}(\Delta l(a) + \delta_o - g_\textrm{1D}(a)) > 0, \quad \textrm{stable equilibrium}, \nonumber \\
            & \frac{d}{da}(\Delta l(a) + \delta_o - g_\textrm{1D}(a)) < 0, \quad \textrm{unstable equilibrium}, \\   
            & \frac{d}{da}(\Delta l(a) + \delta_o - g_\textrm{1D}(a)) = 0, \quad \textrm{critical state} \nonumber.
        \end{align}

\noindent In words, the criterion compares the rate of change of the normal displacement (i.e. $w_\textrm{1D}(a) = \delta_o - g_\textrm{1D}(a)$) at the edge of the contact with respect to the contact radius $dw_\textrm{1D}(a)/da$ to that of the critical length $d\Delta l(a)/da$. This criterion is consistent with the physical picture, since comparison of the extensional displacement of the outer springs to the critical length is what determines detachment. Provided the rate of change of the displacement at the edge exceeds that of the critical length, a small perturbation of the contact radius in either direction $da$  will result in a new stable configuration where $w_\textrm{1D}(a + da) = \Delta l(a + da)$. However, if the perturbation rests at the critical state, a slight perturbation that reduces the contact radius will lead to complete separation of the contact, whereas a perturbation that grows the contact radius will move it back towards stability. 

To determine the critical contact radius it is necessary to consider the two possible types of loading conditions: force-controlled and displacement-controlled.

\subsubsection{Stability condition for displacement-controlled contacts}

Here, the displacement remains constant and therefore the critical state is

    \begin{equation} 
        	\frac{dg_\textrm{1D}(a)}{da}\Bigr|_{a=a_c} = \frac{d\Delta l (a)}{da}\Bigr|_{a=a_c} = \sqrt{\frac{\pi \Delta \gamma}{2 E^*_c a_c}}
    \end{equation}

\subsubsection{Stability condition for force-controlled contacts}

With force held constant, the displacement is allowed to vary. The relation between force and displacement is given by 

    \begin{equation}
        	F = 2E^*_c \int_{0}^{a} [\delta_o - g_\textrm{1D}(x)] \,dx.
    \end{equation}

\noindent Differentiating the above equation under the condition that force must be constant yields

    \begin{equation}
        \begin{split}
                d\int_{0}^{a} [\delta_o - g_\textrm{1D}(x)] \,dx & = da \cdot \frac{\partial}{\partial a} \int_{0}^{a} [\delta_o - g_\textrm{1D}(x)] \,dx + d\delta_o \cdot \frac{\partial}{\partial \delta_o}\int_{0}^{a} [\delta_o - g_\textrm{1D}(x)] \,dx \\
                & = da \cdot [\delta_o(a) - g_\textrm{1D}(a)] + d\delta_o \cdot a \\
                & = -da \cdot \Delta l(a) + d\delta_o \cdot a = 0.
        \end{split}
    \end{equation}

\noindent Rearrangement of the final equality gives 

    \begin{equation}
        	\frac{d\delta_o}{da}\Bigr|_{F=\mathrm{constant}} = \frac{\Delta l(a)}{a}.
    \end{equation}

\noindent With this the critical state condition takes the form

    \begin{equation}
        	\frac{d\Delta l(a)}{da} + \frac{d\delta_o}{da} - \frac{dg_\textrm{1D}(a)}{da} = \frac{d\Delta l(a)}{da} + \frac{\Delta l(a)}{a} - \frac{dg_\textrm{1D}(a)}{da} = 0,
    \end{equation}

\noindent which is equivalent to 

    \begin{equation} 
        	\frac{dg_\textrm{1D}(a)}{da}\Bigr|_{a=a_c} = \left[ \frac{\Delta l(a)}{a} + \frac{d\Delta l(a)}{da} \right]_{a=a_c} = 3\sqrt{\frac{\pi \Delta \gamma}{2 E^*_c a_c}}.
    \end{equation}

Thus we arrive at (\ref{a_crtical}) which combines the stability conditions into a single equation

    \begin{equation*} 
        	\frac{dg_\textrm{1D}(a)}{da}\Bigr|_{a=a_c} = \xi \sqrt{\frac{\pi \Delta \gamma}{2 E^*_c a_c}}, \quad \quad \xi = \begin{cases}
      3, & \textrm{force-control},\\
      1, & \textrm{displacement-control}.
    \end{cases} 
    \end{equation*}

\clearpage

\section{Additional finite element results}

\subsection{Variation of contact quantities with Poisson ratio} \label{Variation of contact quantities with Poisson ratio}

\begin{figure*} [!htb]
 	%\centering
 	\raggedright
 	% Trim{LEFT LOWER RIGHT UPPER}
 	\includegraphics[width=\textwidth, trim = 1cm  18.5cm 13cm 1cm]{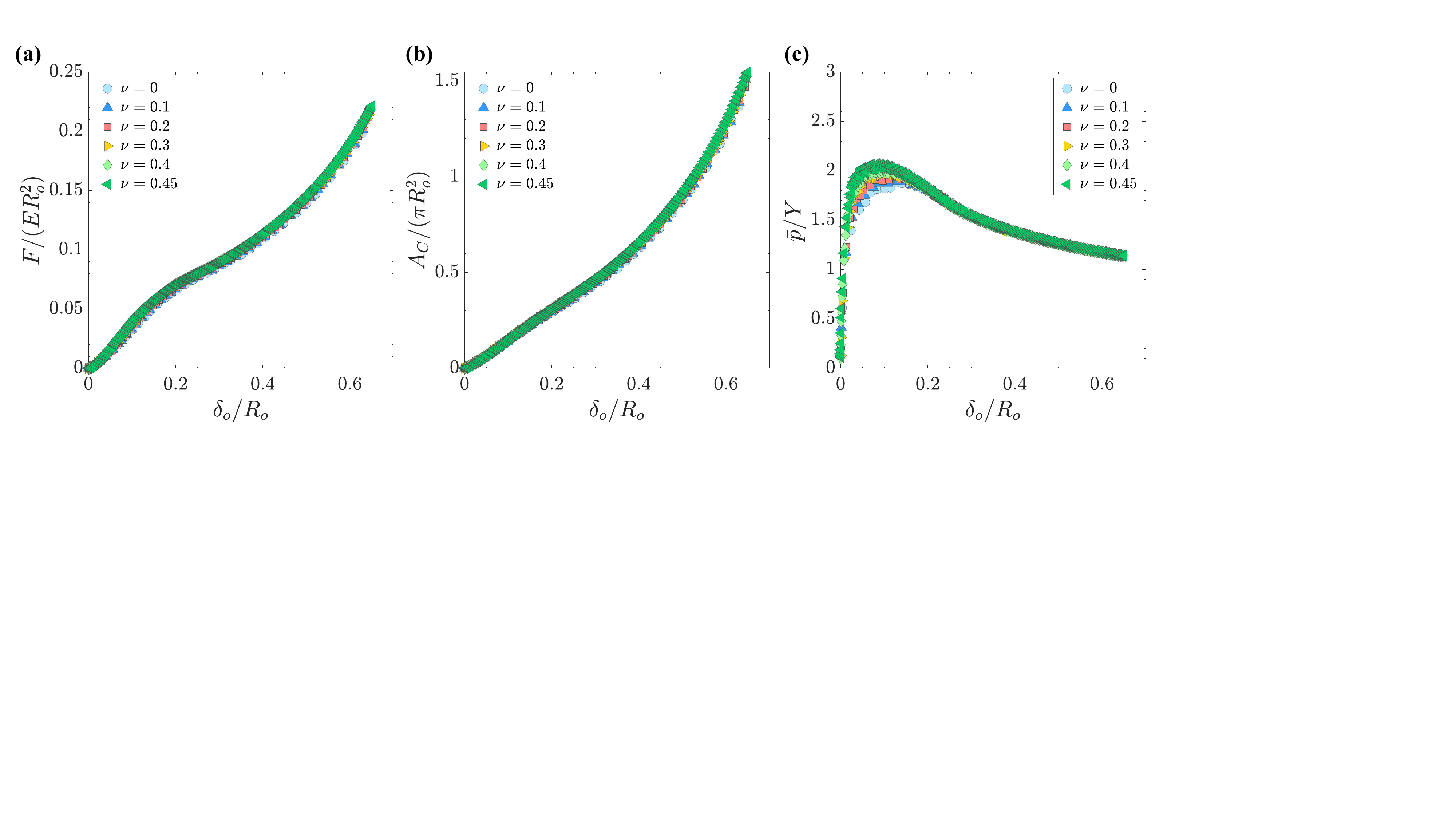}
 	\caption{Finite element method parametric study varying the Poisson ratio in the range $0 \leq \nu \leq 0.45$. (a) Variation of force with displacement. (b) Variation of contact area with displacement. (c) Variation of average pressure with displacement.}
 	\label{fem_nu_variation_results}
 \end{figure*}

The variation of contact force, area, and average pressure with the Poisson ratio are plotted in Fig.~\ref{fem_nu_variation_results}(a)-(c). Changes in the Poisson ratio for the range $0 \leq \nu \leq 0.45$ is found to have negligible effect on all quantities. 

\subsection{Pressure profile variation} \label{Pressure profile variation}

\begin{figure*} [!htb]
 	\centering
 	%\raggedright
 	% Trim{LEFT LOWER RIGHT UPPER}
 	\includegraphics[width=\textwidth, trim = 1cm  2cm 13cm 1cm]{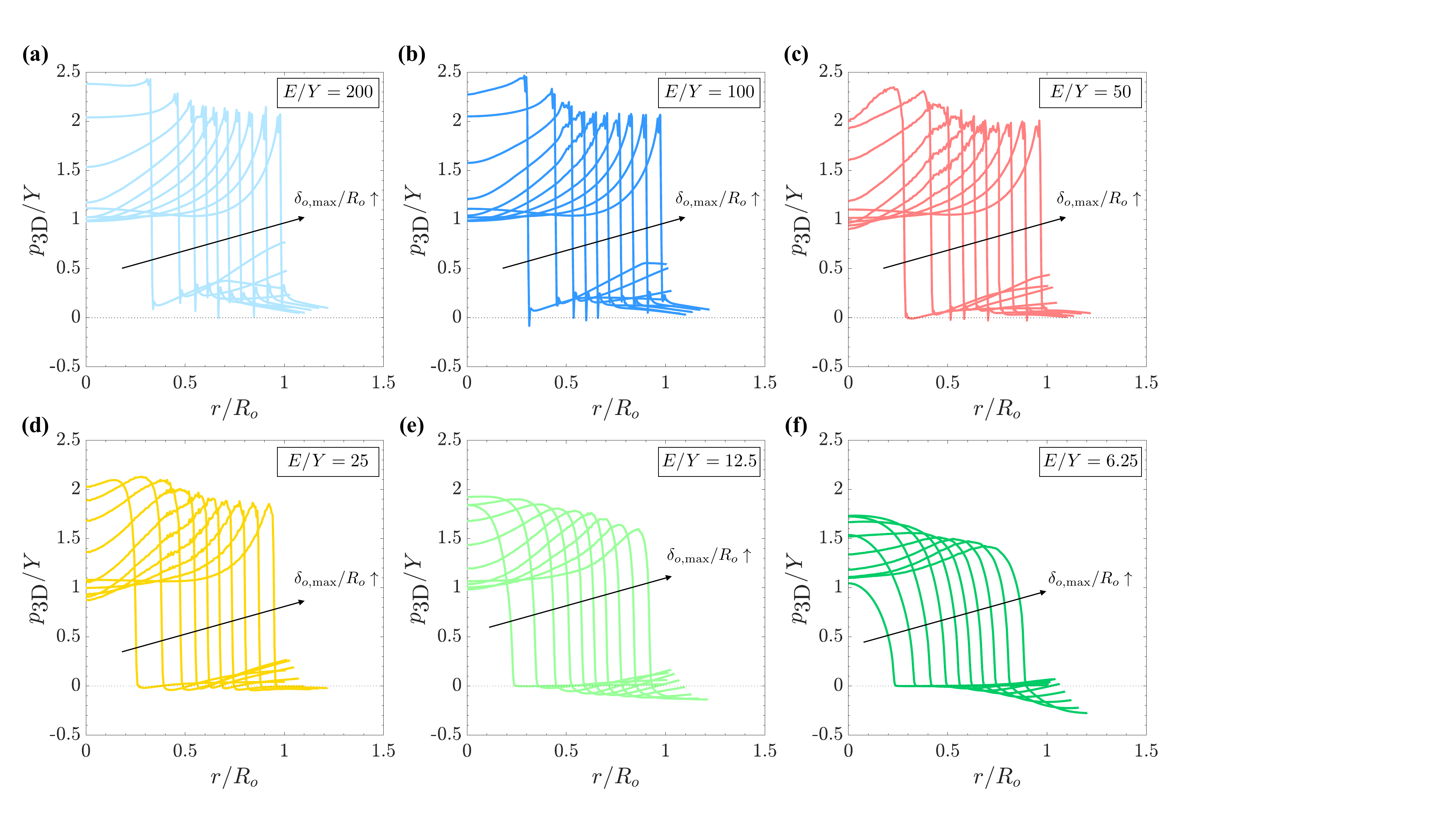}
 	\caption{Normalized pressure profiles after loading to a specified $\delta_{o,\textrm{max}}/R_o$. In each case the individual curves correspond to increasing $\delta_{o,\textrm{max}}/R_o$ moving left to right. For the cases of (a) $E/Y = 200$, (b) $E/Y = 100$, (c) $E/Y = 50$, and (d) $E/Y = 25$ all the pressure profiles correspond to a $\delta_{o,\textrm{max}}/R_o$ that has induced some amount of plastic deformation. For (e) $E/Y = 12.5$ the first pressure profile (i.e. $\delta_{o,\textrm{max}}/R_o = 0.05$) corresponds to a purely elastic deformation whereas all the rest correspond to varying amounts plastic deformation. For (f) $E/Y = 200$ the first two pressure profiles (i.e. $\delta_{o,\textrm{max}}/R_o = 0.05$ and $\delta_{o,\textrm{max}}/R_o = 0.1$ ) correspond to a purely elastic deformation whereas all the rest correspond to varying amounts of plastic deformation.}
 	\label{pressure_profiles}
 \end{figure*}

To determine the pressure profile beneath elastic-perfectly plastic particles contacting a rigid flat a series of finite element simulations is carried out. In these simulations, we consider six different ratios of the Young's modulus to yield stress ranging from $6.25 \leq E/Y \leq 200$ as well as a range of ten deformations spaced linearly from $0.05 \leq \delta_{o,\textrm{max}}/R_o \leq 0.5$. For each $E/Y$ ratio ten simulations are run in which the contact is loaded to the specified maximum deformation, at which point the pressure profile is recorded. The results of these simulations are shown in Fig.~\ref{pressure_profiles}(a)-(f), which displays the normalized pressure profile for specific $\delta_{o,\textrm{max}}/R_o$.

\clearpage

\section{Deriving the differential update form of the apparent radius} \label{Deriving the differential update form of the apparent radius}

To determine an evolution rule for $R$ we imagine an initially spherical elastic perfectly plastic particle compressed between $N$ rigid flats spaced on the surface such that no two contacts impede one another. For simplicity, we assume that the relative distance between the flats and the sphere center to only have ever decreased, albeit not strictly necessary. The deformed particle shape is assumed to well approximated by a sphere of radius $R$ with chopped off spherical caps where the rigid flats intersect\footnote{A similar geometric picture was used successfully in the contact models proposed by Frenning~\cite{frenning2013towards,frenning2015towards}.}. This implies that the current volume of the particle is given as 

    \begin{equation} \label{Vdeformed}
        	 V = \frac{4}{3}\pi R^3 - \sum_{i=1}^{N} \left( \frac{\pi}{3} \delta^2_i[3R - \delta_i] \right),
    \end{equation}

\noindent where $\delta$ is the apparent overlap measured with respect to the apparent radius $R$. To eliminate $\delta$, we recall that it is related to the displacement $\delta_o$ through the following relation

    \begin{equation} \label{delta delta_o relation}
        	 \delta = \delta_o + R - R_o.
    \end{equation}

\noindent This allows (\ref{Vdeformed}) to be rewritten as

    \begin{equation} \label{Vdeformed delta_o}
        	 V = \frac{4}{3}\pi R^3 - \sum_{i=1}^{N} \left( \frac{\pi}{3} (\delta_{o,i} + R - R_o)^2[3R - (\delta_{o,i} + R - R_o)] \right).
    \end{equation}

\noindent The differential of (\ref{Vdeformed delta_o}) can then be taken, noting that the volume can change either due to a change in $R$ or $\delta_o$

    \begin{equation} \label{dV}
        \begin{split}
        	 \Delta V = 4 \Delta R \pi R^2 - \pi \sum_{i=1}^{N} (2 \Delta R R + \Delta \delta_{o,i} [R + Ro -\delta_{o,i}])(\delta_{o,i} + R - R_o) .
        \end{split}
    \end{equation}

\noindent Based on the incompressible nature of plastic deformation we enforce that all volume change must be elastic. Taking compressive volume changes to be positive leads to 

    \begin{equation} \label{dV equals dVe}
        \begin{split}
        	 \Delta V = -\Delta V^e.
        \end{split}
    \end{equation}

\noindent Inserting (\ref{dV}) into (\ref{dV equals dVe}) and solving for $\Delta R$ we arrive at the desired result

\begin{equation} \label{DeltaR appendix}
    \Delta R = \frac{\Delta V^e - \sum_{i=1}^{N} \pi\Delta \delta_{o,i}(2\delta_{o,i} R_o - \delta^2_{o,i} + R^2 - R^2_o) }{2 \pi R \sum_{i=1}^{N}(\delta_{o,i} + R - R_o) - 4 \pi R^2}.
 \end{equation}

\noindent In practice the expression for $\Delta R$ is modified slightly to be 

 \begin{equation}
    \Delta R = \textrm{max} \left[ \frac{\Delta V^e - \sum_{i=1}^{N} \pi\Delta \delta_{o,i}(2\delta_{o,i} R_o - \delta^2_{o,i} + R^2 - R^2_o) }{2 \pi R \sum_{i=1}^{N}(\delta_{o,i} + R - R_o) - 4 \pi R^2}, \; 0 \right],
 \end{equation}

\noindent this is to reflect the fact that plastic deformation is permanent, thus the apparent radius is taken to be strictly increasing.

It should be noted that (\ref{DeltaR appendix}) can become inaccurate under highly confined conditions. In these cases, the fundamental assumption used to derive $\Delta R$---\textit{the deformed volume is well approximated by a sphere with chopped off spherical caps where the rigid flats intersect}---breaks down. This is due to the fact that under high confinement the spherical cap volumes predicted from the intersection with the rigid flats begin to severely intersect one another---a second order geometric effect that is not accounted for in (\ref{DeltaR appendix}). This leads to non-physical and incorrect predictions of $\Delta R$ including negative or anomalously large values.       

\clearpage

\section{Parameters of the MDR contact model} 
\label{Parameters of the MDR contact model}
\begin{table} [ht] 
\begin{center}
        \renewcommand{\arraystretch}{1.25}
 \begin{tabular}{ ||c|l|| } 
		\hline
		$A$ & height of elliptical indenter \\ \hline
            $a$ & contact radius \\ \hline
            $a_\textrm{max}$ & maximum experienced contact radius \\ \hline
            $B$ & width of elliptical indenter \\ \hline
            $\bm{b}$ & branch vector \\ \hline
            $c_A$ & contact area intercept \\ \hline
            $\delta_o$ & displacement \\ \hline
            $\delta_{o,\textrm{max}}$ & maximum experienced displacement \\ \hline
            $\delta$ & apparent overlap \\ \hline
            $\delta_\textrm{max}$ & maximum experienced apparent overlap \\ \hline
            $\delta^e_\textrm{1D}$ & transformed elastic displacement \\ \hline
            $\delta^e_\textrm{1D,max}$ & maximum transformed elastic displacement \\ \hline
            $\delta_R$ & displacement correction \\ \hline
            $\delta_Y$ & yield displacement \\ \hline
            $E$ & Young's modulus \\ \hline
            $E^*_c$ & composite plane strain modulus \\ \hline
            $F_\textrm{MDR}$ & force from MDR \\ \hline
            $\bm{f}$ & force vector \\ \hline
            $G$ & shear modulus \\ \hline
            $\kappa$ & bulk modulus \\ \hline
            $\nu$ & Poisson ratio \\ \hline
            $\bar{p}_H$  & average pressure for Hertz contact \\ \hline
            $p_Y$  & average pressure along hardening curve \\ \hline
            $R_o$ & initial radius \\ \hline
            $R$ & apparent radius \\ \hline
            $\Delta R$ & change in apparent radius \\ \hline
            $V_o$ & initial volume \\ \hline
            $V$ & current volume \\ \hline
            $\Delta V^e$ & change in elastic volume \\ \hline
            $Y$ & yield stress \\ \hline
            $z_R$ & depth of particle center \\ 
		\hline
	\end{tabular}
\end{center}
\caption{Important parameters for evaluation of non-adhesive MDR contact model.}

\label{nonadhesive MDR parameters}
\end{table}

\clearpage

\section{MDR contact model numerical implementation} \label{Sketch of numerical implementation}

\subsection*{Main function}

A sketch of the numerical routine to calculate the forces at $N$ active rigid flat contacts on a single particle using the MDR contact model is outlined below. Parameters that are underlined indicate an array of quantities for each active contact on a particle, for example, $\underline{\delta}_o$ is an array containing the displacement of each active contact on a particle. To index a specific entry of the array the subscript $i$ will be used, for example $\delta_{o,i}$ is the $i$th component of the array $\underline{\delta}_o$.

The known data is the initial radius $R_o$, composite plane strain modulus $E^*_c$, Poisson's ratio $\nu$, bulk modulus $\kappa$, yield stress $Y$, and effective surface energy $\Delta\gamma$. The inputs at the beginning of the step are the current apparent overlap at each contact $\underline{\delta}^{n}$, current maximum apparent overlap at each contact $\underline{\delta}^n_\textrm{max}$, current apparent radius $R^n$, current volume $V^n$, change in apparent overlap at each contact $\Delta \underline{\delta}$, and the change in displacement at each contact $\Delta \underline{\delta}_o$. Although not directly listed as inputs, it is assumed that the contact area intercept $\underline{c}_A$ and yield flag $\underline{Y}_\textrm{flag}$ (i.e. 0 = no yielding yet and 1 = yielding) at each contact are known as they are naturally calculated within the main function. Using the data and inputs, the numerical algorithm is constructed to output the updated apparent overlap at each contact $\underline{\delta}^{n+1}$, updated maximum apparent overlap at each contact $\underline{\delta}^{n+1}_\textrm{max}$, updated apparent radius $R^{n+1}$, updated current volume $V^{n+1}$, and force at each contact $\underline{F}^{n+1}$.

A few quantities that are taken as inputs into the function findUpdatedRadiusAndVol(...) are not given as inputs to the main function including: the updated force vector at each contact $\underline{\bm{f}}^{n+1}$ and updated branch vector at each contact $\underline{\bm{b}}^{n+1}$. The reason for this is that these quantities can be easily determined provided the contact normal vector at each contact $\underline{\bm{n}}$ is known---as is the case for DEM simulations. Writing out their calculation explicitly would distract from the more important aspects of the numerical implementation.

\begin{tcolorbox}[
    %colback={\#e3e9f3},      % Background color
    %colframe={\#006FA9},     % Border color
    boxrule=3pt,        % Border thickness
    rounded corners,      % Sharp corners for the box
    width=\textwidth,    % Width of the box
    title=\textbf{Main function}
]

\noindent \textbf{Data}: $R_o$, $E^*_c$, $\nu$, $\kappa$, $Y$, $\Delta\gamma$ 

\noindent \textbf{Input}: $\underline{\delta}^{n}$, $\underline{\delta}^n_\textrm{max}$, $R^n$, $V^n$, $\Delta \underline{\delta}$, $\Delta \underline{\delta}_o$

\noindent \textbf{Output}: $\underline{\delta}^{n+1}$, $\underline{\delta}^{n+1}_\textrm{max}$, $R^{n+1}$, $V^{n+1}$, $\underline{F}^{n+1}$

\noindent \textbf{begin}  

    \noindent \qquad $\underline{\delta}^{n+1} = \underline{\delta}^{n} + \Delta\underline{\delta}$

    \noindent \qquad $N = \textrm{length}(\underline{\delta}^{n})$

    \noindent \qquad \textbf{for} $i = 1 : N$
    
    \noindent \qquad \qquad \textbf{if} $\delta^{n+1}_i > \delta^n_\textrm{max,i}$ 

        \noindent \qquad \qquad \qquad $\delta^{n+1}_\textrm{max,i} = \delta^{n+1}_i$ 

    \noindent \qquad \qquad \textbf{else}

        \noindent \qquad \qquad \qquad $\delta^{n+1}_\textrm{max,i} = \delta^n_\textrm{max,i}$

    \noindent \qquad \qquad \textbf{end} 

    \noindent \qquad \qquad $\delta = \delta^{n+1}_i$

    \noindent \qquad \qquad $\delta_\textrm{max} = \delta^{n+1}_\textrm{max,i}$

        \noindent \qquad \qquad $R = R^{n}$

    \noindent \qquad \qquad $p_Y = Y\left( 1.75\exp{(-4.4\delta_\textrm{max}/R)+1} \right)$

    \noindent \qquad \qquad $\bar{p}_H = \frac{4E^*_c}{3 \pi \sqrt{R}}\sqrt{\delta}$

\end{tcolorbox}

\begin{tcolorbox}[
    %colback={\#e3e9f3},      % Background color
    %colframe={\#006FA9},     % Border color
    boxrule=3pt,        % Border thickness
    rounded corners,      % Sharp corners for the box
    width=\textwidth,    % Width of the box
    title=\textbf{Main function continued...}
]

    \noindent \qquad \qquad \textbf{if} $Y_\textrm{flag,i} = 0$ and $\bar{p}_H > p_Y$

        \noindent \qquad \qquad \qquad $\delta = \delta_Y$
        
        \noindent \qquad \qquad \qquad $c_{A,i} = \pi(\delta_Y^2 - \delta_Y R)$
        
        \noindent \qquad \qquad \qquad $Y_\textrm{flag,i} = 1$
    
    \noindent \qquad \qquad \textbf{end}

    \noindent \qquad \qquad \textbf{if} $Y_\textrm{flag,i} = 0$ 
    
        \noindent \qquad \qquad \qquad $A = 4R$

        \noindent \qquad \qquad \qquad $B = 2R$

        \noindent \qquad \qquad \qquad $\delta^e_\textrm{1D} = \delta$
        
    \noindent \qquad \qquad \textbf{else}

        \noindent \qquad \qquad \qquad $a_\textrm{max} = \sqrt{(2\delta_\textrm{max} R - \delta^2_\textrm{max}) + c_{A,i}/\pi}$

        \noindent \qquad \qquad \qquad $A = \frac{4p_Y}{E^*_c}a_\textrm{max}$

        \noindent \qquad \qquad \qquad $B = 2a_\textrm{max}$

        \noindent \qquad \qquad \qquad $\delta^e_\textrm{1D,max} = A/2$

        \noindent \qquad \qquad \qquad $F_\textrm{max} = \frac{E^*_c AB}{4}\left[ \arccos\left( {1 - \frac{2\delta^e_{\textrm{1D,max}}}{A}}\right) - \left( 1 - \frac{2\delta^e_{\textrm{1D,max}}}{A} \right) \sqrt{\frac{4\delta^e_{\textrm{1D,max}}}{A} - \frac{4(\delta^e_{\textrm{1D,max}})^2}{A^2}}\right]$

        \noindent \qquad \qquad \qquad $z_R = R - (\delta_\textrm{max} - \delta^e_\textrm{1D,max})$

        \noindent \qquad \qquad \qquad $\delta_R = \frac{F_\textrm{max}}{\pi a_\textrm{max}^2} \left[  \frac{2a^2_\textrm{max} (\nu - 1) - z_R(2\nu - 1)( \sqrt{a^2_\textrm{max} + z^2_R} - z_R)} {2G\sqrt{a_\textrm{max}^2 + z_R^2}} \right]$

        \noindent \qquad \qquad \qquad $\delta^e_\textrm{1D} = \frac{\delta - \delta_\textrm{max} + \delta^e_\textrm{1D,max} + \delta_R}{1 + \delta_R/\delta^e_\textrm{1D,max}}$
        
    \noindent \qquad \qquad \textbf{end} 

    \noindent \qquad \qquad \textbf{if} $\textrm{adhesion} = \textrm{`on'}$ 

        \noindent \qquad \qquad \qquad $F_\textrm{MDR} = \textrm{forceAdhesive}(A,B,\delta^e_\textrm{1D},a)$ 

    \noindent \qquad \qquad \textbf{else}

            \noindent \qquad \qquad \qquad \textbf{if} $\delta^e_\textrm{1D} > 0$
            
                \noindent \qquad \qquad \qquad $F_\textrm{MDR} = \frac{E^*_c AB}{4}\left[ \arccos\left( {1 - \frac{2\delta^e_{\textrm{1D}}}{A}}\right) - \left( 1 - \frac{2\delta^e_{\textrm{1D}}}{A} \right) \sqrt{\frac{4\delta^e_{\textrm{1D}}}{A} - \frac{4(\delta^e_{\textrm{1D}})^2}{A^2}}\right]$

            \noindent \qquad \qquad \qquad \textbf{else} 

                \noindent \qquad \qquad \qquad \qquad $F_\textrm{MDR} = 0$

            \noindent \qquad \qquad \qquad \textbf{end} 

    \noindent \qquad \qquad \textbf{end} 

    \noindent \qquad \textbf{end}

\noindent \qquad $[R^{n+1},V^{n+1}] = \textrm{findUpdatedRadiusAndVol}(R^n,N,\underline{\delta}^{n+1}_o,\Delta \underline{\delta}_o,\underline{\bm{f}}^{n+1},\underline{\bm{b}}^{n+1},V^n)$

\noindent \textbf{end} 

\end{tcolorbox}

\subsection*{Function to update the radius}

To define findUpdatedRadiusAndVol(...) we write an explicit numerical implementation of the multi-neighbor dependent procedure developed in Section~\ref{Closure of the contact law}. It is important to highlight the fact that findUpdatedRadiusAndVol(...) exists outside the \textit{for} loop that runs over each active contact, this is because both outputs of the function are particle properties. 

\begin{tcolorbox}[
    %colback={\#e3e9f3},      % Background color
    %colframe={\#006FA9},     % Border color
    boxrule=1pt,        % Border thickness
    rounded corners,      % Sharp corners for the box
    width=\textwidth    % Width of the box
]

\noindent findUpdatedRadiusAndVol($R^n$,$N$,$\underline{\delta}^{n+1}_o$,$\Delta \underline{\delta}_o$,$\underline{\bm{f}}^{n+1}$,$\underline{\bm{b}}^{n+1}$,$V^n$)

\tcblower

    \noindent \qquad $V_o = 4/3 \pi R^3_o$ 
    
    \noindent \qquad $V^{n+1} = V_o \left[ 1 + \frac{1}{3\kappa V^{n}}\textrm{tr} \left( \sum^N_{i = 1} \bm{f}^{n+1}_i \otimes \bm{b}^{n+1}_i \right) \right]$  

    \noindent \qquad $\Delta V^e = -(V^{n+1} - V^n)$

    \noindent \qquad $\Delta R = \textrm{max} \left[ \frac{\Delta V^e - \sum_{i=1}^{N} \pi\Delta \delta_{o,i}(2\delta^{n+1}_{o,i} R_o - (\delta^{n+1}_{o,i})^2 + (R^n)^2 - R^2_o) }{2 \pi R^n \sum_{i=1}^{N}(\delta^{n+1}_{o,i} + R^n - R_o) - 4 \pi (R^n)^2}, \; 0 \right]$
    
    \noindent \qquad $R^{n+1} = R^n + \Delta R$

\noindent \textbf{end} 

\end{tcolorbox}

\subsection*{Adhesive force function}

To model adhesive contact forceAdhesion(...) requires definition. Unlike the normal contact case the contact radius $a$ must be explicitly tracked during adhesive contact; we assume it is a known input whose value gets updated and within forceAdhesion(...) and passed between steps. The subscripts $\textrm{n.a.}$ and $\textrm{a.r.}$ stand for non-adhesive and adhesive retraction, respectively.

\begin{tcolorbox}[
    %colback={\#e3e9f3},      % Background color
    %colframe={\#006FA9},     % Border color
    boxrule=1pt,        % Border thickness
    rounded corners,      % Sharp corners for the box
    width=\textwidth    % Width of the box
]

\noindent forceAdhesion($A$,$B$,$\delta^e_\textrm{1D}$,$a$)

\tcblower

    \noindent \qquad $g_\textrm{1D}(a) = \frac{A}{2} - \frac{A}{B}\sqrt{\frac{B^2}{4} - a^2}$

    \noindent \qquad $w_\textrm{1D}(a) = \delta^e_\textrm{1D} - g_\textrm{1D}(a)$
    
    \noindent \qquad \textbf{if} $w_\textrm{1D}(a) = 0$

        \noindent \qquad \qquad $a = B/2$
    
        \noindent \qquad \qquad $F^{n+1} = \frac{E^*_c AB}{4}\left[ \arccos\left( {1 - \frac{2\delta^e_{\textrm{1D}}}{A}}\right) - \left( 1 - \frac{2\delta^e_{\textrm{1D}}}{A} \right) \sqrt{\frac{4\delta^e_{\textrm{1D}}}{A} - \frac{4(\delta^e_{\textrm{1D}})^2}{A^2}}\right]$ 
        
    \noindent \qquad \textbf{else}

        \noindent \qquad \qquad $\Delta l = \sqrt{\frac{2 \pi a \Delta \gamma}{E^*_c}}$

        \noindent \qquad \qquad $a_c = \textrm{Solve}\left[ \frac{dg_\textrm{1D}(a)}{da}\Bigr|_{a=a_c} = \sqrt{\frac{\pi \Delta \gamma}{2 E^*_c a_c}}, \; a_c \right]$

    \noindent \qquad \qquad \textbf{if} $w_\textrm{1D}(a) < \Delta l$

        \noindent \qquad \qquad \qquad $\delta^e_\textrm{1D,adh} = g_\textrm{1D}(a)$

        \noindent \qquad \qquad \qquad $F_\textrm{n.a.} = \frac{E^*_c AB}{4}\left[ \arccos\left( {1 - \frac{2\delta^e_{\textrm{1D,adh}}}{A}}\right) - \left( 1 - \frac{2\delta^e_{\textrm{1D,adh}}}{A} \right) \sqrt{\frac{4\delta^e_{\textrm{1D,adh}}}{A} - \frac{4(\delta^e_{\textrm{1D,adh}})^2}{A^2}}\right]$ 

        \noindent \qquad \qquad \qquad $F_\textrm{a.r.} = 2 E^*_c(\delta^e_{1D} - g_\textrm{1D}(a))a$

        \noindent \qquad \qquad \qquad $F_\textrm{MDR} = F_\textrm{n.a.} + F_\textrm{a.r.}$ 

    \noindent \qquad \qquad \textbf{elseif} $w_\textrm{1D}(a) > \Delta l$

        \noindent \qquad \qquad \qquad $a = \textrm{Solve}\left[\delta^e_\textrm{1D} + \Delta l - g_\textrm{1D}(a) = 0, \; a  \right]$

        \noindent \qquad \qquad \qquad \textbf{if} $a < a_c$

            \noindent \qquad \qquad \qquad \qquad $F_\textrm{MDR} = 0$

        \noindent \qquad \qquad \qquad \textbf{else} 

            \noindent \qquad \qquad \qquad \qquad $\delta^e_\textrm{1D,adh} = g_\textrm{1D}(a)$

            \noindent \qquad \qquad \qquad \qquad $F_\textrm{n.a.} = \frac{E^*_c AB}{4}\left[ \arccos\left( {1 - \frac{2\delta^e_{\textrm{1D,adh}}}{A}}\right) - \left( 1 - \frac{2\delta^e_{\textrm{1D,adh}}}{A} \right) \sqrt{\frac{4\delta^e_{\textrm{1D,adh}}}{A} - \frac{4(\delta^e_{\textrm{1D,adh}})^2}{A^2}}\right]$ 

            \noindent \qquad \qquad \qquad \qquad $F_\textrm{a.r.} = 2 E^*_c(\delta^e_{1D} - g_\textrm{1D}(a))a$

            \noindent \qquad \qquad \qquad \qquad $F_\textrm{MDR} = F_\textrm{n.a.} + F_\textrm{a.r.}$

    \noindent \qquad \qquad \qquad \textbf{end}

    \noindent \qquad \qquad \textbf{end} 

    \noindent \qquad \textbf{end} 

\noindent \textbf{end} 

\end{tcolorbox}

\clearpage

\section{Adhesive contact}

\subsection{Concave-convex behavior of the relaxed profiles} \label{Concave-convex behavior of the relaxed profiles}

\begin{figure*} [!htb]
 	\centering
 	%\raggedright
 	% Trim{LEFT LOWER RIGHT UPPER}
 	\includegraphics[width=\textwidth, trim = 1cm  2cm 13cm 1cm]{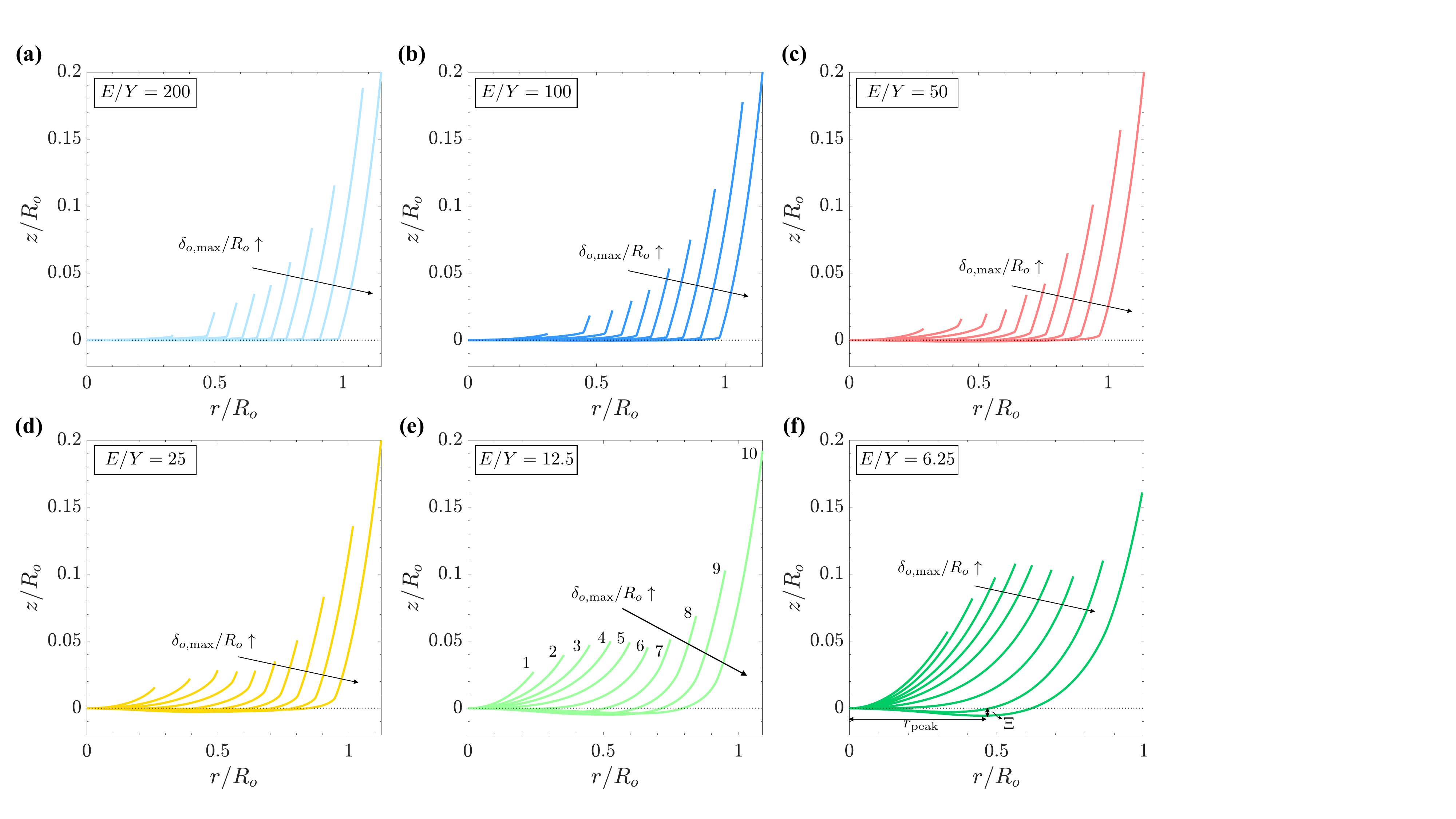}
 	\caption{Normalized relaxed profiles truncated at $r=a_\textrm{max}$ after loading to a specified $\delta_{o,\textrm{max}}/R_o$. In each case the individual curves correspond to increasing $\delta_{o,\textrm{max}}/R_o$ moving left to right. For the cases of (a) $E/Y = 200$, (b) $E/Y = 100$, (c) $E/Y = 50$, and (d) $E/Y = 25$ all the relaxed profiles correspond to a $\delta_{o,\textrm{max}}/R_o$ that has induced some amount of plastic deformation. For (e) $E/Y = 12.5$ the first relaxed profile (i.e. $\delta_{o,\textrm{max}}/R_o = 0.05$) corresponds to a purely elastic deformation whereas all the rest correspond to varying amounts plastic deformation. For (f) $E/Y = 200$ the first two relaxed profiles (i.e. $\delta_{o,\textrm{max}}/R_o = 0.05$ and $\delta_{o,\textrm{max}}/R_o = 0.1$ ) correspond to a purely elastic deformation whereas all the rest correspond to varying amounts of plastic deformation.}
 	\label{relaxed_profiles}
 \end{figure*}

To investigate the unloaded profile shape of an elastic-plastic particle compressed between rigid flats, a series of finite element simulations are conducted. We consider six different ratios of the Young's modulus to yield stress ranging from $6.25 \leq E/Y \leq 200$ as well as a range of ten deformations spaced linearly from $0.05 \leq \delta_{o,\textrm{max}}/R_o \leq 0.5$. For each $E/Y$ ratio, ten simulations are run in which the contact is loaded to the specified maximum deformation and then unloaded completely, at which point the relaxed profile is recorded. The results of these simulations are shown in Fig.~\ref{relaxed_profiles}(a)-(f), which displays the normalized relaxed profiles truncated at $r = a_\textrm{max}$ after loading to a specified $\delta_{o,\textrm{max}}/R_o$.

Focusing on Fig.~\ref{relaxed_profiles}(e) we see that the evolution of the relaxed profile is quite complex. The ten different curves are labeled 1-10, with profile 1 corresponding to $\delta_{o,\textrm{max}}/R_o=0.05$ and profile 10 to $\delta_{o,\textrm{max}}/R_o=0.5$. Profile 1 corresponds to an elastic deformation, whereas for 2-5 we observe that the accumulation of plastic deformation continuously blunts the profiles decreasing their curvature. The nature of this profile evolution has been observed as early as 1948~\cite{tabor1948simple} and more recent efforts have even accounted for its affects on adhesive behavior~\cite{mesarovic2000adhesive}. However, to the author's knowledge there is no mention in the relevant literature of the profile behavior seen in profiles 6-10. For these, we see a markedly different behavior in which the sign of the curvature changes throughout the profile from concave near $r/R_o = 0$ to convex closer to $r/R_o = a_\textrm{max}/R_o$. This concave-convex property is shared between all values of $E/Y$, albeit the size of the peak, or vertical distance from $z=0$ to $z$ at the minima where curvature changes which we denote as $\Xi$, varies depending on $E/Y$. 

\begin{figure*} [!htb]
 	\centering
 	%\raggedright
 	% Trim{LEFT LOWER RIGHT UPPER}
 	\includegraphics[width=\textwidth, trim = 1cm  18.5cm 13cm 1cm]{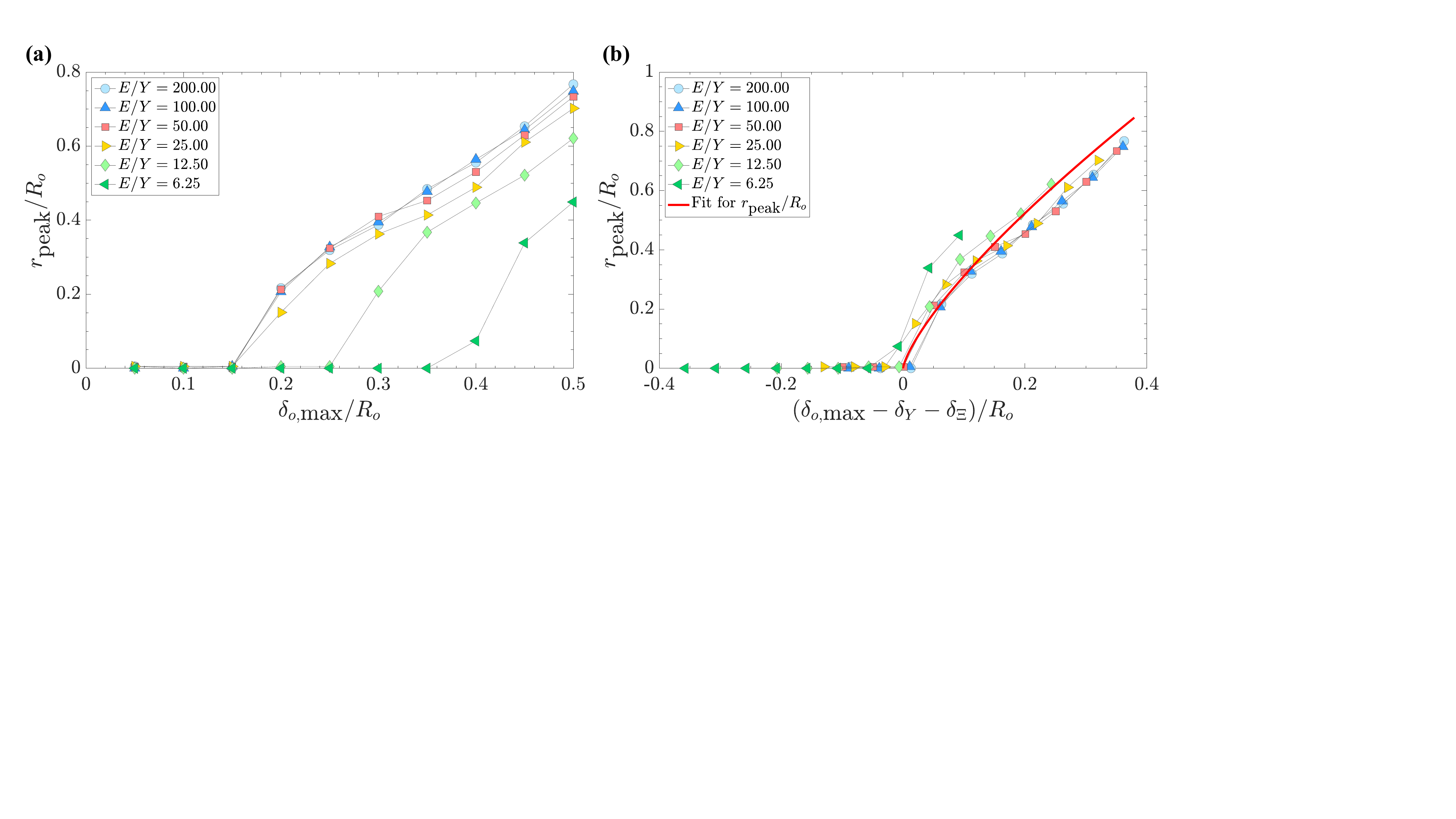}
 	\caption{Evolution of $r_\textrm{peak}$ as a function of $\delta_{o,\textrm{max}}/R_o$. (a) Raw data normalized by $R_o$. (b) Shifting the data along the horizontal axis to collapse it, red solid line is a provided fit for the collapsed data.}
 	\label{peak behavior main body}
 \end{figure*}

To understand when these peaks form and how they evolve, the distance from $r=0$ to $r$ at the minima where curvature changes denoted as $r_\textrm{peak}$, shown in Fig.~\ref{relaxed_profiles}(f), is measured for each case. The behavior of $r_\textrm{peak}$ for all cases is shown in Fig.~\ref{peak behavior main body}(a). Two important observations can be made, first in approaching the rigid-plastic limit (i.e. $E/Y \rightarrow \infty$) the earliest the peaks can form is $\delta_{o,\textrm{max}}/R_o \gtrapprox 0.15$, and second, that increased elastic deformation (i.e. $E/Y \rightarrow 0$) leads to a later onset of the peak formation. A collapse of the data is shown in Fig.~\ref{peak behavior main body}(b) caused by shifting the data along the horizontal axis by the yield displacement $\delta_Y$ (\ref{delta_Y_solve}) and $\delta_\Xi = 0.135R_o$. A general relationship can then be proposed for $r_\textrm{peak}$ as a function of $\delta_{o,\textrm{max}}/R_o$

    \begin{equation} \label{rpeak Ro fit}
       r_\textrm{peak}/R_o = \begin{cases} 
         1.75[(\delta_{o,\textrm{max}} - \delta_Y - \delta_\Xi)/R_o]^{3/4}, & \delta_{o,\textrm{max}} > \delta_Y + \delta_\Xi, \\ \\
      0, & \delta_{o,\textrm{max}} \leq \delta_Y + \delta_\Xi.
        \end{cases} 
    \end{equation}

\noindent where $\delta_Y$ is the yield displacement (\ref{delta_Y_solve}) and $\delta_\Xi = 0.135R_o$. This provided fit is of immediate usefulness because the location of the peaks can be estimated based on only the known input parameters. 

\begin{figure*} [!htb]
 	\centering
 	%\raggedright
 	% Trim{LEFT LOWER RIGHT UPPER}
 	\includegraphics[width=\textwidth, trim = 1cm  2cm 13cm 1cm]{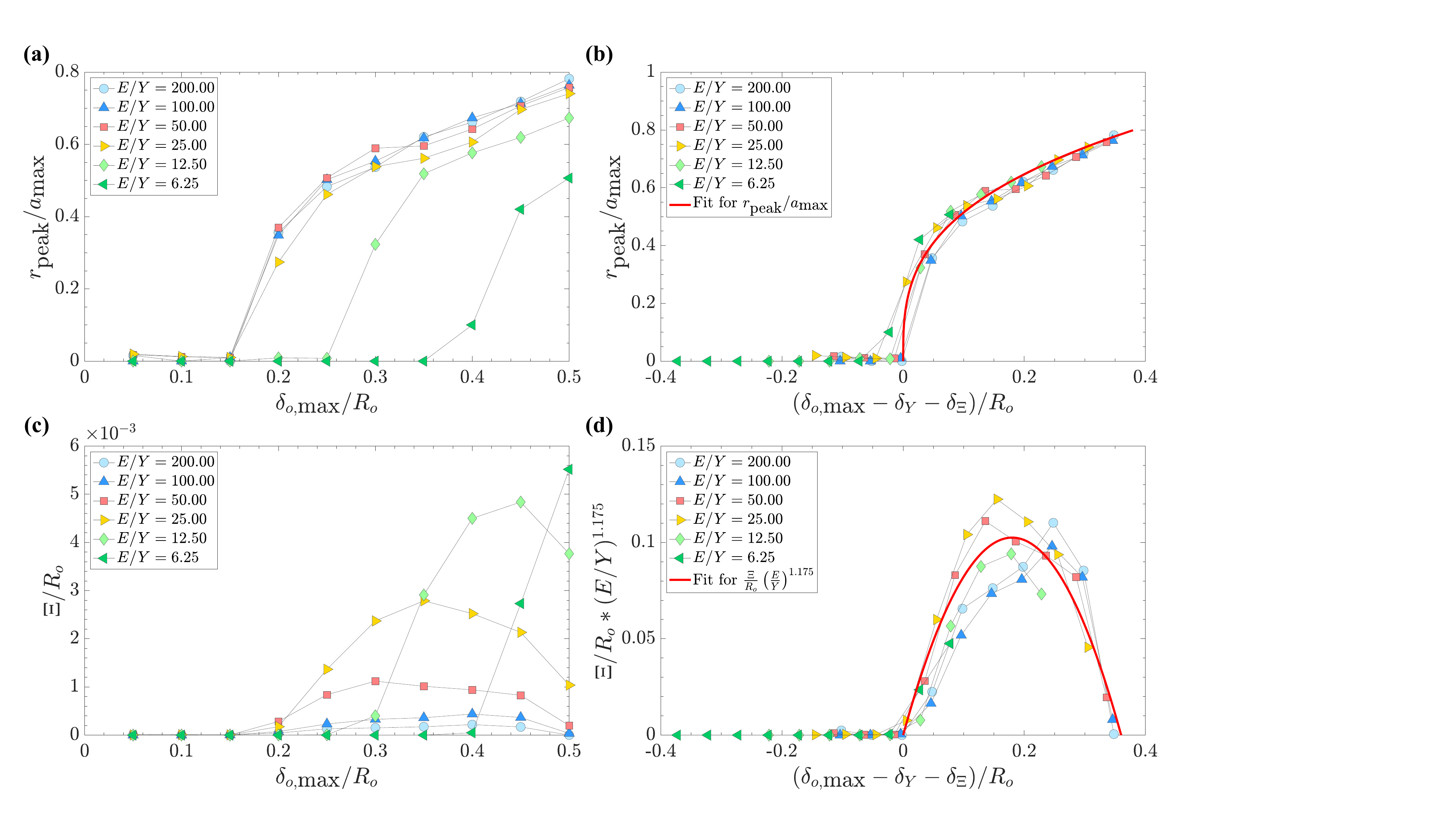}
 	\caption{Behavior of the concave region that develops due to large plastic deformation as a function of $\delta_{o,\textrm{max}}/R_o$. (a) Evolution of $r_\textrm{peak}/a_\textrm{max}$. (b) Collapse of data from (a) by shifting the horizontal axis in the negative direction by $\delta_Y$ and $\delta_\Xi = 0.15R_o$. The provided fit is the solid red line. (c) Evolution of $\Xi/R_o$. (d) Collapse of data from (c) by the same horizontal shift and scaling the vertical axis by $(E/Y)^{1.175}$. The provided fit is the solid red line.}
 	\label{peak behavior appendix}
 \end{figure*}

An even better collapse of the data can be achieved when $r_\textrm{peak}$ is normalized by the maximum experienced contact radius before unloading $a_\textrm{max}$ as shown in Fig.~\ref{peak behavior appendix}(b), with an associated fit of  

    \begin{equation} \label{rpeak amax fit}
       r_\textrm{peak}/a_\textrm{max} = \begin{cases} 
         1.1[(\delta_{o,\textrm{max}} - \delta_Y - \delta_\Xi)/R_o]^{1/3}, & \delta_{o,\textrm{max}} > \delta_Y + \delta_\Xi, \\ \\
      0, & \delta_{o,\textrm{max}} < \delta_Y + \delta_\Xi,
        \end{cases} 
    \end{equation}

\noindent where $\delta_\Xi = 0.135R_o$ once again. However, this fit requires knowledge of $a_\textrm{max}$ which cannot be determined a priori without simulating the contact.

 Turning now to $\Xi$, its behavior is shown in Fig.~\ref{peak behavior appendix}(c) where $\Xi$ is normalized by $R_o$. The data can be reasonably collapsed by shifting the curves along the horizontal axis in the negative direction by $\delta_Y$ and $\delta_\Xi = 0.15R_o$ to produce Fig.~\ref{peak behavior appendix}(d). A fit for the collapsed data can then be proposed

    \begin{equation} \label{Xi fit}
       \frac{\Xi}{R_o} \left( \frac{E}{Y} \right)^{1.175} = \begin{cases} 
         0.1025 - \frac{0.1025}{0.18}[(\delta_{o,\textrm{max}} - \delta_Y - \delta_\Xi)/R_o - 0.18]^2, & \delta_Y + \delta_\Xi < \delta_{o,\textrm{max}} < 0.36  , \\ \\
      0, & \textrm{otherwise}.
        \end{cases} 
    \end{equation}

\subsection{Conditions for accuracy of the adhesive MDR contact model} \label{Conditions for accuracy of the adhesive MDR contact model}

Adhesive behavior is directly linked to profile shape. In the presented MDR contact model it is assumed that the mapped 1D indenter is always elliptical in shape, mapping this back to the 3D space gives an indenter that shares a convex profile throughout. It has been shown in Section~\ref{Adhesion in the fully-plastic regime} that the true profile contains both concave and convex regions, this naturally leads to the following question:

\begin{itemize}
    \item \textit{Can a profile that is purely convex ever display similar adhesive behavior to that of a profile that is concave near the center and then transitions convex at the edges?} 
\end{itemize}

In general this may be a difficult question to answer, however for the case at hand there are two key relations between the differing profiles namely, for a given displacement they will produce the same contact radius and average pressure. This restricts the geometric form for the MDR indenter making the comparison more well-constrained. 

In addressing the posed question, it is necessary to establish a ground truth to compare the adhesive MDR contact model against. Given that the true indenter profile determined by FEM after plastic deformation is a not associated with any simple geometric shape, no analytical forms describing the adhesive behavior exist. Instead, the adhesive behavior can be determined numerically by using a discretized form of the adhesive MDR scheme described in Section~\ref{Normal contact with adhesion}. This will provide adhesive results corresponding to the \textit{real} relaxed profiles that are consistent with the JKR theory of adhesion---establishing a ground truth. The first step in this process requires mapping the profiles obtained from the FEM simulations to the 1D transformed space. During the transformation, the fundamental shape of the indenter is preserved, that is, purely convex profiles map to purely convex profiles and concave-convex profiles map to concave-convex profiles that are compressed or stretched slightly depending on the original shape. 

We pause to discuss an important point regarding the applicability of the adhesive MDR to concave-convex indenters. Application of the MDR requires that the contact be axisymmetric and that the contact area remains compact (i.e. continuous). The former is trivially satisfied due to the axisymmetric nature of the FEM simulations. The latter is not guaranteed due to the concave-convex nature of the indenters determined by FEM. To see this imagine contact with a concave-convex indenter in the non-adhesive case, initially an annular contact ring would form that fills in to a circular contact after enough compression. However, in the presence of adhesion this issue vanishes, provided that the indenter has initially been displaced enough to form a compact contact area. This is because the springs inside the outer peaks adhere to the surface guaranteeing a compact contact area even after decompression. Furthermore, the criterion for springs breaking is only applied to the outer edge where there is a jump discontinuity in the normal displacement profile $w_\textrm{1D}(x)$. Thus, the adhesive contact separates radially inward from the outer edge until it reaches the peak at which point the contact is always unstable and full separation occurs, ensuring a compact contact always exists.  

\begin{figure*} [!htb]
 	\centering
 	%\raggedright
 	% Trim{LEFT LOWER RIGHT UPPER}
 	\includegraphics[width=\textwidth, trim = 1cm  2cm 13cm 1cm]{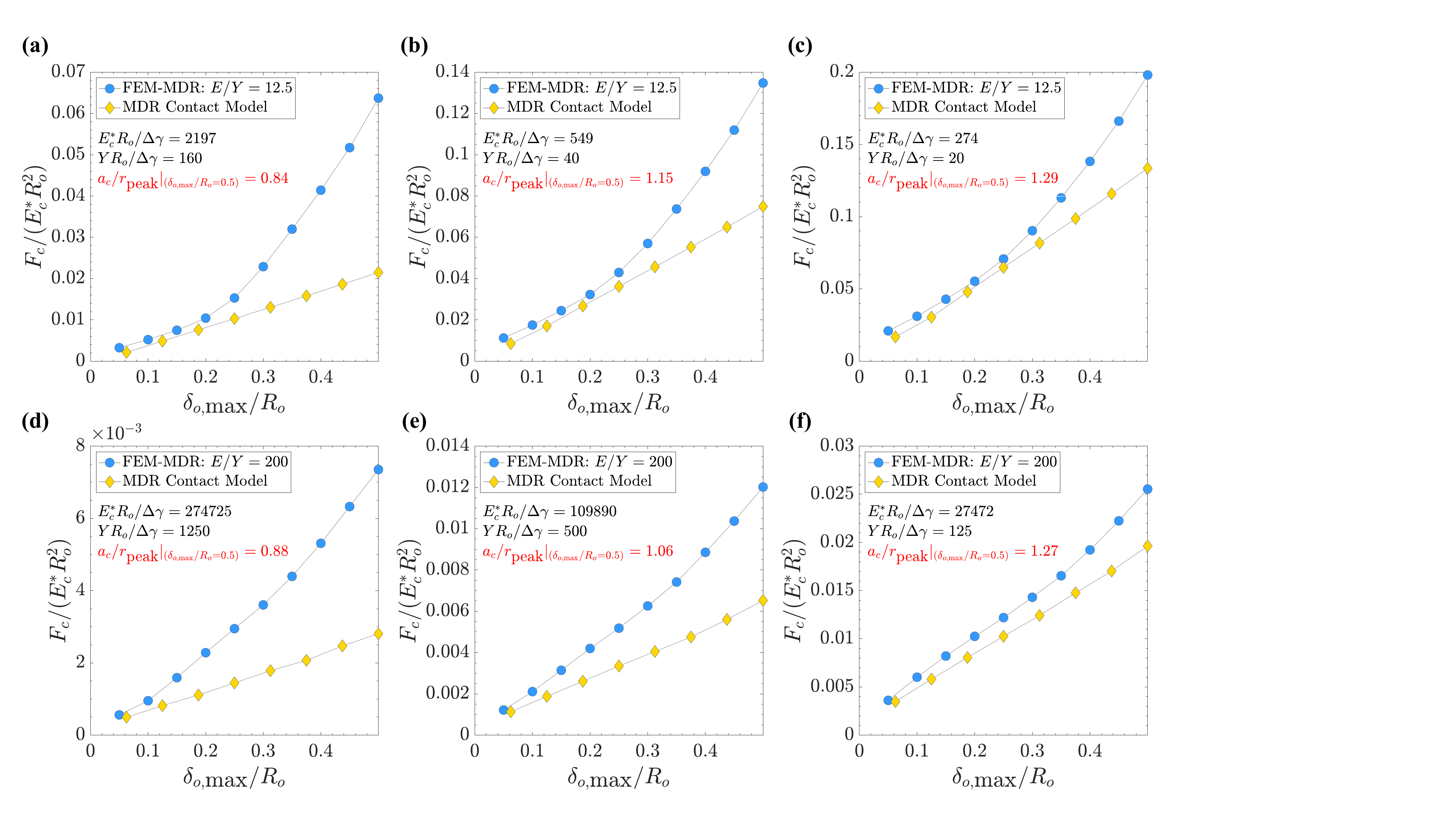}
 	\caption{Plots of the normalized critical force versus the normalized maximum displacement. Two curves are shown for comparison with the FEM-MDR indicated with blue circles and the adhesive MDR contact model shown with gold diamonds. The top row (a)-(c) corresponds to a fixed $E/Y = 12.5$, with an increasing value of $\Delta\gamma$ moving left to right. The bottom row (d)-(f) corresponds to a fixed $E/Y = 200$, again with an increasing value of $\Delta\gamma$ moving left to right. In both the top and bottom rows improving agreement is seen with increasing $\Delta\gamma$. This improvement is directly correlated to an increase in $a_c/r_\textrm{peak}$ (evaluated at $\delta_{o,\textrm{max}}/R_o$ = 0.5) indicated in red text on each plot. Each plot also contains its corresponding values for the dimensionless groups $Y R_o/\Delta \gamma$ and $E^*_c R_o/\Delta\gamma$, whose significance becomes clear later in the section.}
 	\label{critical_force}
 \end{figure*}

With the FEM profiles mapped, force-displacement curves for a given profile are created by considering different displacements followed by subsequent decompressions to the equilibrium position (i.e. the outer springs have reached the critical extensional length $\Delta l$). From the resulting force-displacement curves the critical force $F_c$ defined as the maximum tensile force sustained by the adhesive contact can be determined. This process is denoted as FEM-MDR. Analogous experiments are carried out using the adhesive MDR contact model, whereby the contact is loaded to the same $\delta_{o,\textrm{max}}/R_o$ associated with a specific FEM profile, then unloaded until separation occurs, taking note of $F_c$. From this information, the behavior of the critical force as a function of the maximum displacement, which tracks plastic deformation, can be measured. The specific results for the FEM-MDR and adhesive MDR contact model associated with $E/Y = 12.5$ and $E/Y = 200$ are shown in Fig.~\ref{critical_force}(a)-(c) and Fig.~\ref{critical_force}(b)-(f), respectively. 

For both $E/Y =12.5$ and $E/Y = 200$, we see that the match between the FEM-MDR and adhesive MDR contact model improves moving left to right. This improvement is caused by changing only a single parameter, the effective surface energy $\Delta\gamma$. The important consequence of increasing the effective surface energy is that the critical contact radius at separation $a_c$ for both the FEM-MDR and adhesive MDR contact model grows. Focusing specifically on $a_c$ for the MDR contact model we can track its value in relation to the radial location of $r_\textrm{peak}$, in particular we look at the value of $a_c/r_\textrm{peak}$ evaluated at $\delta_{o,\textrm{max}}/R_o = 0.5$. For both Fig.~\ref{critical_force}(a) and (d), we see that $a_c/r_\textrm{peak} < 1$, which results in early agreement of the predicted critical force before formation of the peaks but larger discrepancy after their formation. Moving to Fig.~\ref{critical_force}(b) and (e), we see that $a_c/r_\textrm{peak}$ takes on the values of $1.15$ and $1.06$, respectively. Correspondingly the disagreement in regards to the predicted critical force at larger $\delta_{o,\textrm{max}}/R_o$ decreases. Finally in Fig.~\ref{critical_force}(c) and (f), we see further increase in the values of $a_c/r_\textrm{peak}$, which again results in even better agreement between the two curves. This trend intuitively make sense for two reasons. First, as established, $a_c$ of the concave-convex indenter is always greater than or equal to $r_\textrm{peak}$, therefore for $a_c$ of the purely convex indenter to ever agree it must also exceed $r_\textrm{peak}$. Next, the more the critical radius exceeds the peaks the less important the specific details of the profile within the contact area become. 

These results lead to the following important conclusions: 

\begin{enumerate}
\item If $a_c/r_\textrm{peak} > 1$ the critical pull-off force predicted by the purely convex MDR contact model profile can be taken to reasonably well represent the related true critical pull-off force of the concave-convex indenter.
\item The more $a_c/r_\textrm{peak}$ exceeds $1$ the better the purely convex MDR contact model profile predicts the critical pull-off force of the true concave-convex indenter.
\end{enumerate}

\noindent If the contact was purely elastic the criterion for the validity of the adhesive MDR contact model could simply end at $a_c/r_\textrm{peak} > 1$. However, the materials under consideration are elastic-perfectly plastic, meaning that yielding needs to be accounted for. By design the contact model is not equipped to handle plastic deformation induced by a tensile state, imposing the requirement that no yielding may occur in tension. We recall at this point that the stress profile predicted by the JKR theory of adhesion tends towards an infinite tensile stress at the edge of the contact, immediately contradicting the requirement just stated. Therefore we adopt the fundamental assumption of linear elastic fracture mechanics (LEFM)~\cite{anand2023introduction}---that provided small-scale yielding (SSY) holds (i.e. the plastic zone size is small in comparison to the characteristic dimensions) then the solutions from linear elasticity can still be applied. The typical characteristic length scales in LEFM consist of a crack length and two distances to the free boundaries. In the case of an adhesive contact, there are only two characteristic length scales: the particle radius $R_o$ and contact radius $a$. No crack length is present, a result that falls naturally out of the expression for the plastic zone size which shows no dependency on either of the two characteristic lengths of the contact. In developing the criterion for SSY of the adhesive contact we drop $R_o$ as a characteristic length since under confined conditions the particles form continuous chains of length much greater than $a$. Additonally, we focus on the critical contact radius $a_c$ in place of the contact radius since it is the minimum allowable contact radius. Hence, the criterion for SSY becomes

\begin{equation} \label{SSY criterion}
    \textrm{if  } 2a_c \geq 15r_\textrm{Ip}, \textrm{  then SSY holds,}
\end{equation}

\noindent where $r_\textrm{Ip}$ is the plastic zone size. To develop an expression for $r_\textrm{Ip}$ we note that it can be written as a function of $E^*_c$, $\Delta \gamma$, and $Y$\footnote{Selection of these independent variables is not immediately obvious and is the result of prior knowledge about the functional form for $r_\textrm{Ip}$.}

\begin{equation} \label{rIp dimensional}
    r_\textrm{Ip} = f_1(E^*_c,\Delta \gamma,Y).
\end{equation}

\noindent Applying dimensional analysis to (\ref{rIp dimensional}) immediately results in the following form for $r_\textrm{Ip}$

\begin{equation} \label{rIp}
    r_\textrm{Ip} = \frac{E^*_c \Delta\gamma}{\pi Y^2}.
\end{equation}

\noindent Justification of the factor of $1/\pi$ is given alongside a more detailed derivation for $r_\textrm{Ip}$ in~\ref{Derivation of the plastic zone size}. From (\ref{rIp}) we see that the plastic zone size is indeed independent of any characteristic length.  

Thus, the criterion for validity of the adhesive MDR contact model can be succinctly written with only two inequalities

\begin{equation} \label{adhesion criterion}
    \textrm{if  } a_c/r_\textrm{peak} > 1 \textrm{ and } 2a_c/r_\textrm{Ip} \geq 15, \textrm{  then the adhesive MDR contact model is valid.}
\end{equation}

The conditions of (\ref{adhesion criterion}) are sufficient to understand whether the adhesive MDR contact model is going to predict reasonable results. However, because $a_c$ and $r_\textrm{peak}$ both depend on the amount of plastic deformation it becomes a difficult task to predict a priori, before simulating, whether the adhesive contact model will be valid. To achieve this predictive rather than reactive capability we apply dimensional analysis to reveal that the two crucial ratios: $a_c/r_\textrm{peak}$ and $a_c/r_\textrm{Ip}$ depend on the same dimensionless groups

\begin{equation} \label{ac rpeak dimensional}
    \frac{a_c}{r_\textrm{peak}} = f_i \left( \frac{\delta_{o,\textrm{max}}}{R_o},\frac{E^*_c R_o}{\Delta\gamma},\frac{YR_o}{\Delta\gamma} \right),
\end{equation}

\begin{equation} \label{ac rIp dimensional}
    \frac{a_c}{r_\textrm{Ip}} = f_{ii} \left( \frac{\delta_{o,\textrm{max}}}{R_o},\frac{E^*_c R_o}{\Delta\gamma},\frac{YR_o}{\Delta\gamma} \right).
\end{equation}

\noindent A study can then be constructed that sweeps over the various combinations bounded by the generally accepted limiting values of mechanical and geometric properties of natural material~\cite{ashby1995mechanical}. For each combination (\ref{adhesion criterion}) is evaluated to determine if the adhesive MDR contact model is valid. If it fails the test, the mode of failure is also cataloged. 
 
\begin{figure*} [!htb]
 	\centering
 	%\raggedright
 	% Trim{LEFT LOWER RIGHT UPPER}
 	\includegraphics[width=\textwidth, trim = 1cm  2cm 13cm 1cm]{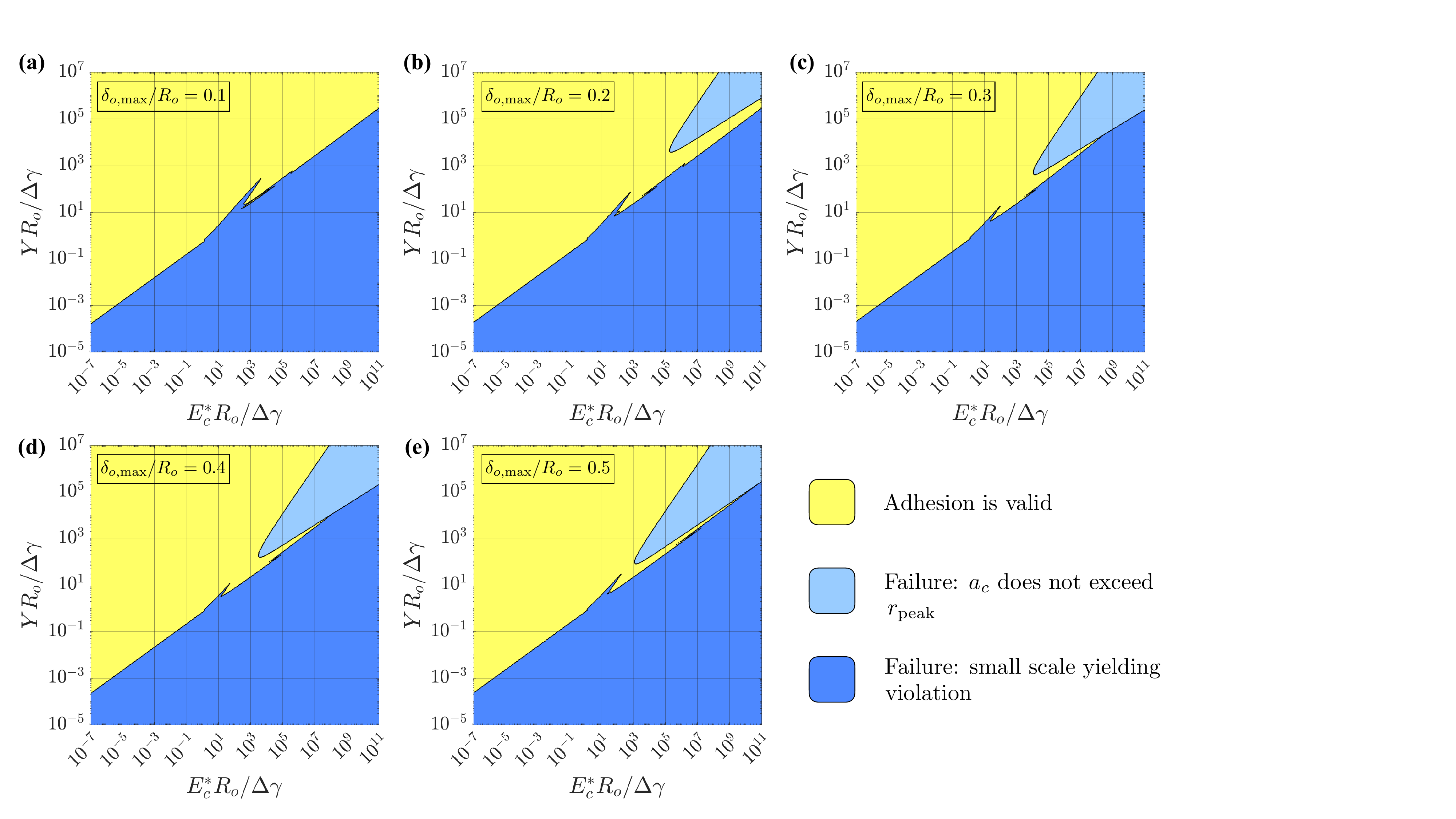}
 	\caption{Adhesion phase diagrams to determine if the adhesive MDR contact model will predict reasonable results. The vertical and horizontal logscale axes correspond to $Y R_o/\Delta \gamma$ and $E^*_c R_o/\Delta\gamma$, respectively. The color code is shared between all plots with yellow meaning adhesion is valid and with blue that it is invalid. The shade of blue identifies the exact mode of failure. The difference between plots (a)-(e) is the fixed value of $\delta_\textrm{max}/R_o$ that they correspond to, with (a) starting at $0.1$ going up by increments of $0.1$ to (e) at $0.5$.}
 	\label{adhesion_phase_diagrams}
 \end{figure*}

 The results of this study lead to the creation of adhesion phase diagrams detailing approximately when the adhesive MDR contact model is valid, as shown in Fig.~\ref{adhesion_phase_diagrams}(a)-(e). In all graphs the vertical and horizontal logscale axes correspond to $Y R_o/\Delta \gamma$ and $E^*_c R_o/\Delta\gamma$, respectively. The difference in each plot comes from the fixed value of $\delta_{o,\textrm{max}}/R_o$ considered beginning with $0.1$ for Fig.~\ref{adhesion_phase_diagrams}(a) and ending with $0.5$ 
 for Fig.~\ref{adhesion_phase_diagrams}(e). Each of the plots contains the same color code to indicate whether the adhesive contact model is valid (yellow) or invalid (blue). For the invalid case there are two shades of blue used to provide insight into the reasoning: the lighter blue is failure due to $a_c/r_\textrm{peak} < 1$ and the darker blue is failure due to small scale yielding. To use the phase diagrams first estimate the expected maximum displacement $\delta_{o,\textrm{max}}/R_o$ and match it to the closest corresponding plot in Fig.~\ref{adhesion_phase_diagrams}. Next, locate the point on the plot corresponding to $Y R_o/\Delta \gamma$ and $E^*_c R_o/\Delta\gamma$ for the given material to see whether the adhesive MDR contact model is valid.

\subsection{Derivation of the plastic zone size} \label{Derivation of the plastic zone size}

The derivation provided here for the plastic zone size $r_\textrm{Ip}$ is analogous to the presentation provided for linear elastic fracture mechanics in~\cite{anand2023introduction}. To begin, we imagine following the procedure outlined in Section \ref{Normal contact without adhesion}, whereby an adhesive indenter is compressed to a contact radius of $a$ and then retracted until the outer springs reach their critical extensional length $\Delta l$. We note that the normal stress $\sigma_\textrm{3D}$ (where we take tension as positive) along an adhesive contact, much like the force, can be seen as the summation of a non-adhesive (n.a.) and an adhesive retraction (a.r.) contribution. 

\begin{equation}
    \sigma_\textrm{3D}(r) = \sigma_\textrm{3D,n.a.}(r) + \sigma_\textrm{3D,a.r.}(r).
\end{equation}

\noindent For the JKR theory of adhesion it is known that the adhesive retraction contribution corresponds exactly to the normal stress profile beneath an adhesive cylindrical punch of radius $a$ displaced in tension by a distance $\Delta l$

\begin{equation}
    \sigma_\textrm{3D,a.r.}(r) = \frac{E^*_c \Delta l}{\pi\sqrt{a^2-r^2}}.
\end{equation}

\noindent Because $\sigma_\textrm{3D,n.a.}(r)$ is required to vanish near the edge of the contact, whereas $\sigma_\textrm{3D,a.r.}(r)$ tends to an infinite tensile stress, it is reasonable to approximate the normal stress profile near the edge of the contact as

\begin{equation}
    \sigma_\textrm{3D}(r) \approx \frac{E^*_c \Delta l}{\pi\sqrt{a^2-r^2}}. 
\end{equation}

\noindent Taking $r = a-\Delta$, where $\Delta$ is a distance much smaller than $a$ we can write  

\begin{equation}
    \sigma_\textrm{3D}(a-\Delta) = \frac{E^*_c \Delta l}{\pi\sqrt{a^2-(a-\Delta)^2}}, 
\end{equation}

\noindent which can then be converted to

\begin{equation}
    \sigma_\textrm{3D}(a-\Delta) = \frac{E^*_c \Delta l}{\pi\sqrt{2a\Delta-\Delta^2}}. 
\end{equation}

\noindent In the limit as $\Delta$ goes to zero the second order term drops leaving 

\begin{equation} \label{sigma3D limit small Delta}
    \lim_{\Delta \to 0} \sigma_\textrm{3D}(a-\Delta) = \frac{E^*_c \Delta l}{\pi\sqrt{2a\Delta}}. 
\end{equation}

\noindent Recalling the expression for the critical extensional length $\Delta l$ (\ref{Deltal}) and substituting it into (\ref{sigma3D limit small Delta}) we arrive at

\begin{equation} \label{sigma3D SSY form}
    \sigma_\textrm{3D}(a-\Delta) = \frac{\sqrt{2E^*_c\Delta\gamma}}{\sqrt{2\pi\Delta}}. 
\end{equation}

\noindent Equating (\ref{sigma3D SSY form}) to the yield stress and solving for $\Delta$ gives the expression for the plastic zone size

\begin{equation} 
    r_\textrm{Ip} = \frac{E^*_c\Delta\gamma}{\pi Y^2}. 
\end{equation}

A slightly more rigorous analysis can be carried out that does not completely ignore $\sigma_\textrm{3D,n.a.}(r)$, but rather accounts for it approximately by assuming its value is constant and equal to the average normal stress caused in compression $p_c = F_\textrm{n.a.}/(\pi a^2)$. Taking this into account leads to the slightly modified expression for $r_\textrm{Ip}$

\begin{equation} 
    r_\textrm{Ip} = \frac{E^*_c\Delta\gamma}{\pi (Y+p_c)^2}, 
\end{equation}

\noindent which predicts a smaller plastic zone size since $p_c$ taken to be strictly greater than zero.

\end{document}